\documentclass[a4paper,12pt, openany]{book}

\usepackage{graphicx}
\usepackage[affil-it]{authblk}
\usepackage{amsmath,amssymb,mathrsfs,eucal,amsthm,braket}			
\usepackage{enumerate}
\usepackage{fancyhdr}

\begin{document}
\title{Decoherence and definite outcomes}
\author{Klaus Colanero}
\affil{Universit\`a degli Studi di Firenze}
\date{}

\maketitle

\chapter*{}
\null
\vfill
\begin{center}
This work by Klaus Colanero is licensed under a Creative Commons by Attribution license: http://creativecommons.org/licenses/by/3.0/ \\
\bigskip
\includegraphics[scale=1,keepaspectratio=true]{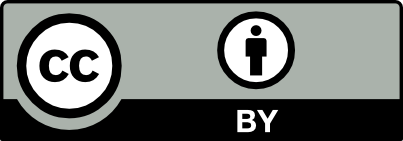} \\
\end{center}

\tableofcontents

\chapter*{Acknowledgements}

I would like to thank my supervisors, Professor Roberto Casalbuoni and Professor Elena Castellani, for their helpful suggestions and constructive criticism.

I would like to acknowledge insightful and clarifying discussions with Dr. Max Schlosshauer, Prof. Claus Kiefer and Prof. Osvaldo Pessoa.

Credit is also due to the teachers of the various Logic courses who, with their competence and teaching skills, allowed me to overcome my initial impression of having landed in the wrong place. I now have a stronger foundation and better tools to study Philosophy of Science.

Special thanks go to my classmates who benevolently welcomed this much older lad among their peers.

I thank my housemates who gracefully tolerated my studying style peculiarities and occasionally shared some philosophical discussion on the chief systems of the world.

Finally, I feel obliged to thank my parents and my sisters who trusted I was still a sane person when undertaking these further studies well beyond my years of youth.

\chapter*{}

\emph{``The continuing dispute about quantum measurement theory...is between people who view with different degrees of concern or complacency the following fact: so long as the wave packet reduction is an essential component, and so long as we do not know exactly when and how it takes over from the Schr\"odinger equation, we do not have an exact and unambiguous formulation of our most fundamental physical theory.''} \\

\hfill J.S.~Bell

\chapter{Introduction}

In the last few years the general notion of \emph{decoherence} in quantum mechanics has become increasingly common among physicists, philosophers of physics and quantum information scientists. And rightly so, because it represents both a further application of the predictive and explicative power of quantum theory, and an attempt to break the stalemate situation with respect to the interpretation of quantum mechanics.
Powerful as it might be, however, the decoherence programme has not solved the measurement problem yet. Specifically, and contrary to some claims, it has not solved the definite outcomes problem, better known as the problem of the wave function collapse.
Due to the wide scope of the decoherence programme, which could be summarized as the attempt to recover the classical phenomena from quantum physics, the problem of definite outcomes happens to be often confused with other, loosely related, issues.
Such a confusion is the main motivation of this study.

This thesis has three aims: 
\begin{itemize}
 \item to clarify in detail the relation between the decoherence mechanism and the problem of definite outcomes,
 \item to dispel common misconceptions about the measurement problem in quantum mechanics, and
 \item to present some recent alternative approaches in the quest for a satisfactory solution of the definite outcomes problem.
\end{itemize}

Since the last decade, it has become quite common, especially among physicists, to think that the successes of the decoherence programme in explaining the emergence of classical phenomena have solved the measurement problem.
Such an idea is unfortunately incorrect because, while the decoherence mechanism accounts very well for the disappearing of interference in macroscopic systems and provides an enlightening understanding for the appearance of preferred robust pointer states, it does not offer, by itself, an explanation for the apparent collapse of the state vector.

At a time when the decoherence programme is becoming well defined both in its scope and in its physical and mathematical foundations, it seems necessary to identify in detail the place of definite outcomes within the decoherence framework. This appears all the more relevant considering the promising and fascinating developments in experimental techniques for macroscopic quantum state control on the one hand, and in quantum information and computation on the other.

Also, it is still difficult to overstate the importance of reducing misconceptions about the measurement problem in quantum mechanics.
After about eighty years since the formulation of the Schr\"odinger equation and the unprecedented success of quantum theory, it is quite peculiar that the problem of definite outcomes is still considered by most either a metatheoretical issue to be dealt with by means of a consistent interpretation, or a problem already solved for all practical purposes by the decoherence programme.
As a matter of fact, not all practical issues are addressed by the decoherence programme.
At the same time it is legitimate and reasonable to try and look for solutions to the problem within physics itself, either through an even finer grained analysis of system--apparatus--environment dynamics, or through the introduction of new physical laws whose consequences can be subjected to experimental test.

With regard to the physics based approach to the problem, theoretical constraints to its viability are examined and recent proposals for feasible experimental tests are introduced.

What is not in the scope of this work is a complete review of the results and successes of the decoherence programme in explaining the emergence of classical phenomena from quantum dynamics.
Instead, as the title suggests, the focus will be specifically on the relation between decoherence and definite outcomes.
In the same spirit, the different interpretations of quantum mechanics are not presented in a comprehensive manner, but they are analyzed from the specific point of view of the definite outcomes problem and with respect to the theoretical and experimental results of the decoherence programme.

In pursuing the aims listed above, philosophical assumptions and implications of each approach are explicitly analyzed in order to have a clear distinction between physical, logical and metaphysical issues involved.
Weaknesses and merits of the various theoretical and interpretive approaches are highlighted.
 
This thesis is organized as follows. In Chapter~\ref{ch:the_problem} the definite outcomes problem is defined and analyzed. Recent experiments on superposition states of macroscopic systems are examined and their implications for the standard interpretation pointed out.
The basic ideas of the decoherence formalism are also introduced, in order to provide a clear foundation for the subsequent considerations.

In Chapter~\ref{ch:interpretations} the most commonly adopted interpretations are reviewed with particular attention to the issue of definite outcomes and the application of the decoherence programme. In order to underline the different approaches with respect to state vector reduction, the interpretations are grouped in theory extending solutions and interpretive solutions.

Chapter~\ref{ch:decoherence} is devoted to the theory of decoherence. Attention is put on identifying the essential and general aspects within the great formal diversity in the application of the theory to different systems. The analysis is again focused on clarifying connections and implications for the definite outcomes problem. Misconceptions with regard to merits and limitations of the decoherence programme are addressed.

Environment-induced and gravity-induced collapse proposals are discussed in Chapter~\ref{ch:future}. 
The impossibility of a ``for all practical purposes'' solution of the definite outcomes problem, based on an environment-induced collapse, is discussed.

\chapter{The problem}
\label{ch:the_problem}

From a very general and interpretation agnostic point of view, the problem of definite outcomes can be described as follows. According to quantum theory a physical system can be in any of the infinitely many possible states represented by a vector in the Hilbert space associated to the system. As a matter of fact, any time the system is observed it is found in one of only a few possible states: the ones corresponding to a definite value of the physical quantity which is effectively being measured through the observation. If, for example, the state of a spin $1/2$ system is $\alpha \ket{\uparrow} + \beta \ket{\downarrow}$, an observer measuring the value of the spin will either find the system in the up state, $\ket{\uparrow}$, or in the down state, $\ket{\downarrow}$. On the contrary, even though it is not clear what would constitute the observation of states such as $\alpha \ket{\uparrow} + \beta \ket{\downarrow}$, it appears that the system, upon observation, is never found in this kind of state.

Given such a state of affairs, one has to work out whether the explanation of the definite outcomes lies in (1) the addition of one or more postulates to the basic building blocks given by the unitary and deterministic Schr\"odinger evolution, in (2) a suitable interpretation of the relation between the mathematical ingredients of the theory and what happens in the world, or even (3) directly within the unitary evolution of the wave function of the universe.

Obviously, a suitable interpretation is always necessary for any physical theory to be able to connect the formalism to the world, but such an interpretation might also include propositions which are later found to be redundant because they can be derived by the mathematical formalism and by the other interpretative postulates. In this regard, as it will become clear in the following sections, a typical example is Bohr's postulate that the measurement apparatus is governed by classical physics and not by quantum mechanics.

In the following section, the above mentioned issues will be presented in a formal and detailed manner. Moreover, the basic ideas at the heart of two different approaches to the measurement problem will be presented.

\section{Superposition states and measurement outcomes}

While in classical mechanics the state of a system is represented by a pair of conjugate variables $\{Q, P\}$, e.g. position and momentum of an object, in quantum mechanics it is represented by a vector, $\ket{\psi}$, in the Hilbert space associated to the system.
The vector $\ket{\psi}$ can be written as a superposition of a set of vectors which forms a complete basis for that Hilbert space. The eigenstates of a Hermitian operator constitute such a complete set and thus we can for example write:
\begin{equation}
    \ket{\psi} = \sum_k \alpha_k \ket{E_k} = \int \psi (x) \ket{x} dx ,
\label{psi_hermitian_basis}
\end{equation}
where $\ket{E_k}$ is the $k$th eigenvector of the total energy operator, i.e. the Hamiltonian operator of the system, and $\ket{x}$ is the eigenvector of the position operator corresponding to position $x$. The coefficients $\alpha_k$ and $\psi(x) dx$ contain the information about the relative weight and phase of each basis vector. In Eq.(\ref{psi_hermitian_basis}) the state vector $\ket{\psi}$ has been expressed in two different basis in order to show that it can analogously be expressed in terms of a set of eigenvectors of any Hermitian operator.

An arbitrary state $\ket{\psi}$ will always correspond to a superposition in some vector basis. If the system is, for example, in an eigenstate $\ket{E_i}$ of the Hamiltonian, $\hat H$, in general it will not be in an eigenstate of the momentum operator, $\hat p$, but it will be in a superposition of momentum eigenstates, $\int \braket{p|E_i} \ket{p} dp$.

As it has been mentioned, it is necessary that the mathematical formalism be complemented with an interpretation, which specifies how to relate state vectors and the operations on them with the actual properties of the physical system of interest.
As a basic interpretative prescription it seems natural to connect the eigenstates of a self-adjoint operator to a well defined property of the associated physical system. That is, more explicitly, if the state of the system is described by the eigenstate $\ket{p}$ of the momentum operator, then we expect it to possess a well defined value of the momentum $p$. Or, in other words, that the value $p$ of the physical quantity momentum is an objective property of the system. Such a prescription corresponds to the so called eigenstate-eigenvalue link, or e-e-link, and it appears consistent with experimental observations.

The first non-trivial issue, and a main aspect of the definite outcomes problem, is that a priori it is not clear what physical properties should be associated to a system described by a generic superposition. For example, it is also not clear a priori what should be the meaning of measuring the energy of a system which is not in an eigenstate of the Hamiltonian operator, such as $\sum_k \alpha_k \ket{E_k}$.

In order to build a clear picture of the problem it is helpful to consider one of the simplest, yet powerfully explicative, quantum mechanical systems: a spin $1/2$ system.
The two eigenstates of the $S_z$ operator, $\ket{+}_z$ and $\ket{-}_z$, constitute a complete set of basis vectors for the two-dimensional Hilbert space associated to the system.

If the system, say a single electron \cite{chesnel_young-type_2007}, is prepared in such a way that it is in one of the eigenstates of $S_z$, then it is more than reasonable to expect that a Stern-Gerlach apparatus \cite{gerlach_magnetische_1922}, set up in the $z$-direction, will provide the correct outcome for the eigenvalue of $S_z$, that is, the one corresponding to the actual state of the electron.
But what if the state of the electron is $\alpha \ket{+}_z + \beta \ket{-}_z$? What should we expect to see on the screen which records the electron position after passing through the Stern-Gerlach apparatus?
At this stage, without additional interpretative prescriptions, the answer is not obvious. And this is the first issue at the heart of the measurement problem in quantum mechanics.

\subsection{Inconsistent epistemic probability interpretation}
\label{subsec:inconsistent_probabilities}
As a matter of fact, performing the above mentioned experiment with electrons in a superposition state, one sometimes observes a spot on the position corresponding to $\ket{+}_z$, other times a spot on the position corresponding to $\ket{-}_z$. Remarkably, it also happens that the relative frequencies of hits on the upper and lower spots converge to $|\alpha|^2$ and $|\beta|^2$ respectively.
This observation immediately suggests a straightforward but inconsistent interpretation of superpositions: states for which the value of the observable of interest is not known, but only the probabilities of occurrence are known. Such an epistemic probability interpretation is inconsistent, because in general it does not produce correct predictions for the average values of observables. This becomes particularly clear through the formalism of density matrices, which allows to deal with pure states as well as actual statistical mixtures of states in the same framework.
In such a framework the state of a system is described by a density matrix $\rho$ which, analogously to the state vector, contains all the information about the system. The average value of any observable $\hat A$ results from the trace operation on the matrix $\hat A \rho$.

According to the straightforward, naive, interpretation of superposition states, there is a $|\alpha|^2$ probability that the system is in the spin-up state and a $|\beta|^2$ probability that it is in the spin-down state. Such a situation is described by the following statistical mixture density matrix:
\begin{equation}
\rho_{\mbox{mix}} = \left(
\begin{array}{cc}
|\alpha|^2 & 0\\
 0 & |\beta|^2
\end{array}
\right)
    \label{statistical_mixture_density_matrix}
\end{equation}
We can calculate the average value of the observable associated to $S_z$ by computing $\mbox{Tr}(S_z\rho_{\mbox{mix}})$.
The operator $S_z$ in the chosen basis is represented by the following diagonal matrix:
\begin{equation}
S_z = \frac{\hbar}{2}\left(
\begin{array}{cc}
1 & 0\\
 0 & -1
\end{array}
\right) .
    \label{S_z_matrix}
\end{equation}
And its expectation value is
\begin{equation}
<S_z> = \mbox{Tr}(S_z\rho_{\mbox{mix}}) = \sum_i (S_z\rho_{\mbox{mix}})_{ii} = \frac{\hbar}{2} \left( |\alpha|^2 - |\beta|^2 \right).
    \label{trace_rho_mix}
\end{equation}
The same result is obtained by computing $<S_z>$ with the pure state density matrix $\rho_{\mbox{pure}}$, related to the pure state $\ket{\psi} = \alpha \ket{+}_z + \beta \ket{-}_z$ through the following definition:
\begin{equation}
\rho_{\mbox{pure}} = \ket{\psi} \bra{\psi} = \left(
\begin{array}{cc}
|\alpha|^2 & \alpha \beta^* \\
 \alpha^* \beta & |\beta|^2
\end{array}
\right) .
    \label{pure_state_density_matrix}
\end{equation}
Things are different, however, if other observables are considered. A straightforward choice is to compare the average value of the $S_x$ operator (or analogously $S_y$):
\begin{equation}
S_x = \frac{\hbar}{2}\left(
\begin{array}{cc}
0 & 1\\
 1 & 0
\end{array}
\right) .
    \label{S_x_matrix}
\end{equation}
It is easy to verify that the average value in the mixed state, which follows from the naive, epistemic probabilities, interpretation is 
\begin{equation}
<S_x>_{\mbox{mix}} = \mbox{Tr}(S_x\rho_{\mbox{mix}}) = 0, 
\label{S_x_mix_average}
\end{equation}
while, on the other hand, the average value in the pure state $\alpha \ket{+}_z + \beta \ket{-}_z$ is
\begin{equation}
<S_x>_{\mbox{pure}} = \mbox{Tr}(S_x\rho_{\mbox{pure}}) = \frac{\hbar}{2} \alpha\beta^* + \alpha^*\beta =\frac{\hbar}{2} \mbox{Re}(\alpha\beta^*).
\label{S_x_pure_average}
\end{equation}

As it happens, carefully performed experiments are consistent with the latter result, not with the first. The naive, epistemic probabilities, interpretation, is inconsistent.

\subsection{Non-diagonal terms and interference}
\label{subsec:non_diagonal_terms}
The above considerations, which are standard in any introductory course in quantum mechanics, are presented here in order to stress from the very beginning the central place and the non-trivial nature of the definite outcomes problem. Namely, even if one simply accepts that superposition states do not belong to the world of observations, the relation between state vectors and the actual observed phenomena is a matter of careful interpretation.

Before proceeding to define and address the core issue of the definite outcomes problem, it is instructive to make one more remark about the spin $1/2$ system considered above: the difference between the two $<S_x>$ results originates from the presence of the non-diagonal terms in the pure state density matrix. Such terms, which in a Stern-Gerlach type experiment manifest themselves in a relatively subtle 
way in the averages $<S_x>$ and $<S_y>$, show their effect drammatically in interference experiments, such as with the double slit setup.

In a double slit interference experiment the state of the system, a particle emerging from a barrier with two openings, can be described by the state vector $\ket{\psi} = c_1 \ket{\psi_1(t)} + c_2 \ket{\psi_2(t)}$, where $\ket{\psi_1(t=0)}$ and $\ket{\psi_2(t=0)}$ represent the state of a particle well localized around the first and the second slit respectively. As it is shown in quantum mechanics textbooks, the square modulus of the wave function $\psi(x,t)$, which as in Eq.(\ref{psi_hermitian_basis}) is the coefficient of the position eigenvector $\ket{x}$, acquires a typical, modulated, pattern with several minima and maxima, instead of just the two peaks one would expect from particles passing through one or the other slit. Such a pattern is directly related to the mixed terms, $c_1 c_2^*\braket{\psi_2(t)|x}\braket{x|\psi_1(t)}$ and $c_1^* c_2\braket{\psi_1(t)|x}\braket{x|\psi_2(t)}$, which appear in the expression for $|\psi(x,t)|^2$. These also correspond to the non-diagonal terms in the density matrix.

The above considerations, regarding the non-diagonal, or interference, terms, though not evidently central to the discussion on the definite outcomes problem, constitute one of the main issues addressed by the decoherence programme. Also they are too often either considered the only measurement problem or mixed up with the issue of definite outcomes to produce a confused picture of the measurement problem.
It is thus crucial to have a clear understanding of this aspect of quantum mechanical states.

In the following section the general approach to the description of the measurement process within the  quantum mechanical framework is presented, along with the issues it raises.

\section{Von Neumann measurement theory}
\label{sec:von_neumann}
The process of measuring the properties of a physical system is perhaps the most subtle phenomenon to be dealt with in physics, at least because ultimately it has to deal, one way or another, with the properties and features of an observer. Even with such complex problems, however, it makes sense and is always fruitful to adopt an analytical approach which starts from undisputable or extremely reasonable assumptions. Regardless of the peculiarities of the observer, (1) one such assumption is that the measurement process  of a quantum system involves the interaction between the system and the measurement apparatus. Moreover (2) the apparatus is supposed to be composed of smaller parts whose behaviour is described by the laws of quantum mechanics, specifically the Schr\"odinger equation. Finally, (3) the quantum system to be observed interacts with the macroscopic apparatus through its constituent quantum parts.

The just mentioned ``extremely reasonable'' assumptions are at the basis of the Von Neumann scheme for modelling the measurement process.
According to this scheme there is a quantum system $Q$ with an associated Hilbert space $\mathcal{H}_Q$ and a Hamiltonian $H_Q$, and there is a measurement apparatus $M$ designed to measure the value of an observable $\hat A$ of the system $Q$. The dynamics of the macroscopic apparatus $M$ also depends, through the Schr\"odinger equation, on a quantum Hamiltonian $H_M$, and its associated Hilbert space is denoted by $\mathcal{H}_M$. The measurement takes place when the two systems, $Q$ and $M$, interact by means of an interaction Hamiltonian $H_I$ which can be considered different from zero only for a finite and relatively short time interval, the measurement time.
The total system, comprised of $Q$ and $M$, is described by a state vector $\ket{\Psi}$ belonging to the Hilbert space given by the tensor product of the two spaces, $\mathcal{H}_Q \otimes \mathcal{H}_M$.

The apparatus features a pointer from which the outcomes can be read out. The state of the apparatus can be represented by a vector $\ket{\Phi} = \sum_{X,\vec{y}} c_{X,\vec{y}} \ket{X,\vec{y}}_M$, where $X$ is the eigenvalue of the pointer position operator and $\vec{y}$ are the eigenvalues related to the other observables necessary to completely describe the state of the apparatus.

As a first and basic requirement, the apparatus should be able to reveal the e-e-link, that is: for each  distinct eigenstate of the operator $\hat A$, the pointer will be in a distinct eigenstate of the position operator $\hat X$.
Before the measurement, system $M$ is supposed to be in a ``ready'' state $\ket{X_0}_M = \sum_{\vec{y}} c_{X_0,\vec{y}} \ket{X_0,\vec{y}}_M$. Consequentely, if the state of $Q$ is $\ket{a_1}_Q$, the total system state vector before the measurement is $\ket{\Psi}_{in} = \ket{a_1}_Q \ket{X_0}_M$. After the measurement it should be $\ket{\Psi}_{fin} = \ket{a_1}_Q \ket{X_1}_M$. In other words it is necessary that the interaction Hamiltonian $H_I$ produces the following transition:
\begin{equation}
\ket{a_k}_Q \ket{X_0}_M \Longrightarrow \ket{a_k}_Q \ket{X_k}_M ,
    \label{Von_Neumann_eelink}
\end{equation}
where $\ket{X_k}_M$ represents a specific eigenstate of $\hat X$ which can be univocally associated to $\ket{a_k}_Q$.
This is possible if the interaction Hamiltonian is of the following type:
\begin{equation}
H_I = g(t) \hat P \hat A ,
    \label{Von_Neumann_H_I}
\end{equation}
where $\hat A$ is the operator associated to the observable to be measured, $\hat P$ is the adjunct operator of $\hat X$, and $g(t)$ is a real-valued function defined as
\begin{equation}
g(t) = \left\{
\begin{array}{ll}
g & t_0 < t < t_0 + \tau \\ 
0 & \mbox{otherwise}
\end{array}
\right.
    \label{Von_Neumann_g}
\end{equation}
During the measurement the state evolution is governed by the total Hamiltonian, $H = H_Q + H_M + H_I$, through the unitary evolution operator $U(t) = \exp{-\frac{i}{\hbar} H t}$. If, however, the interaction is chosen strong enough to dominate, for short measurement times $\tau$, the time evolution, then the effect of the measurement interaction can be calculated, with good approximation, by neglecting the contribution of $H_Q+H_M$:
\begin{equation}
\ket{\Psi(t_0)}=\ket{a_k}_Q \ket{X_0}_M \Longrightarrow \ket{\Psi(t_0+\tau)}= e^{-\frac{i}{\hbar} H_I \tau} \ket{\Psi(t_0)} .
    \label{Von_Neumann_Ut}
\end{equation}
Noting that for conjugate operators, like $\hat X$ and $\hat P$, holds that
\begin{equation}
e^{-\frac{i}{\hbar} \hat P a} \ket{X} = \ket{X+a}
    \label{translation_operator}
\end{equation}
and applying the operator $\exp{-\frac{i}{\hbar} g \hat P \hat A \tau}$ first to $\ket{a_k}$, then the final state $e^{-\frac{i}{\hbar} H_I \tau} \ket{\Psi(t_0)}$ can be calculated as
\begin{equation}
e^{-\frac{i}{\hbar} g \hat P \hat A \tau} \ket{a_k}_Q \ket{X_0}_M = \ket{a_k}_Q e^{-\frac{i}{\hbar} g \hat P a_k \tau} \ket{X_0}_M = \ket{a_k}_Q \ket{X_0+a_k g\tau}_M .
    \label{Von_Neumann_H_I_eelink}
\end{equation}
The pointer has moved from $X_0$ to $X_k = X_0+a_k g\tau$

The above derivation shows that a measurement apparatus can be designed in such a way that, for each measured system eigenstate $\ket{a_k}$, the measurement process brings the pointer from the ``ready'' state $\ket{X_0}_M$ to a distinct pointer position eigenstate $\ket{X_0+a_k g\tau}_M$. It is possible to conventionally associate the eigenvalue $a_k$ of the operator $\hat A$ to this pointer eigenstate. The apparatus is thus consistent with the e-e-link requirement.

Having addressed the interaction mechanism between system and apparatus, it is now possible to derive the final state of the apparatus after the measurement on a generic state of the system $Q$, $\ket{\psi}_Q = \sum_k c_k \ket{a_k}$. Thanks to the linearity of the time evolution operator $U(t)$, it is straightforward to find
\begin{equation}
\ket{\psi}_Q \ket{X_0}_M \Longrightarrow \sum_k e^{-\frac{i}{\hbar} g \hat P \hat A \tau} c_k \ket{a_k}_Q \ket{X_0}_M = \sum_k c_k \ket{a_k}_Q \ket{X_0+a_k g\tau}_M .
    \label{Von_Neumann_superposition}
\end{equation}
The apparatus is predicted to evolve from the measurement ready position to a superposition of pointer eigenstates!
The superposition of the initial state of system $Q$ propagates to the macroscopic apparatus $M$ through entangling, caused by the measurement interaction itself. 

As a matter of fact, however, situations in which the pointer of a macroscopic apparatus appears in a superposition state seem never to occur. On the contrary, definite pointer positions are always observed.
This, cutting the frills, is the central issue about the definite outcomes problem.

\subsection{Standard Copenhagen position}
\label{subsec:Copenh_decoh}
Highlighting the two main results presented so far, the definite outcomes problem can be stated in more detail as follows:
\begin{itemize}
    \item computation of statistical average values of observables show that superposition states are indeed different from classically not well defined states, such as statistical mixtures, but
    \item such states, predicted to occur in macroscopic as well as microscopic systems, are never directly observed in single measurements.
\end{itemize}

Before proceeding further, it is necessary to take into account another indirectly relevant phenomenon: the apparent lack of interference in experiments with macroscopic systems. In the previous section it has been stated that carefully performed experiments agree with the calculations based on a pure state density matrix, e.g. the result of Eq.(\ref{S_x_pure_average}). Experiments on systems comprised of more than a few tens of atoms, however, provide a more puzzling picture of the physical phenomena. Contrary to what is consistently found for photons and electrons, statistical average values of observables turn out to agree with the predictions of a statistical mixture of definite states, like for example in Eq.(\ref{S_x_mix_average}). Likewise, interference experiments with systems larger than a few atoms do not show interference patterns. 

The so called Copenhagen interpretation of quantum mechanics is the first successful attempt at building a consistent framework to relate the results of quantum theory to the experiments' outcomes. Here by successful attempt it is meant consistent within the experimental accuracy reached at the time the interpretation was proposed.
The Copenhagen interpretation can be summarized in the following statements:
\begin{enumerate}[(i)]
    \item The state vector $\ket{\psi}$ contains all the information about the system.
    \item The time evolution of the vector $\ket{\psi}$ is governed by the Schr\"odinger equation.
    \item An observer can access the system only through a macroscopic apparatus which is subject to the laws of classical physics.
    \item The measurement of the observable $\hat A$ changes the state of the system from $\ket{\psi}$ to $\ket{a_k}$, one of the eigenstates of $\hat A$, breaking the unitary evolution of the Schr\"odinger equation.
    \item The probability that the measurement process brings the system in the state $\ket{a_k}$ is given by $|c_k|^2$, the square modulus of the $k$th coefficient in the expansion $\ket{\psi} = \sum_i c_i \ket{a_i}$.
\end{enumerate}
Notice that statement (v) implies that, when the system is in an eigenstate of an observable $\hat A$, it will be found in that state with probability $1$. As a consequence the corresponding eigenvalue is an objective property of the system.

The lack of interference for macroscopic objects and the problem of definite outcomes are addressed by means of statements (iii) and (iv). Postulate (iii) tells why quantum effects are not observed macroscopically, and postulate (iv) why specific, one-valued, outcomes are observed instead of the predictions of Von Neumann measurement scheme.
In other words, it is postulated that the world is governed by different laws: quantum mechanics for the microscopic world, and classical physics for the macroscopic, directly accessible, world. Moreover, even though the microscopic system can actually be in a superposition state, in its interaction with a macroscopic apparatus, perhaps the observer herself, it will necessarily ``collapse'' to an eigenstate corresponding to the observed physical quantity.
The boundary between microscopic and macroscopic regime is not specified.

In the next chapter the Copenhagen interpretation will be analyzed in more detail, but it can be safely said that, as puzzling and vague as it may seem, it is not in contradiction with the experimental observations performed at the time it was proposed.

The first question that naturally comes to mind is whether statement (iii) correctly describes the physical state of affairs about macroscopic systems, or whether it is an approximation of the actual quantum dynamics of complex, many particle systems. Equivalently, from a more formal point of view, it is meaningful to ask whether statement (iii) is necessary for a complete interpretation of quantum theory, or it can actually be derived through the careful application of the other four statements to the study of macroscopic systems.

The above questions have been addressed theoretically starting from the 1970's \cite{zeh_interpretation_1970, zeh_toward_1973} and experimentally only after the 1990's \cite{borde_molecular_1994, schollkopf_nondestructive_1994} due to the technical difficulties involved. 
The basic idea at the heart of the theoretical approach is that the closed system idealization, usually employed in modelling the system under study, is often unsuitable. The reason is that, in the general case, the interactions with particles in the environment, such as photons, electrons or atoms, and even the interaction with the internal degrees of freedom, greatly affect the dynamics of the system and thus cannot be neglected. 
The wide claim, that the whole decoherence programme is committed to prove, is that the interaction of a 
system with a large number of particles, whose dynamical details are not completely controlled by the observer, modifies the system's behaviour in such a way that it agrees with the laws of classical physics. Such a general aim is often described as the explanation for the emergence of classicality from the quantum world.

As already mentioned, however, a complete investigation of the decoherence programme is not in the scope of the present work. On the contrary, in order to clarify the connections between decoherence and definite outcomes, it can be more fruitful to focus on the relevant aspects. One of them is the explanation of the lack of interference in macroscopic objects. The relevance of such issue to the definite outcomes problem lies in the fact that, if macroscopic superpositions are allowed, then the Von Neumann measurement scheme is valid at least within some, yet to be defined, domain.

\subsection{General decoherence proposal}
\label{subsec:general_decoherence_approach}

In Chapter~\ref{ch:decoherence} the decoherence approach will be presented in detail, but it is useful to anticipate here its success in explaining the fast decay to zero of the non-diagonal terms in the density matrix of the system under investigation. This result is arrived at by starting the study from the pure state density matrix of the ``whole'' physical system: system plus environment. As already mentioned, in Section~\ref{subsec:inconsistent_probabilities} the average value of an observable $\hat O$ can be computed by performing the trace of the matrix $\hat O \rho$. In the case of a complex system, that is a system whose Hilbert space $\mathcal{H}$ is the tensor product of several subsystems Hilbert spaces, e.g. $\mathcal H = \mathcal{H}_a \otimes \mathcal{H}_b$, if the observable $\hat O$ is associated to subsystem $b$ only, the trace operation can be performed in steps:
\begin{equation}
<\hat O> = \mbox{Tr}\left( \hat O \rho \right) = \mbox{Tr}_b \left(\mbox{Tr}_a (\hat O \rho) \right) = \mbox{Tr}_b \left( \hat O \rho_{a r} \right),
    \label{reduced_density_matrix}
\end{equation}
where $\rho_{ar}$ indicates the matrix obtained by performing a partial trace over the degrees of freedom of system $a$ only. The matrix thus obtained, called reduced density matrix, does not, in general, correspond to a pure state. Instead it describes the not well defined state of a subsystem as it appears when averaging the degrees of freedom of the other, unobserved, subsystems. 
This mathematical result suggests that, the apparent statistical mixture state of non-microscopic physical systems, is related to its density matrix being actually a reduced density matrix of a larger system, which may for example include the environment.
The main and most striking success of the decoherence programme is due to the application of the above idea. 
By averaging out the environment degrees of freedom from the whole system pure state density matrix, it can be shown that, under certain realistic conditions, the non-diagonal terms of the reduced matrix rapidly decay to zero.

As a clarifying example a system constituted by a spin-1/2 and an environment can be considered. If the spin-1/2 system is initially prepared in a state $\ket{\Sigma}$ and the state of the environment is $\ket{\varepsilon}$, the state vector of the total system will be $\ket{\Sigma}\otimes \ket{\varepsilon}$. 
Assuming that $\ket{\Sigma}$ is a superposition of $\hat S_z$ eigenstates, that is $\ket{\Sigma}=\alpha_i \ket{+}_z +\beta_i \ket{-}_z$, the system-environment interaction will modify the total system state vector as follows:
\begin{equation}
\ket{\Psi_i} = \left(\alpha_i \ket{+}_z +\beta_i \ket{-}_z\right)\ket{\varepsilon} \Longrightarrow \ket{\Psi_f} = \alpha_f \ket{+}_z \ket{\varepsilon_+} +\beta_f \ket{-}_z \ket{\varepsilon_-} .
    \label{eq:spin_environment_entanglement}
\end{equation}
The interaction between two systems will generally produce an entangled state. In the equation above $\ket{\varepsilon_+}$ and $\ket{\varepsilon_-}$ are two different environment states which are not necessarily orthogonal.

The density matrix of the total system can thus be represented in Dirac notation as:
\begin{equation}
\rho = \ket{\Psi_f}\bra{\Psi_f} = \left( \alpha_f \ket{+}_z \ket{\varepsilon_+} +\beta_f \ket{-}_z \ket{\varepsilon_-} \right) \left( \bra{\varepsilon_+} \bra{+}_z \alpha_f^* + \bra{\varepsilon_-} \bra{-}_z \beta_f^* \right)
    \label{eq:spin_env_density_matrix}
\end{equation}
Performing a partial trace of $\rho$, on the environment degrees of freedom, consists in computing $\sum_i \bra{\epsilon_i}\rho\ket{\epsilon_i}$, where the $\ket{\epsilon_i}$ constitute a complete basis set for the Hilbert space of the environment. Thus the expression for $\rho_{\varepsilon r}$ ($= \mbox{Tr}_\varepsilon (\rho)$) becomes:
\begin{equation}
\rho_{\varepsilon r} = \sum_i \left(\alpha_f \ket{+}_z \langle \epsilon_i| \varepsilon_+ \rangle +\beta_f \ket{-}_z \langle \epsilon_i| \varepsilon_- \rangle\right) \left(\langle \varepsilon_+ | \epsilon_i \rangle \bra{+}_z \alpha_f^* + \langle \varepsilon_- | \epsilon_i \rangle \bra{-}_z \beta_f^* \right) .
    \label{eq:spin_env_reduced_rho}
\end{equation}
Expanding the diadic form and considering that $\sum_i \ket{\epsilon_i}\bra{\epsilon_i} = \mathbb{I}$,  one obtains
\begin{eqnarray}
\rho_{\varepsilon r} & = & |\alpha_f|^2 \ket{+}_z \bra{+}_z  + |\beta_f|^2 \ket{-}_z \bra{-}_z + \nonumber \\
& & \alpha_f^* \beta_f \langle \varepsilon_+ | \varepsilon_- \rangle \ket{-}_z \bra{+}_z  + \alpha_f \beta_f^* \langle \varepsilon_- | \varepsilon_+ \rangle \ket{+}_z \bra{-}_z  .
    \label{eq:spin_env_reduced_rho_simp}
\end{eqnarray}
This reduced density matrix describes the state of the spin-1/2 subsystem which, in general, is not a pure state, i.e. it cannot be expressed as $\ket{\psi}\bra{\psi}$ for any vector $\ket{\psi}$ belonging to the spin-1/2 Hilbert space. 

In matrix representation $\rho_{\varepsilon r}$ is:
\begin{equation}
\rho_{\varepsilon r} = \left(
    \begin{array}{cc}
|\alpha_f|^2 & \alpha_f \beta_f^* \langle \varepsilon_- | \varepsilon_+ \rangle\\
 & \\
 \alpha_f^* \beta_f \langle \varepsilon_+ | \varepsilon_- \rangle & |\beta_f|^2
    \end{array}
\right) .
    \label{eq:spin_env_reduced_rho_matrix}
\end{equation}
If the inner product $\langle \varepsilon_- | \varepsilon_+ \rangle$ equals zero, the above matrix is diagonal and represents a simple statistical mixture of spin-up and spin-down in the proportion given by $|\alpha_f|^2$ and $|\beta_f|^2$.

As it can be seen in Eq.~(\ref{eq:spin_env_reduced_rho_matrix}), the non-diagonal terms contain the information about the phases of the coefficients $\alpha_f$ and $\beta_f$, which clearly cannot be extracted from the diagonal ones alone. 
Such loss of information about the phases is referred to as loss of coherence, from quantum optics terminology, hence \emph{decoherence}.

The main success of the decoherence programme consists in showing that, when a large number of degrees of freedom are involved, such inner product very rapidly tends to zero. This is clearly the case for environment state vectors which represent the state of a large number of subsystems.
As an introduction to the the decoherence process, its main stages can be summarized as follows:
\begin{enumerate}
    \item Before the interaction takes place, spin system and environment are not entangled. The non-diagonal (interference) terms of the reduced density matrix are unaffected: $\langle \varepsilon_- | \varepsilon_+ \rangle = \langle \varepsilon | \varepsilon \rangle = 1$. The spin-1/2 subsystem retains full coherence.
    \item The interaction entangles the states of the two subsystems as in Eq~(\ref{eq:spin_environment_entanglement}).
    \item The Schr\"odinger dynamics of the large number of systems in the environment results in a rapid evolution of $\ket{\varepsilon_+}$ and $\ket{\varepsilon_-}$ towards two nearly orthogonal vectors. The non-diagonal terms in Eq.~(\ref{eq:spin_env_reduced_rho_matrix}) tend to zero.
\end{enumerate}
The transition from situation (1) to situation (3) can explain why for macroscopic systems, which are composed of a large number of subsystems, the outcomes distribution does not reflect the underlying quantum statistics and appears instead as due to a classical statistical ensemble.

The same mechanism should apply to microscopic systems which are not well screened from the interaction with  surrounding particles and fields, and should thus account for the fact that the outcomes of some experiments on microscopic systems correspond to a diagonal density matrix like Eq.~(\ref{statistical_mixture_density_matrix}), instead of Eq.~(\ref{pure_state_density_matrix}).

In summary, the main result of the decoherence approach suggests that postulating classical physics laws for macroscopic systems (statement (iii) of the Copenhagen interpretation) is unnecessary and that there is no fundamental difference between the laws for microscopic systems and those for macroscopic ones.

On the other hand it is evident that this achievement of the decoherence programme does not address the issue of the state vector reduction (statement (iv)), and thus of the emergence of definite outcomes. This becomes even clearer when considering the physical meaning of the formal operation of averaging over the unobserved degrees of freedom. In fact computation of average values implies that one has assumed the Born rule (statement (v)) which in turn implies the adoption of the projection postulate (statement (iv)) \cite{jr._can_1997}.

Nevertheless the decoherence mechanism, by predicting the occurrence of macroscopic superpositions, even though for very short times, maintains the relevance of the Von Neumann measurement scheme and opens up again the question of whether or not the unitary evolution breaks at some stage and of what would be such a stage.

\section{Observation of macroscopic superpositions}
\label{sec:macroscopic_superpositions}
The analysis carried out so far shows the main link between the standard interpretation, the decoherence approach and the problem of definite outcomes: the occurrence of macroscopic superpositions.

First, the standard interpretation postulates the collapse of the wave function of the system when observed. Secondly, the decoherence mechanism shows that the classical behaviour of a macroscopic apparatus can in principle be derived from its quantum mechanical description.
Finally, taking into account these two points, if macroscopic systems can actually be observed in superposition states, then only one of two conclusions follows: 
either (1) there is no actual collapse of the state vector, or (2), if a wave function collapse does actually occur, then it cannot be attributed to the classical nature of the apparatus.

In recent years there has been an increasing number of experiments aimed at testing the quantum mechanical nature of macroscopic systems. To this author's knowledge none of them is explicitely designed for testing the collapse mechanism of the state vector. This can be justified by the fact that direct verification of the collapse of a single system wave function is beyond current experimental techniques and indirect study of the collapse mechanism still very difficult \cite{schlosshauer_experimental_2006,marshall_towards_2003,van_wezel_nanoscale_2012}.
Valuable insights about the problem of definite outcomes can nevertheless be extracted from a careful analysis of the experiments.
This is not always a straightforward exercise as it is also not always clear how to derive the implications of such experiments for other foundational issues. Aim of this section is to clarify what can be concluded with regard to the definite outcomes from the analysis of some relevant experiments.

As mentioned, the experiments considered in this section are all designed for the general aim of verifying the predictions of quantum mechanics with respect to macroscopic systems and to improve the control on the system-environment interaction, and consequently on the system quantum behaviour.
For the purpose at hand such experiments can be divided in three groups: double-slit type experiments, quantum state control, and macroscopic superposition realization.
Such a classification is certainly quite arbitrary, but it can be helpful in order to clarify the different information that can be extracted from various experiments performed with a similar purpose.

\subsection{Double-slit type experiments}
\label{sec:double_slit_exp}

Young's double-slit experiment is by most considered the clearest example of the quantum behavior of an object. By many it is also considered the most beautiful experiment of nineteenth century \cite{double-slit_exp}.

In the traditional, Bohr inspired, complementarity view, double-slit interference experiments are designed to show the wave-particle duality of microscopic systems. In the context of the present investigation, however, it is clear that such a position cannot be assumed a priori because it is based on a specific interpretative solution to the definite outcomes problem.
It is instead meaningful to consider interference experiments just for what they are: tests of the predictive power of quantum mechanics.
According to the Schr\"odinger equation, the wave function of an isolated physical system, initially prepared in an eigenstate of the momentum of its center of mass, in the presence of a barrier with a double slit, will evolve to form a pattern of maxima and minima, analogous to optical interference phenomena.
Regardless of whether wave function collapse occurs or not, if such interference pattern is experimentally observed, then it is necessary to infer that the system does not obey classical physics.

Already in 1930 Estermann and Stern \cite{i._estermann__1930} successfully performed a diffraction experiment with $\mbox{He}$ atoms and $\mbox{H}_2$ molecules, thus showing that the predictions of wave mechanics hold for composite, though microscopic, systems too. However only in recent years, improved control on environment perturbation and system preparation have allowed physicists to realistically attempt interference experiments with nearly macroscopic systems.

In 1999 Arndt et al. from the group of Anton Zeilinger \cite{arndt_wave-particle_1999} achieved one of the first great leaps in size. They successfully observed the diffraction of $\mbox{C}_{60}$, fullerene molecules, through a $100 \mbox{nm}$ grating. The experimentally observed diffraction pattern is in excellent agreement with the theoretical one, obtained with a de Broglie wavelength, $\lambda = h/Mv$, computed considering $M$ as the total mass of the $\mbox{C}_{60}$ molecule. The good quantitative agreement thus obtained, in the words of the authors themselves, is indication that \emph{each $\mbox{C}_{60}$ molecule interferes only with itself}. In other words the effect is not due to interference of individual smaller components, such as atoms, nuclei or electrons.

A relevant aspect of this experiment is the analysis of possible decoherence effects. The successful realization of molecular diffraction implies that phase coherence is not lost during the molecule flight from the source to the detector. In the whole system--apparatus setup there are three main possible causes of decoherence: photon emission due to the molecule's excited internal degrees of freedom, photon absorption from the environment blackbody radiation, and scattering of background molecules. The authors show that, in the case of this particular setup, all of them give a negligible contribution to decoherence.

A noteworthy detail in the article is the explanation of why photon emission or absorption do not remarkably affect phase coherence. The authors correctly state that only photons with a wavelength shorter than the distance between neighboring slits can, \emph{with a single scattering}, completely destroy the interference pattern. In fact, that could occasionally correspond to the detection of the molecule through one particular slit. As Bohr would say, that would reveal the particle nature of the molecule and destroy interference.
From a foundational and interpretative point of view, however, such an explanation has to be handled with care. First because the ``which-way'' detection argument, though factually correct, supposes the collapse of the wave function; second and more importantly because the above argument is not directly related to the decoherence mechanism mentioned in subsection~\ref{subsec:Copenh_decoh}, which explains the decay of the non-diagonal terms while maintaining the overall superposition.
To the authors' credit it has to be said that the above distinctions are implicit in their article when they state that \emph{decoherence is however also possible via multi-photon scattering}, citing classical works on decoherence \cite{joos_emergence_1985,zurek_decoherence_1991,joos_decoherence_2003}.

With regard to the measurement problem the result by Arndt et al. is relevant mainly for its demonstration of an almost macroscopic system in a superposition state. It constitutes, in fact, a further evidence of the difficulty in justifying the standard `orthodox' separation of the world in two kingdoms: a quantum--microscopic regime and a classical--macroscopic one, as implied by statement~(iii) in the previous section presentation.
In order to have an idea of how macroscopic the system is, it is meaningful to notice that the de Broglie wavelength of the $\mbox{C}_{60}$ molecules used in this experiment is 400 times smaller than the diameter of the molecule.

In 2003 the same group \cite{hackermuller_wave_2003} further improved on the size of the interfering molecules by performing a similar experiment with fluorofullerene, $\mbox{C}_{60}\mbox{F}_{48}$, which is about twice as massive as $\mbox{C}_{60}$. The latest and most impressive demostration of this kind is however the one performed by Gerlich et al. in 2011 \cite{gerlich_quantum_2011}. The experiment shows interference of molecules composed of up to $430$ atoms.
In the authors' own words the experiment proves \emph{the quantum wave nature and delocalization of
compounds} with a maximal size of up to $60$\AA, masses up to $6910$AMU and de Broglie wavelengths down to $1$pm. Such a wavelength is, in this case, $6000$ times smaller than the size of the object. This contrasts sharply with the common wisdom, according to which interference effects are always negligible when the size of the object is larger than its de Broglie wavelength.
The article also stresses that the interference observed is due to the wavelength of the whole molecule as a single entity, in contrast to experiments with `macroscopic' Bose-Einstein condensates, where interference is expected because the wavelength of the BEC is essentially the same as the relatively large one of the single constituent atoms.
With respect to the issue of macroscopic delocalization, the authors highlight the fact that the grating's width corresponds, in a wave-particle complementarity view, to a path separation of almost two orders of magnitude larger than the size of the molecules.

What this latest experiment shows, as the earlier ones of its kind and more strikingly, is that decoherence and definite outcomes are two distinct phenomena. In successful interference experiments decoherence is effectively prevented, while, evidently, the state of each single interfering object collapses in interaction with the detector. This point will be considered in more detail in Section~\ref{sec:facts_and_misconceptions}.

\subsection{Quantum state control experiments}
\label{quantum_state_control}

Achieving interference with larger and larger molecular compounds is certainly a strong indication of the absence of a break up scale for quantum theory, but molecules, even as large as proteins, are not what one imagines when thinking of macroscopic objects.
Micromechanical systems considered in recent years are much closer to the common notion of macroscopic system, particularly for their being visible at least with an optical microscope.

Probably the most ambitious programme to date for preparing a macroscopic system in a superposition state is described in the paper by Marshall et al. \cite{marshall_towards_2003} and further analyzed by Kleckner et al. \cite{kleckner_creating_2008}.
The authors propose the use of a special Michelson interferometer, where one of the cavity mirrors is replaced by a tiny mirror attached to a micromechanical cantilever, as to create and observe superposition and entanglement of the photon-cantilever system. The design of the experiment allows not only testing of various decoherence mechanisms, but also of some wave function reduction models: the authors in particular are interested in a gravity induced collapse proposal by Penrose \cite{penrose_gravitys_1996} (see Section~\ref{sec:gravity_induced_collapse}).
Both papers show that, on the one hand, the experiment is feasible in principle with current state-of-the-art technology, on the other hand, that an excellent control on the system-environment interaction has to be achieved in order to meet the conditions for unambiguous observation of quantum phenomena. Chief among these conditions is the ability to maintain the cantilever in its vibrational ground state when not interacting. Because of their relatively large mass, in micromechanical oscillators the energy gap between different vibrational states is particularly small (corresponding to resonance frequencies of the order of tens of megahertz) and standard cryogenic methods for mechanical systems are not enough to achieve temperatures below the excitation energies of the oscillator \cite{aspelmeyer_quantum_2010}. 

Motivated by foundational issues as well as by technical application opportunities, several groups have in recent years taken up the challenge of realizing quantum micromechanical systems.
Directly relevant for the above proposed experiment are the works by O'Connel et al. \cite{oconnell_quantum_2010} and by Teufel et al. \cite{teufel_sideband_2011}.
In both experiments the micromechanical system is a cantilever and both consist of about $10^{12}$ atoms, close to the $10^{14}$ atoms of the Bouwmeester proposal \cite{kleckner_creating_2008}.
Both micromechanical oscillators are coupled to a microwave resonant circuit which acts as the measurement apparatus (whose signal is to be amplified). Remarkably, from the point of view of quantum computation, in the 2010 experiment~\cite{oconnell_quantum_2010} the microwave resonant circuit is designed to realize a two-level system, and thus store a quantum bit of information. This is achieved by using a superconducting quantum interference device: a SQUID.

The two experiments differ substantially in the details of the actual implementation of the system-apparatus interaction and in the measurement readout.
They both succeed however in two common goals: cooling of the oscillator to its ground state and controlled single quantum excitation.
The setup of O'Connel et al. allowed them to also observe the exchange of a quantized excitation between the oscillator and the SQUID, in agreement with the quantum mechanism of Rabi oscillations.
As a further evidence of the quantum mechanical nature of the cantilever behavior, the authors successfully placed the system in a harmonic oscillator coherent state, as indicated by the good agreement of the theory with the qubit response.
On the other hand, the design by Teufel et al., thanks to the tunable coupling strength between their mechanical resonator and a microwave cavity, can improve remarkably the coherence time of the system state, extending it to over $100 \mu \mbox{s}$. This is much larger than typical decoherence times of SQUID based qubits and thus allows performing quantum operations on such a mechanical system.

In short, aside from the enormous scientific and technical achievement these experiments constitute, their immediate relevance, with respect to the measurement problem in general and the definite outcomes in particular, consists in demonstrating that naked-eye visible mechanical objects satisfy quantum physics. In fact the oscillator of Ref.~\cite{oconnell_quantum_2010} is about $50 \mu \mbox{m}$ in two of the three dimensions and thus noticeable to the naked eye; the one of Ref~\cite{teufel_sideband_2011} is about $20 \mu \mbox{m}$, slightly below the human eye capability.

\subsection{Macroscopic superposition realization}
\label{sec:macroscopic_superposition_realization}

As presented above, direct or indirect evidence of superposition states have been obtained for macromolecules and micromechanical oscillators. There is another class of physical systems which is particularly suitable to test the occurence of macroscopic superposition states: superconducting quantum interference devices, commonly referred to with the acronym of SQUIDs.
A superconducting quantum interference device, in its most elementary form, consists of a superconducting loop cut by a small piece of insulating material and driven by an external magnetic flux. The junction between the superconductor and the insulator is called a Josephson junction and it is narrow enough to allow Cooper pairs tunneling thus keeping the whole loop superconducting. Typical values of the supercurrent in a SQUID are in the microampere range, corresponding to the collective motion of millions of Cooper pairs.
The Josephson junction generates a potential energy term in the Hamiltonian that describes the supercurrent and thus makes it possible to design non trivial configurations such as, for example, with a double well potential.

Experimentally the basic idea is to prepare and observe the supercurrent flowing in the ring in a superposition of clockwise and anticlockwise states. 
Among the first breakthroughs in this regard, it is worth mentioning two indipendent experiments reported in 2000: one by Friedman et al. \cite{friedman_quantum_2000} and one by van der Wal et al. \cite{wal_quantum_2000}.
As was the case for the micromechanical oscillators, besides several differences in the implementation, the two experiments share all the main features: a radio frequency superconducting loop as the observed system, an inductively coupled d.c.~SQUID for the measuring apparatus, and the realization of a double well potential in the coordinate which describes the magnetic flux $\Phi$ through the ring.
By varying the externally applied magnetic flux $\Phi_{ext}$, it is possible to modify the double well potential and to make it more or less symmetrical. For asymmetrical double well, there exist two low energy eigenstates, say $\ket{L}$ and $\ket{R}$, which correspond to two opposite current flows and two different energies.
When $\Phi_{ext} = 1/2 \Phi_0$, where $\phi_0$ is the elementary quantum of flux across the ring, the double well is symmetrical and the two low energy eigenstates, instead of becoming degenerate as one would classically expect, remain distinct in energy, but change into a symmetrical and antisymmetrical superposition of clockwise and anticlockwise flows: $1/\sqrt{2} (\ket{L} + \ket{R})$ and $1/\sqrt{2} (\ket{L} - \ket{R})$.

The two groups demonstrated a superposition of opposite supercurrent states by observing the variation of the energy of the two low energy eigenstates as a function of the externally applied magnetic field by verifying the persistence of the energy gap, and by finding excellent agreement with the quantum theoretical prediction.
The observation of such macroscopic superpositions is indirect, as is also the case for the other experiments presented so far. The observation is indirect not only according to the common meaning of being mediated by the theory through the `direct' observation of some other physical quantity, but in the sense that it does not correspond to individual measurements. It instead consists in comparing the average over a large number of individual measurements with the theoretical predictions \cite{wal_quantum_2000}.
In this regard a very interesting series of experiments has been performed by Tanaka et al. \cite{tanaka_dc-squid_2002, takayanagi_readout_2002}. They demonstrate reliable `single-shot' measurements of magnetic flux (by observing the switching current) in SQUID systems of the same type as the two previous experiments.
To this end they increased the coupling between the observed two-level system and the d.c.~SQUID, thus achieving a larger output signal of the measuring d.c.~SQUID, while at the same time reducing fluctuations and noise. The drawback of this setup is a shorter decoherence time, because of the increased interaction with the many degrees of freedom of the other parts of the measuring apparatus attached to the d.c.~SQUID.
This means that this particular setup, at least within their current level of control of decoherence effects, is not suitable for quantum computation applications. This is because the system state loses coherence in a time too short to complete quantum gates operations.
From the point of view of the foundational issues of macroscopic superpositions and wave function reduction, on the other hand, the experiments by Tanaka et al. are remarkable because they allow \emph{the first direct observation of a macroscopic quantum superposition}. Here `direct observation' means corresponding to a single measurement whose result is interpreted by means of the theory, in contrast to an `indirect observation' reconstructed from an average value over a large number of measurements.

The direct observation of a superposition of clockwise and anticlockwise supercurrent states is described as follows. With the strong coupling setup mentioned above, in the symmetric double well configuration, the magnetic flux (switching current) measured by the apparatus, i.e. the probe constituted by the d.c.~SQUID, does not correspond to the system state $\ket{L}$ or $\ket{R}$, occurring randomly with a frequency given by the Born rule. Instead it deterministically corresponds to one of the two low level energy eigenstates in which the observed system is initially placed: $\ket{0} = 1/2 (\ket{L} - \ket{R})$ or $\ket{1} = 1/2 (\ket{L} + \ket{R})$. The measurement in this case is not a projection on one of the states with a definite current flow, but on one of the two energy eigenstates of the two level system. 
For a generic value of the applied external flux, $\Phi_{ext}$, the energy eigenstates can be written as $a \ket{L} + b \ket{R}$. Tanaka et al. show \cite{takayanagi_readout_2002, semba_quantum_2009} that the single measurement by the apparatus corresponds to a determination of $|a|^2$ (or equivalently $1-|b|^2$).

Interestingly, the authors' theoretical analysis \cite{semba_quantum_2009} predicts, through numerical calculations, that, in the case of very strong coupling which induces a strong decoherence, the measurement outcomes would correspond again to $\ket{L}$ or $\ket{R}$ in a probabilistic manner.

In conclusion, what these single measurement observations imply, with regard to the definite outcomes problem, is that macroscopic superpositions can be directly accessible to the observer, that is, they are not destined to remain a state of affairs which exists only as long as it is not observed.

Certainly most readers will have already thought of the analogy between the superconducting two level system and a spin $1/2$ system in a magnetic field: particularly of the analogy between the states $\ket{L}$, $\ket{R}$, $\ket{0}$, $\ket{1}$ and the eigenstates of the spin operators $\hat S_z$ and $\hat S_x$. From this point of view the result by Tanaka et al. is nothing new: it simply corresponds to measuring the value of $\hat S_x$ and stating that the electron is in a superposition state of $\ket{+}_z$ and $\ket{-}_z$. The difference between the two cases lies in the scale of the superposition: one electron, or one atom, in one case and millions of electrons in a coherent superposition in the other case. The importance of the achievement does not lie in the `striking psychological factor', but in the difficulty of observing a superposition in a macroscopic spin system, as shown by Simon and other authors \cite{raeisi_coarse_2011}.

\section{Summary of main issues}
\label{sec:facts_and_misconceptions}

At the end of this analytic presentation it is useful to explicitely define the main concepts that will be used throughout this thesis and to sum up the conclusions that can already be drawn from the theoretical and experimental results presented so far.

First of all, as it should be already evident, throughout this thesis the terms `state vector reduction' and `wave function collapse' are used strictly to indicate the non-unitary transition of a single quantum system (not an ensemble) from a state $\sum_i \ket{a_i}$ to a state $\ket{a_k}$. The two terms here \emph{do not} refer to the change in statistical distribution of measurement outcomes: from a distribution reflecting the presence of quantum interference to a classical statistics one.

The expression `definite outcomes' is used to refer to the individual, single-valued, measurement outcomes. State vector reduction, either objective or subjective (see Chapter~\ref{ch:interpretations}), is invoked in order to account for the lack of direct observation of superposition states in a single measurement.

In this thesis the term `decoherence' is used in its narrower meaning of loss of information about the coefficients' phases of the quantum superposition. The decoherence mechanism accounts for such a loss by predicting the decay to zero of the non-diagonal terms of the reduced density matrix.

Having clarified the basic notions, it is possible to draw some basic conclusions and to identify the main issues that have to be addressed.

First and most importantly from the discussion developed so far it should be clear that the decoherence mechanism by itself does not explain the occurrence of definite outcomes. On the contrary, the assumption, that definite outcomes occur and that they are distributed according to the Born rule, allows the meaningful use of the reduced density matrix formalism.

If, for example, one considers a single spin-1/2 system and describes its state by means of a density matrix, then phase decoherence consists in the following transition:
\begin{equation}
\left(
\begin{array}{cc}
|\alpha|^2 & \alpha \beta^* \\
 \alpha^* \beta & |\beta|^2
\end{array}
\right) \xrightarrow[decoherence]{}
\left(
\begin{array}{cc}
|\alpha|^2 & 0 \\
 0 & |\beta|^2
\end{array}
\right).
  \label{eq:spin_decoh_trans}
\end{equation}
On the other hand, the transition occuring upon performing a measurement on $\hat S_z$ can be of two types, depending on whether decoherence is slower or faster than the measurement process.  Assuming, for example, that the measurement outcome is $S_z = +\hbar/2$ and that the system loses coherence at a slow rate, then the transition will be:
\begin{equation}
\left(
\begin{array}{cc}
|\alpha|^2 & \alpha \beta^* \\
 \alpha^* \beta & |\beta|^2
\end{array}
\right) \xrightarrow[collapse]{}
\left(
\begin{array}{cc}
1 & 0 \\
 0 & 0
\end{array}
\right).
  \label{eq:spin_def_out_no_decoh}
\end{equation}
Instead, if decoherence has occured in the time between the system preparation and the end of the measurement process, then the transition will be from the decohered density matrix to the `collapsed' one:
\begin{equation}
\left(
\begin{array}{cc}
|\alpha|^2 & 0 \\
 0 & |\beta|^2
\end{array}
\right) \xrightarrow[collapse]{}
\left(
\begin{array}{cc}
1 & 0 \\
 0 & 0
\end{array}
\right).
  \label{eq:spin_def_out_with_decoh}
\end{equation}

The decoherence process of Eq.~(\ref{eq:spin_decoh_trans}) is explained by realizing that the spin-1/2 system is actually embedded in a larger system and that the two-by-two density matrix should be derived from the density matrix describing the state of the larger system.
One of the main aims of this thesis is clarifying the relation between the process leading to Eq.~(\ref{eq:spin_decoh_trans}) and the transition described in Eqs.~(\ref{eq:spin_def_out_no_decoh})~and~(\ref{eq:spin_def_out_with_decoh}).
In this regard one relevant issue is how to account for the empirical validity of the Born rule. Since it is critically used in the decoherence theory, it requires an independent account in order to avoid a circular explanation. Derivations of the Born rule from general principles such as the one by Gleason \cite{gleason_measures_1957} and the one by Zurek \cite{zurek_decoherence_2003} will be discussed in Section~\ref{sec:relative_state_interpretation} and Chapter~\ref{ch:decoherence}.

The distinct nature of the decoherence process and of the wave function collapse is also evident from interference experiments that demonstrate collapse without loss of coherence. Of course one might remark that a particle loses its phase coherence upon detection by a macroscopic apparatus, but clearly this happens \emph{after} the detection. Otherwise, according to the decoherence mechanism, we would not obtain the interference pattern, but only the classically expected spots.

To make the point clearer it is worthwhile to consider a classic double slit interference experiment. In order to correctly describe the dynamics of the interfering particles, the apparatus and the interaction with environment particles should also be taken into account. If decoherence occurs before collapse, either due to the macroscopic apparatus or to the rest of the environment, then only two lines will be left on the apparatus screen. In fact in this case, if the reduced density matrix is expressed in the position eigenbasis, only the two diagonal terms corresponding to the classical positions are different from zero (see Section~\ref{subsec:non_diagonal_terms}).
Viceversa, if the predicted interference pattern builds up, it necessarily follows that decoherence has not yet occurred when the apparatus records the particles positions. Once the state of the particle has collapsed to an eigenstate of position then decoherence will most likely occur due to interaction with the huge number of subsystems comprising the apparatus. This latter decoherence process, however, is unrelated to the relevant part of the interference experiment. In fact, by this time the particle position has already been stably recorded by the apparatus which will finally present the interference pattern.

An important conclusion that can be drawn from the experiments presented in Section~\ref{sec:macroscopic_superpositions} is that macroscopic systems too are quantum mechanical and can be in superposition states. It has to be acknowledged that a $50 \mu\mbox{m}$ drum, such as the micromechanical oscillator in Ref.~\cite{oconnell_quantum_2010}, is not a full sized musical instrument, but it is clear that there is no experimental evidence for a fundamental cut-off scale for the validity of quantum mechanics.
An immediate consequence is that the definite outcomes problem cannot be explained by a naive appeal to a classical physics regime.

With regard to the issue of macroscopic superposition states, even though recent experiments are in remarkable agreement with theoretical predictions, an object has never been directly observed in a delocalized state. Even in the experiment of Tanaka et. Al. \cite{takayanagi_readout_2002}, where a macroscopic supercurrent has been `directly' observed in a superposition state, the apparatus pointer (indicating the switching current) was localized around a well defined position. 
The insights gained from the whole decoherence programme and from these latest experiments suggest that
the state vector reduction problem may be strictly related to the localization properties of macroscopic systems. This, at least, is a hypotesis that cannot be neglected before careful examination.

In Chapters~\ref{ch:interpretations}~and~\ref{ch:future} various attempts are presented with the aim of deriving single measurement outcomes from physical laws, thus avoiding the ad-hoc collapse postulate.

\chapter{Definite outcomes in different interpretations}
\label{ch:interpretations}

In order to reach the three aims stated in the Introduction, a critical overview of the way definite outcomes are addressed, in the most commonly adopted quantum mechanics interpretations, is necessary.
In fact, to clarify the relation between the decoherence mechanism and the occurrence of definite outcomes, one has to understand how a specific interpretation addresses the apparent collapse of the wave function.
Moreover, several common misconceptions actually consist in misunderstandings about the interpretation. And, clearly, in order to appreciate the merits and weaknesses of alternative approaches, it is  necessary to know the more established ones.

Before starting to delve into the different interpretations, it is useful to make clear why a physical theory needs an interpretation at all and to observe that this is not a peculiarity of quantum mechanics. 
In fact any physical theory must include a set of statements which specify the meaning, with respect to the physical world, of the basic formal objects of the theory. This is necessary for a theory to be \emph{physical}. 
In classical physics, for example, it is necessary to state what is meant by a `material point', a `force', a `field' etc...
Likewise quantum mechanics interpretations answer the following type of questions: What is the wave function? What is an operator? What is the relation between an eigenvalue and the properties of the physical system?

A clear difference between interpretative statements in classical and quantum physics lies in the fact that basic formal concepts in classical physics relate to more direct and familiar experiences.
Also, quantum physics raises a number of additional and specific interpretative issues such as the occurrence of definite outcomes and the emergence of a macroscopic \emph{classical world}. This latter expression refers collectively to a number of properties which are expected in a classical physics system: localization (i.e. no particle interference), non-quantized energy, deterministic evolution, independent measurements.

For the purpose of the present work it is convenient to classify the various interpretations according to the way the definite outcomes problem is addressed. Consequently, interpretations can be grouped in two types: (1) theory extending solutions and (2) purely interpretative solutions.
As the names imply, interpretations belonging to the first type aim at explaining the state vector collapse through the addition of some physical law to the basic, no collapse, quantum mechanics, while those belonging to the second type aim at solving the problem through an appropriate interpretation of what is observed in terms of the unitary evolution of the state vector of the universe.

According to this classification criterium there should actually be a third type of intepretation, which approaches the problem without adding new physics laws or specific interpretative prescriptions, and which, instead, looks for a solution within the already known physics laws, either by studying the influence of gravity on the wave function evolution \cite{diosi_models_1989, penrose_gravitys_1996}, or in a `For All Practical Purposes' (\emph{FAPP}) fashion \cite{jr._can_1997}. This third type of approach is still very little explored and is the topic of the last chapter of this thesis.

In the presentation that follows the decoherence mechanism will only be considered to highlight its relation with each interpretation: particularly what specific aspects are clarified by decoherence and how its predictions can constitute a consistency test for a particular interpretation.

For brevity, and to facilitate the analysis and comparison of the different interpretations, it is useful to state at the beginning what they have in common. This can be summed up in two broad statements:
\begin{itemize}
    \item The state of the physical system is described, in part or completely, by a state vector.
    \item The time evolution of the state vector is governed by the Schr\"odinger equation, with the possible addition of a stochastic term.
\end{itemize}
Starting from this basic common ground, the following discussion will focus on the strenghts and weaknesses of the two different types of approach to the solution of the definite outcomes problem.

\section{Theory extending solutions: adding new principles}
\subsection{Copenhagen interpretation}
\label{Copenhagen_interpretation}

The Copenhagen interpretation, in its Bohr inspired version, has been already introduced in Section~\ref{subsec:Copenh_decoh}. The knowledgeable reader is asked forgiveness for the fact that, throughout this thesis, the term `Copenhagen interpretation' is liberally used to refer to a whole set of different interpretations with non-trivial differences. They can however be grouped together from the point of view of the ultimate fate of the wave function upon observation of the system properties.

According to this class of interpretations, the state vector completely describes the state of the system and evolves unitarily according to the Schr\"odinger equation as long as the system is not observed. The specific aspect of the Copenhagen interpretation lies in the postulates regarding the observation or measurement of the system. It is explicitely postulated that the measurement of an observable changes the state of the system from a generic state $\ket{\psi}$ to one of the eigenstates of the observable, and the probabilities for different outcomes obey Born's rule. In other words the state vector reduction is \emph{postulated}: it is taken as a fact of nature.
This means that it does not need to be formally explained. However, the various versions of the Copenhagen interpretation aim at giving a plausible motivation to the collapse postulate.

As stated in Section~\ref{subsec:Copenh_decoh}, Bohr's explanation of the origin of the wave function collapse relies mainly on his other postulate regarding the classical nature of macroscopic systems. In light of the analysis carried out in Section~\ref{sec:macroscopic_superpositions}, Bohr's division of the physical phenomena in two clearly separated domains cannot be upheld without conflicting with recent experiments on macroscopic systems in superposition states.
For the sake of correctness it has to be said that Bohr's arguments in this regard are very sophisticated ones based on the notion of complementarity \cite{kristian_bohr_2007, osnaghi_origin_2009}. Nevertheless they are in contrast with recent evidence of the validity of quantum mechanics at the macroscopic scale.
Ultimately the problem with Bohr's position consists in trying to explain the collapse of the wave function through the classical properties of macroscopic systems, while experimental evidence suggests the opposite: an explanation for the classical behavior should be derived from the quantum formalism. 

To address these difficulties within the framework of the Copenhagen interpretation several proposals have been made to mark the boundary where the unitary evolution breaks down and the wave function collapse occurs.
A well known proposal is the one by Von Neumann, who suggested that the reduction of the state vector is linked to the presence of a conscious observer \cite{bacciagaluppi_role_2012}. In his view the measurement process consists of two steps: (1) the system-apparatus interaction governed by the Schr\"odinger equation, as presented in Section~\ref{sec:von_neumann}, and (2) the `reading' by a conscious observer which, in some way, results in a definite outcome. Von Neumann leaves ample room for interpreting how observer consciousness is linked to definite outcomes: it could be through a direct cause-effect relation or the link could be epistemic, i.e. acquisition of information by the observer can only result in a well defined value. 

Though quite vague, Von Neumann proposal has the merit of providing a first clear framework for analyzing the measurement process, thus putting it in the domain of testable theories. In fact the first step of Von Neumann scheme, often referred to as pre-measurement, is standard part of any current approach to the measurement process.

An alternative, very radical, argument in support of the postulate of the collapse of the wave function is proposed by Omn\`es \cite{omnes_interpretation_1994}. He starts by observing that the possibility of describing physical phenomena by some general causal laws should not be taken for granted, as anyone with a basic knowledge of philosophy of science knows. According to Omn\`es, the greatest achievement of quantum theory is perhaps having lead our human understanding of the physical world to one of its limits. The occurrence of definite outcomes is perhaps one of the things that happen in the world without being intelligible. And it should be taken as a basic fact. Such an argument may appear either too easy or too radical, but is actually reasonable and in perfect agreement with all observations.

It is clear that the main weakness of the Copenhagen based interpretations are the arguments in support of the collapse postulate. Besides this, however, and except for Bohr's claim of the existence of two physical domains, they constitute a consistent framework for the description of the physical world by means of the quantum mechanical formalism. In particular, the different versions of the Copenhagen interpretation are a suitable foundation for the decoherence programme in its general aim of obtaining the classical properties of macroscopic systems from quantum theory. 
On the one hand, the reduced density matrix formalism requires the Born rule in order to meaningfully represent the state of the object subsystem in the total object-apparatus-environment system. The Born rule is most easily understood as a statistical law that governs the frequency of each measurement outcome. In the Copenhagen interpretation both the Born rule and the occurrence of definite outcomes are postulated and this provides a straightforward justification for the average over the unobserved degrees of freedom of the total system.
On the other hand, the decoherence mechanism explains a number of phenomena which do not obviously follow from the postulates of quantum mechanics.
In this context, the successful results of the decoherence programme become confirmations of the validity of the Copenhagen version of quantum theory.

\subsection{Bohmian mechanics and hidden variables}

A completely alternative approach to the issue of the wave function collapse is given by the de Broglie-Bohm theory, or Bohmian mechanics, which constitutes a deterministic version of quantum theory.
Bohmian mechanics eliminates the collapse postulate and provides a completely causal account of the quantum mechanical phenomena.

The theory is based on the following postulates \cite{goldstein_bohmian_2012}:
\begin{enumerate}[(i)]
    \item The fundamental variables describing the motion of the system are the positions, $q_k$, of the particles.
    \item The wave function does not completely describe the state of the system.
    \item The square modulus of the wave function, $|\psi(q_k)|^2$, corresponds to the probability distribution of the particles position.
    \item The time evolution of the wave function is governed by the Schr\"odinger equation.
    \item The motion of the fundamental variables depends on the wave function through the guiding equation.
\end{enumerate}
The guiding equation is a first order differentail equation in the particles positions:
\begin{equation}
\frac{dq_k}{dt} = \frac{\hbar}{m_k} \mbox{Im} \left[ \frac{\psi^*(q_1...q_N)\partial_{q_k}\psi(q_1...q_N)}{\psi^*\psi}\right] .
\label{eq:guiding_equation}
\end{equation}
Such an apparently obscure equation becomes immediately intelligible when noting that the right hand side is proportional to the probability current density $J_k$:
\begin{equation}
\frac{\hbar}{m_k} \mbox{Im} \left[ \frac{\psi^*\partial_{q_k}\psi}{\psi^*\psi}\right] = \frac{J_k}{\rho}
\label{eq:current_density}
\end{equation}
and by recalling that the classical relation between current and velocity is $\vec{J} = \rho \vec{v}$.

It can be shown that Bohm's clever idea produces a theory whose predictions are experimentally indistinguishable from the predictions of the standard Copenhagen interpretation.
The main motivations of the de Broglie-Bohm interpretation consist in removing the collapse postulate, providing a causal description of physical phenomena, and showing that the quantum theoretical description of the physical properties of a system is complete.
And it actually succeeds in reaching these goals.

Regardless of any particular interpretation, the collapse of the wave function is generally linked to the measurement of some quantity.
The measurement process in Bohmian mechanics consists in the interaction between the system under investigation and the measurement apparatus. Such a process involves the wave function of the whole larger system: observed system and apparatus. As for the Von Neumann scheme, a superposition wavefunction, for the investigated system before the measurement, produces an entangled wave function for the whole system-apparatus. The difference here is that, according to de Broglie-Bohm interpretation, each single particle actually has a well defined position. Thus, on the one hand, the total wave function remains a superposition, while, on the other hand, the actual objects, system and apparatus, continue to have well defined properties, such as position, momentum or energy.
Since, in general, the entanglement remains in the global wave function, the motion of the pointer after the measurement might also be highly non-classical because of the effect of the entangled wave function on the particles positions through the guiding equation. The lack of empirical observation of such non-classical macroscopic dynamics can be explained by means of the decoherence mechanism.

The absence of wave function collapse and the use of the total wave function of a system which includes the observed subsystem, the apparatus and any other particle necessary to make the whole system reasonably isolated, potentially the whole universe, may induce someone to assimilate de Broglie-Bohm interpretation to a relative state interpretation. This is however not appropriate because, in Bohmian mechanics, particles and their positions are primitive notions, while in relative state interpretations they are not.

It is important to notice that Bohmian mechanics is not in contrast with the results of experiments aimed at testing the violation of Bell's inequalities. Starting from the one by Aspect onwards \cite{aspect_experimental_1981, aspect_experimental_1982}, these experiments imply that \emph{local} hidden variable theories are incompatible with the observed statistics. But Bohmian mechanics is clearly non-local, as it can be inferred by inspecting the guiding equation, Eq.~(\ref{eq:guiding_equation}), for an $N$ particles system. The velocity of the $k$th particle, $\dot q_k$, depends on the position of all the other particles, through the wave function $\psi(q_1...q_N)$.
Thus, as Bell himself states \cite{bell_speakable_2004,goldstein_bohmian_2012}, one of the merits of Bohmian mechanics is to explicitly show the non-local nature of quantum mechanics.
This observation should also dispel the impression that Bohm's theory is an attempt at reformulating quantum theory as an improved version of newtonian mechanics. The wave function, which produces non-locality, is an object completely extraneous to classical physics.

de Broglie-Bohm theory, though empirically indistinguishable from standard Copenhagen quantum mechanics, can be considered more appealing with respect to a number of aspects. First of all, by assigning a guiding role to the wave function, superposition states are no more ghostly objects which manifest themselves only indirectly, but have the same status of eigenfunctions: they all manifest themselves indirectly through the particles' dynamics.
Obviously, the notorious problem of Schr\"odinger's cat, is removed, because at every instant the cat is either dead or alive. As previously mentioned, the cat might in principle show some non-classical behavior due to the guidance of the entangled wave function, but this, most of the time, should be taken care of by the decoherence mechanism. In other words, Bohmian mechanics does not rule out a successful interference experiment with Schr\"odinger cats, if decoherence is kept under control as in the experiments presented in Section~\ref{sec:macroscopic_superpositions}, but it does rule out an interference pattern obtained with one cat only.

Other attractive aspects of Bohmian mechanics are the intrinsic completeness of the physical properties description and a more straightforward interpretation of the classical limit.\\

A main weakness of the de Broglie-Bohm interpretation lies in the difficulty of producing a Lorentz invariant formulation necessary for consistency with special relativity.
From a more metaphysical point of view another weakness may be the not so clear ontological status of the wave function. Strictly according to the formulation introduced above, the wave function is a companion mathematical object of the particle and it tells the particle where to go. At the same time it contains information about the \emph{epistemic} uncertainty in the position of each single particle.

For practical purposes one can consider the wave function as a purely epistemic object, only related to the observer's knowledge of the system state. Consequently, the observer has to be considered completely outside of the physical measurement process.
According to such a point of view, after the measurement, one can replace the pre-measurement wave function with the appropriate eigenfunction of the measured observable. In this case, however, the collapse postulate is automatically reintroduced, thus erasing one of the most appealing features of Bohmian mechanics.

The reduced density matrix formalism finds a perfectly fitting framework in Bohmian mechanics, because, in such an interpretation, the mathematical operation of tracing over environment basis states has the meaning of statistical average over the unobserved degrees of freedom. As a consequence all the standard decoherence formalism is applicable in Bohmian mechanics.

\subsection{Objective collapse theories}
\label{objective_collapse_theories}
A third possible approach within the theory extending solutions is given by the `objective collapse theories'. The basic idea behind such class of solutions is that the wave function dynamics is not completely determined by the Schr\"odinger equation. Instead, a collapse inducing term, which appears at random times, should be added to the unitary evolution.

Historically such an approach resulted from the attempt at explaining the state vector reduction as due to the interaction between the subsystem to be observed and an open-system comprised of apparatus and environment \cite{ghirardi_collapse_2011}. The interaction with the open-system was assumed to generate \emph{random processes at random times} \cite{ghirardi_collapse_2011} leading, among other effects, to the wave function collapse. The attempt at modelling such an effect was made either by adding some ad-hoc localization process \cite{fonda_l._evolution_1973} or by replacing the Schr\"odinger equation with a stochastic differential equation \cite{pearle_reduction_1976, gisin_quantum_1984}. Such theories were in any case not meant to be considered a fundamental description of natural phenomena.

In 1986 Ghirardi, Rimini and Weber proposed a collapse theory \cite{ghirardi_unified_1986} which relates the collapse inducing terms to an actual physical law not derivable from the standard quantum mechanical interaction between system and environment.
The basic assumptions of the original GRW theory can be listed as follows:
\begin{enumerate}[(i)]
   \item The wave function completely describes the physical system.
   \item The wave function of each elementary physical system is subjected to random localization processes, in addition to the unitary Schr\"odinger evolution.
   \item The random localization processes are a fundamental law of nature. 
   \item Localization is more likely to occur at points where the square modulus of the wave function is larger.
   \item Localization occurs at random times, according to a Poisson distribution for the frequency. 
\end{enumerate}

By developing in detail the above points, a precise collapse theory can be formulated.

An objective collapse theory attempts to explain the empirical observation of definite outcomes as a consequence of an actual spatial localization of the wave function. From this point of view it is reasonable to assume that the wave function contains all the information about the system. In fact, if the theory can provide a one-to-one link between the appearance of a well localized object and the formation of a well localized wave function, then there is no need to postulate the objective existence of other properties of the system outside of those which can be derived from the wave function.
Thus, the assumption, that all there is to know about the system is contained in the wave function, is legitimate in the framework of objective collapse theories.

The second assumption is self-explanatory. As in the Copenhagen-style interpretations, unitary time evolution is explicitely broken. Here, however, the departure from unitarity is not due to the measurement process, but to a natural process which occurs regardless of whether or not there is someone observing the system. This is expressed by assumption~(iii).\\ 
The localization process instantaneously transforms the elementary system's wave function $\psi(\vec{q})$ in a new function which is sharply peaked around a random position $\vec{X}$. 
This, in the general case of a system composed of $N$ particles, is assumed to occur according to the following scheme:
\begin{equation}
    \Psi(\vec{q_1},...,\vec{q_N}) \Longrightarrow C \Psi(\vec{q_1},...,\vec{q_N}) K e^{-\frac{1}{2d^2}|\vec{q_i}-\vec{X}|^2} ,
\label{eq:GRW_localization}
\end{equation}
In the above equation $C$ is simply the normalization constant of the new wave function, $d$ is the half-width of the gaussian collapse function, and $K$ is a normalization term for the collapse probability which introduces a non-linearity in the localization process, as it will be explained shortly.
It is important to stress that, in the context of an objective collapse theory, the parameter $d$ is a fundamental constant to be inferred from the experiments.

Eq.~(\ref{eq:GRW_localization}) shows that each random localization process acts on a single particle and in general does not imply the localization of all individual subsystems. Nevertheless such a mechanism can account for the fast localization of macroscopic systems as it can easily be understood by considering the weirdest delocalized state for a macroscopic system: the one for which the wave function of the whole system is peaked around two well separated points in space. Such a situation can be described by a wave function of the following form:
\begin{equation}
    \Psi(\vec{q_1},...,\vec{q_N}) = \psi_U(\vec{q_1})...\psi_U(\vec{q_N}) + \psi_D(\vec{q_1})...\psi_D(\vec{q_N}) ,
\end{equation}
where $\psi_U(\vec{q})$ and $\psi_D(\vec{q})$ are larger than zero only in a small region around one point, $\vec{q} = \vec{X_U}$ and $\vec{q} = \vec{X_D}$ respectively.
If one of the constituent particles of the macroscopic system, say particle $m$, undergoes a localization process around point $\vec{X_U}$, the whole wave function $\Psi(\vec{q_1},...,\vec{q_N})$ gets multiplied by $K e^{-\frac{1}{2d^2}|\vec{q_m}-\vec{X_U}|^2}$, yielding a new wave function $\Psi_{loc}$:
\begin{equation}
    \Psi_{loc} \propto \psi_U(\vec{q_1})...\psi_U(\vec{q_N})K e^{-\frac{1}{2d^2}|\vec{q_m}-\vec{X_U}|^2} + \psi_D(\vec{q_1})...\psi_D(\vec{q_N})K e^{-\frac{1}{2d^2}|\vec{q_m}-\vec{X_U}|^2} .
\label{eq:collapse_delocalized_state}
\end{equation}
Since $\psi_D(\vec{q_1})...\psi_D(\vec{q_N})$ is zero around $\vec{q} = \vec{X_U}$, and the gaussian collapse function is negligible at points where $\psi_D(\vec{q_1})...\psi_D(\vec{q_N})$ is non zero, the second term of the sum in Eq.~(\ref{eq:collapse_delocalized_state}) can be neglected and the whole $\Psi_{loc}$ becomes effectively localized around $\vec{X_U}$:
\begin{equation}
    \Psi_{loc} \propto \psi_U(\vec{q_1})...\psi_U(\vec{q_N})K e^{-\frac{1}{2d^2}|\vec{q_m}-\vec{X_U}|^2} .
\end{equation}

From the above it can be inferred that the localization processes of the GRW theory do not produce wave function collapse of general many body systems, but they do provide an effective localization of those systems which are classically defined as rigid bodies. This is consistent with empirical observations, since, while macroscopic rigid bodies do not appear delocalized, systems with a large number of particles, such as for example photons emitted by a laser, can actually be found in stable delocalized states.

Assumption (iv) needs to be made precise by stating explicitely how the spatial probability distribution for the localization processes depends on the wave function. This should be done in such a way that, within the theory, the Born rule can be reproduced to a good degree of approximation, given that so far it has not been contradicted by experiments.
To this end the GRW theory prescribes that the probability distribution, $P(\vec{X})$, of collapse processes  is given by the square modulus of the collapsed wave function:
\begin{equation}
    P(\vec{X}) = \int d\vec{q_1}...d\vec{q_N} |C|^2 \Psi^*(\vec{q_1},...,\vec{q_N})\Psi(\vec{q_1},...,\vec{q_N}) K^2 e^{-2\frac{1}{2d^2}|\vec{q_i}-\vec{X}|^2} .
\label{eq:collapse_distribution}
\end{equation}
The condition $\int d\vec{X} P(\vec{X}) = 1$ is satisfied by a suitable value of the term $K$, which consequently becomes a functional of the localized wave function integral. In this way a non-linearity is introduced in the wave function evolution: some non-linearity is in fact necessary for an actual collapse of the wave function \cite{ghirardi_collapse_2011}.

Finally, assumption (v) deals with the issue of the frequency of localization processes. 
The Poisson distribution guarantees that an elementary particle's wave function sooner or later will undergo a localization, but it is extremely unlikely that such a process occurs immediately after the system is prepared in a specific state. The mean frequency $\bar{f}$ of the distribution is the most critical parameter of the theory for two reasons. First, its value has to be large enough to allow microscopic systems to evolve for long times according to the Schr\"odinger equation, while at the same time it cannot be too large in order to allow very fast localization of macroscopic systems.
Second, average localization times can actually be measured with most recent experimental techniques \cite{van_wezel_nanoscale_2012}, consequently freedom of choice for the value of $\bar{f}$ is further constrained.

The difference between objective collapse theories and the standard Copenhagen theory does not only lie in the presence of a physical collapse mechanism, but also in a different ontology for the wave function.
An epistemic interpretation of the wave function does not make much sense in objective collapse theories. In fact, according to such an interpretation the wave function is an object whose square modulus yields the probability to find the system with a specific value of a certain property. In an objective collapse theory, however, such a specific value is determined by an actual, observer independent, modification of the wave function and the Born probabilities are an approximate consequence of the collapse probability $P(\vec{X})$.
Preserving the epistemic nature of the wave function would imply that the collapsed wave function $\Psi_{loc}(\vec{q_1},...,\vec{q_N})$, not being strictly a Dirac delta function, still admits a wide range of possible position outcomes. Clearly this interpretation would eliminate the need of postulating an objective collapse mechanism. Instead such a mechanism is postulated exactly in the attempt to \emph{identify} the appearance of a definite position of an object with the narrow shape of its wave function.

Shifting the probabilities from the measurement outcomes to the occurrence of a physical phenomenon, such as the wave function collapse described above, requires a change in the interpretation of the wave function.
In the context of an objective collapse theory it is possible to consider the square modulus of the wave function as the actual mass distribution in space. In this way the wave function acquires the status of a real physical entity. In some interpretations reality is also given to the `beables' behind the localization process. The collapse events can, in fact, be thought of as being produced by the appearance of \emph{hittings} or \emph{flashes} in space-time, which are postulated as new fundamental physical entities.

A major merit of objective collapse theories consists in providing precise predictions for collapse times and width. This makes the theory, at least in principle, falsifiable by experiments with regard to the issue of definite outcomes.
In particular, experiments achieving superposition of macroscopic systems' states, as those analyzed in Section~\ref{sec:macroscopic_superpositions}, can be designed to try and rule out the existence of a collapse mechanism as the one described in this section. If, for example, experiments were to show that reducing decoherence inducing factors allows the survival of coherent superposition states for increasingly larger times, then the hypotesis of an objective collapse mechanism of the above type would be extremely difficult to uphold.

With regard to the decoherence programme, its formalism is applicable to objective collapse theories as well, but with one peculiarity: localization times for the unobserved degrees of freedom have to be much shorter than the decoherence time of the system under investigation. In fact, if the typical localization time of the environment particles predicted by the GRW theory were longer than the decoherence time obtained with the decoherence formalism, then the partial trace of the total density matrix would not describe the statistical average over the unobserved degrees of freedom. This is because in order to interpret the quantum mechanical internal product $\bra{\psi}\hat{O}\ket{\psi}$ as the statistical average of $\hat{O}$ over many measurements, one has to apply the Born rule. And to apply the Born rule one has to assume that wave function collapse has occurred. Thus, if the theory predicts a localization process which is slower than the decoherence process, the decoherence formalism cannot be interpreted in the usual way. 
This peculiarity of objective collapse theories is another aspect which can be in principle subjected to experimental investigation.

\section{Interpretative solutions: understanding what we see}

\subsection{Relative state interpretation}
\label{sec:relative_state_interpretation}

The first and best known proposal for describing physical phenomena by means of only the state vector and the Schr\"odinger equation, without additional postulates, is the relative state interpretation by Hugh Everett III \cite{everett_theory_2011}.

The two ideas at the heart of Everett's proposal are (1) that quantum mechanics describes the evolution of a closed system which includes the observer, and (2) that measurement outcomes are definite with respect to each single observer state that appears in the entangled state vector resulting from the Von Neumann measurement scheme.
To undestand these two ideas it is convenient to clearly state Everett's assumptions. This also makes the comparison with other interpretations easier.
Everett's theory is built on the following postulates:
\begin{enumerate}[(i)]
   \item The wave function completely describes the physical system.
   \item The Schr\"odinger equation applies to the wave function of a closed system which includes the observer.
   \item The only actually closed system is the whole universe.
   \item A specific observer, i.e. someone possessing all the expected classical properties of an everyday observer, corresponds to only one of the observer states appearing in the linear superposition of the wave function of the universe.
   \item All the different systems described by each component of the linear superposition of the universal wave function do actually exist at the same time. 
\end{enumerate}
With regard to the problem of definite outcomes, the key interpretative idea is given by postulate (iv), but in order to fully understand Everett's proposal it is necessary to proceed by steps.
The state of the whole universe is completely described by a state vector $\ket{\Psi}$. The time evolution of $\ket{\Psi}$ is deterministic and completely specified by the Schr\"odinger equation. 
A reasonable scenario, before a measurement takes place, is described by the following state vector:
\begin{equation}
\ket{\Psi_{bf}} = \ket{\psi_{bf}}\ket{\phi_{bf}}\ket{\epsilon_{bf}}
    \label{eq:everett_before_measurement}
\end{equation}
where $\ket{\psi_{bf}}$ is the state of the subsystem to be observed, $\ket{\phi_{bf}}$ the state of the observer (including the measurement apparatus) who has all the classical properties one would expect, i.e. negligible uncertainty on position and momentum, and $\ket{\epsilon_{bf}}$ the state of the rest of the universe. Immediately after an interaction that, for example, measures the energy, the state will be of the form:
\begin{equation}
\ket{\Psi_{aft}} = \sum_k c_k \ket{E_k}\ket{\phi_k}\ket{\epsilon_{aft}} .
    \label{eq:everett_after_measurement}
\end{equation}
Here the state vector $\ket{\phi_k}$ represents both the measurement apparatus, with the pointer indicating the value $E_k$ on a graduated scale, and the observer registering such a value. For the sake of keeping the argument simple, it is assumed here that the state of the rest of the universe, $\ket{\epsilon_{aft}}$, does not become entangled with the object-observer subsystem during the measurement process (the entanglement may occur at later times).
The crucial aspect of Everett's proposal consists in interpreting the superposition $\sum_k c_k \ket{E_k}\ket{\phi_k}$ not as representing an observer in an entangled state, but different observers registering different values of the energy of the measured system. In other words an object in the superposition state $\ket{\psi_{bf}}$ is in the state $\ket{E_m}$ relative to the observer in the state $\ket{\phi_m}$, while it is in the state $\ket{E_n}$ relative to the observer in the state $\ket{\phi_n}$.

Such an interpretation of the wave function evidently does not require any collapse postulate, since no collapse occurs in the universal wave function. The bulk of the problem is however shifted to two different aspects of the theory: (1) the ontology of a universe with an apparently variable number of observers, and (2) the justification for the emergence of \emph{actual} observer states by appealing only to the theory's postulates. Several theories, all sharing the above listed assumptions, have been developed to address the first issue: bare theory, many worlds, many minds, many histories...\cite{barrett_everetts_2011} With regard to the main issue of this thesis it is however more relevant to address the second problem, which is common to all versions of the relative state theory, rather than analyzing the differences between them.

As it has been pointed out, the collapse postulate is avoided by assigning a different meaning to a superposition like the one in Eq.~(\ref{eq:everett_after_measurement}). In order for this programme to work out successfully, a number of questions have to be addressed:
\begin{enumerate}
    \item How to account for the empirically accurate Born rule?
    \item How to resolve the ambiguity in the choice of the vector basis on which the superposition is written?
    \item What can prevent one observer state to interfere with another observer state?
\end{enumerate}
The first question is the most puzzling one. This is not simply due to the fact that Everett style theories are completely deterministic and that all outcomes actually occur. In fact, it becomes immediately clear that outcome probabilities have to be related to each individual observer's experience (the memory of an observer corresponding to a specific Everett branch). Since, in general, at each measurement a splitting occurs, it is not unreasonable for a specific individual observer to ask what is the probability that she or her `descendant' will end up in one branch or another. The problem is that, since each possible branch actually occurs, there is no reason why a specific observer should not associate equal likelihood to all of them.

This issue has been addressed through a wide range of approaches. In his original work \cite{everett_theory_2011}, Everett did not specify in detail how the problem could be solved, but it appears that his idea for tackling it was based on the fact that if probabilities have to be associated to observers' experiences, then Born's prescription should be followed because it is the only choice which is invariant under the Schr\"odinger dynamics \cite{barrett_everetts_2011}.

Several proposals have been made to derive the Born rule by appealing only to Everett's postulates. The most recent ones are the proposals by Saunders~\cite{saunders_derivation_2004}, Wallace~\cite{wallace_everettian_2003}, and Zurek~\cite{zurek_probabilities_2005}. Saunders and Wallace build on a work by Deutsch~\cite{deutsch_quantum_1999} and attempt to explain the Born rule by means of a game based argument in which an observer is rewarded according to the measurement outcome. In their argument a reward is given to the  player depending on the outcome of a Stern-Gerlach experiment: say $p_0$ and $p_1$ for spin-down and spin-up respectively.
They define \emph{the minimum payment a rational player would accept not to play the game} \cite{alastair_i.m._everett_2009} as $V(\theta) = w_0(\theta) p_0 + w_1(\theta) p_1$, where $\theta$ is the angle of the SG apparatus with respect to the $z$ axis and the system is prepared with $S_z = +1$. Symmetry arguments show that $w_0(\theta) = \cos^2(\theta/2)$ and $w_1(\theta) = \sin^2(\theta/2)$, that is they equal the square modulus of the coefficients $c_0$ and $c_1$ of the state $\ket{+}_z$ in the vector basis rotated by an angle $\theta$.

Zurek's proposal is based on a symmetry possessed by entangled systems under a class of transformations which he calls \emph{envariant} transformations, a shorthand for environment-assisted invariant \cite{zurek_probabilities_2005}. Zurek's derivation will be presented in detail in Chapter~\ref{ch:decoherence}.

Despite the elegance and ingenuity of these derivations, recently several authors have shown \cite{patrick_van_esch_born_2007, alastair_i.m._everett_2009} that they are always based on some extra assumption which then needs to be added to Everett's postulates. 

This fact takes away one of the appealing aspects of relative state theories: being based on fewer assumptions.

The issue of the vector basis choice is also specific of Everett type theories. The problem can be stated as follows. After a measurement of the energy has taken place, the universal state vector $\ket{\Psi_{aft}}$ is given by Eq.~(\ref{eq:everett_after_measurement}), which can be interpreted as representing $N$ observers, as many as the energy eigenstates, registering $N$ different values. This same state can also be expressed as a different superposition in terms of a different orthonormal basis of the Hilbert space. Given that the total Hilbert space $\mathcal{H}$ corresponds to the tensor product $\mathcal{H}_s \otimes \mathcal{H}_O \otimes \mathcal{H}_\epsilon$, the state $\ket{\Psi_{aft}}$ can be written as:
\begin{equation}
\ket{\Psi_{aft}} = \sum_{i,j,k} a_i b_j c_k \ket{\phi_s(i)} \ket{\phi_O(j)} \ket{\phi_\epsilon(k)} ,
    \label{eq:everett_general_basis}
\end{equation}
where $\ket{\phi_s(i)}$, $\ket{\phi_O(j)}$, and $\ket{\phi_\epsilon(k)}$ are basis vectors of the Hilbert spaces $\mathcal{H}_s$, $\mathcal{H}_O$, and $\mathcal{H}_\epsilon$ respectively. An Everettian reading of the above expression describes each individual observer $\ket{\phi_O(j)}$ registering at the same time a set of different values of the observable associated to the basis of eigenvectors $\ket{\phi_s(i)}$. The definite outcomes problem seems to be back in the theory. What criteria should be used to distinguish an actual observer state from states which do not actually represent an observer?

In the spirit of Everett's proposal, the simplest criterium would be to consider as actual observer states only those univocally associated to a definite value of an observable. This implies that the actual observer states can be read directly only when the state vector is expressed in the Schmidt basis:
\begin{equation}
\ket{\Psi_{aft}} = \sum_{l} \alpha_l \ket{\xi_{s}(l)} \ket{\xi_{O}(l)} \ket{\xi_{\epsilon}(l)} ,
    \label{eq:everett_Schmidt_basis}
\end{equation}
that is, when there exists a vector basis for the Hilbert space $\cal H$ whose vectors $\ket{\Xi (l)}$ can be expressed as $\ket{\xi_{s}(l)} \otimes \ket{\xi_{O}(l)} \otimes \ket{\xi_{\epsilon}(l)}$.
Fortunately there is a theorem which ensures the uniqueness of the Schmidt decomposition \cite{elby_triorthogonal_1994}. Its existence, however, is not guaranteed if the Hilbert space is not built from the bottom up as a tensor product of smaller Hilbert spaces.

It appears that a relative state formulation of quantum mechanics could work with the prescription that the existing branches of the universe have to be read from the istantaneous Schmidt decomposition of the universal state vector. Unfortunately several studies \cite{barvinsky_preferred_1995} have demonstrated that there is no guarantee that the observer states resulting from the istantaneous Schmidt decomposition describe a system with classical or semiclassical properties. An alternative mechanism for selecting the correct observer states is required.

A possible solution comes from the decoherence approach. Instead of looking for the istantaneous Schmidt decomposition of the universal state vector, one may ask whether the interaction with the environment can provide a mechanism for an effective factorization of the terms in Eq.~(\ref{eq:everett_general_basis}). 
A positive answer comes from Zurek's idea \cite{zurek_pointer_1981, zurek_environment_induced_1982} that actual, everyday, measurement outcomes correspond to system-observer states which are not modified by the interaction with the environment. That is, once the pointer of a measurement apparatus indicates a certain value of the observable, it will remain in that position for a time long enough to at least allow recording by the observer. Moreover, the observer measurement record is supposed to be stable. All this suggests that there are system-observer states which are not modified by the interaction with the environment. As a consequence, such states, which can be indicated as $\ket{\chi_S(k)}\ket{\chi_O(k)}$, will undergo the following transition when interacting with the environment:
\begin{equation}
    \left( \sum_k \ket{\chi_S(k)}\ket{\chi_O(k)}\right) \ket{\epsilon_{aft}} \Longrightarrow \sum_k \ket{\chi_S(k)}\ket{\chi_O(k)} \ket{\chi_\epsilon(k)} .
\label{eq:Everett_robust_pointer_states}
\end{equation}

The environment interaction entangles the \emph{robust} states $\ket{\chi_S(k)}\ket{\chi_O(k)}$ with a set of environment states, in a fashion completely analogous to the Von Neumann measurement mechanism.
At this point the right-hand side expression of Eq.~(\ref{eq:Everett_robust_pointer_states}) can be interpreted, according to Everett's prescription, as describing a universe with as many branches as the number of robust pointer states, each one with an observer who has a definite and stable record of the system measured property. The detailed implementation of the robust pointer states idea will be discussed in Chapter~\ref{ch:decoherence}.

The third and last issue that needs to be addressed for a relative state interpretation to be viable is how to account for the non-interference between different Everett branches. Even though the relative state interpretation considers each term of the sum in Eq.~(\ref{eq:Everett_robust_pointer_states}) as referring to a distinct branch, or world, the Schr\"odinger dynamics may cause the state of one branch to be affected by the state of another branch. As a matter of fact, this is not what happens in the world of human observers, where macroscopic systems appear to obey the laws of classical physics. It seems quite straightforward to appeal to the decoherence mechanism to explain the lack of interference between different branches, but special care must be taken. As shown in Section~\ref{subsec:general_decoherence_approach}, the Born rule is implicit in the interpretation of reduced density matrices. An account of the emergence of probabilities in relative state theories is thus necessary for explaining the lack of interference between different branches by means of decoherence arguments.

\chapter{The decoherence approach}
\label{ch:decoherence}

In Chapter~\ref{ch:the_problem} the basic ideas behind the decoherence mechanism have been presented. The purpose of the present chapter is to analyze how such ideas can be expressed in the quantum mechanical formalism so that, starting from a quantum pure state, a classical statistics mixture can be obtained.

The very first consideration behind the development of the decoherence theory consists in realizing that the standard description of the measurement process is too idealized. Specifically, it does not take into account the fact that, in measuring the properties of a microscopic system, an amplification process is always involved. The information about the system properties, recorded by the ``probe'' of the measurement apparatus, has to be  amplified in order to be accessible by the human observer. Such an amplification can only be achieved through the interaction of a large number of apparatus subsystems either with the recording probe or directly with the system.

The key concepts at the heart of the decoherence theory emerge from an analysis of these just mentioned considerations. As a guiding aid through the many aspects of the theory, it is useful to list out such key notions:
\begin{itemize}
\item The measured system is not isolated: \\
Measurement interaction, macroscopic apparatus, and possible external environment perturbations cannot be neglected.
\item Entanglement: \\
The linearity of the Schr\"odinger evolution entangles the states of interacting systems.
\item Huge number of component subsystems:\\
The macroscopic apparatus (or the external environment), being composed of many interacting subsystems, features an almost continuous energy spectrum.
The tiny magnitude of the energy gaps increases the likelihood that environment relative states become orthogonal. \\
The information about the system's state relative phase is rapidly dispersed among the many degrees of freedom. \\
Coherence recurrence times become comparable to Poincar\'e recurrence times.
\item Energy dissipation and decoherence:\\
System-apparatus interaction produces both energy dissipation and loss of system state coherence. The latter occurs much faster than the first.
\end{itemize}
The detailed study of these key notions in specific quantum systems allows to construct a more realistic description of the measurement process.
Such decoherence based description is able to clarify many of the issues related to the transition from quantum to classical phenomena.

Currently, stating a set of general rules, applicable to any physical system, does not appear possible for the decoherence mechanism. 

A case by case mathematical treatment is thus required. However, the fact, that very different systems for which the above four concepts are applicable manifest decay of off-diagonal terms, is guarantee of the general nature of the decoherence mechanism.

One important result that emerges quite naturally from the decoherence approach is the existence of robust pointer states. This can explain why macroscopic records of quantum measurements remain stable in time, instead of changing under the perturbation of the environment.
Of particular relevance are the implications of environment selected robust pointer states in different quantum mechanics interpretations.

In what follows the process of decoherence is presented through some simple models that have become standard in the literature on the topic.
The focus is on what the decoherence results do or do not imply with regard to the measurement problem.
To this end decoherence theory is analyzed in its two basic aspects: the formalism and the mechanism. Particular care is put in pointing out features common to all systems and in identifying the \emph{essence} of decoherence.

\section{Decay of interference terms}
\label{sec:decay_interference_terms}

In Section~\ref{subsec:general_decoherence_approach} the main aspects of the decoherence process have been introduced by means of a spin-1/2 system in interaction with an unspecified environment. The two crucial stages of the process are:
\begin{enumerate}
\item The system-environment interaction that entangles the states and produces the following transition
\begin{equation}
\ket{\Psi_i} = \left(\alpha_i \ket{+}_z +\beta_i \ket{-}_z\right)\ket{\varepsilon} \Longrightarrow \ket{\Psi_f} = \alpha_f \ket{+}_z \ket{\varepsilon_+} +\beta_f \ket{-}_z \ket{\varepsilon_-} .
    \label{eq:spin_environment_entanglement_chapter4}
\end{equation}
\item The unitary evolution of the environment states $\ket{\varepsilon_+}$ and $\ket{\varepsilon_-}$ towards approximate orthogonality.
\end{enumerate}
How these two stages can actually take place in a specific physical system is the topic of the present section

\subsection{Decoherence mechanism}
\label{sec:decoherence_mechanism}

\subsubsection{Spin decoherence}

A reasonable and simple idea to model an environment interacting with a spin-1/2 system is to think of it as composed of a large number of two-level systems. The simplicity of this model, first proposed by Zurek \cite{zurek_pointer_1981, zurek_environment_induced_1982}, allows to detect some essential aspects of decoherence without the complexity of more realistic models.
Assuming the environment composed of $N$ two-level systems, labeled by an index $k$, each component's state vector can be expressed as:
\begin{equation}
\ket{\epsilon}_k = a_k \ket{0}_k + b_k \ket{1}_k ,
\label{eq:two_level_sys_state}
\end{equation}
with $\ket{0}_k$ and $\ket{1}_k$ an orthonormal basis for the $k$-th subsystem Hilbert space.
The state of the whole environment, $\ket{\varepsilon}$, can thus be represented as:
\begin{equation}
\ket{\varepsilon} = \ket{\epsilon}_1 \otimes \ldots \otimes \ket{\epsilon}_j \otimes \ldots \otimes \ket{\epsilon}_N .
\label{eq:spin_env_state}
\end{equation}

Obviously, in order for the system to be perturbed by the environment, an interaction term, $\hat H_{s\varepsilon}$, has to be present in the total Hamiltonian.
The entangled vectors $\ket{\varepsilon_+}$ and $\ket{\varepsilon_-}$ may vary greatly depending on the specific form of $\hat H_{s\varepsilon}$.
It will become clear, however, that such variations do not qualitatively affect the decoherence mechanism: the two vectors will become nearly orthogonal even though the final states may be very different in each particular case.

The interaction of a spin-1/2 with other two-level systems can be described quite generally by a Hamiltonian of the following form:
\begin{equation}
\hat H_{s\varepsilon} = \sum_{k=1}^N \hat H_{s\epsilon_k} ,
\label{eq:spin_env_int}
\end{equation}
where the terms of the sum are
\begin{equation}
\hat H_{s\epsilon_k} = \left( \ket{+}\bra{+} - \ket{-}\bra{-} \right) \otimes g_k \left( \ket{0}_k\bra{0}_k - \ket{1}_k\bra{1}_k \right) \prod_{j \neq k} \otimes \hat I_{j} .
\label{eq:spin_env_int_k}
\end{equation}
Implicit in the above Hamiltonian is the assumption that the spin-1/2 interacts with one `environment particle' at a time. This, besides being quite a realistic approximation, should not be crucial for the occurrence of decoherence. At the same time it greatly simplifies calculations.

Having specified the interaction Hamiltonian, it is possible to derive the time dependence of the system-environment state vector. For simplicity, the spin-1/2 system and environment self-Hamiltonians will be considered constant (and equal to zero).
Considering the initial state
\begin{equation}
\ket{\Psi(0)} = \ket{\Sigma(0)} \otimes \ket{\varepsilon(0)} = \left(\alpha \ket{+} +\beta \ket{-} \right) \prod_k \otimes \left(a_k \ket{0}_k + b_k \ket{1}_k \right) ,
\label{eq:spin_dec_init_state}
\end{equation}
the state at time $t$ will be:
\begin{equation}
\ket{\Psi(t)} = e^{-i\hat H_{s\varepsilon}t} \ket{\Psi(0)} = \alpha \ket{+} \ket{\varepsilon_+(t)} +\beta \ket{-} \ket{\varepsilon_-(t)} .
\label{eq:spin_dec_time_t_state}
\end{equation}
Such an expression was already introduced in Eq.~(\ref{eq:spin_environment_entanglement}), but here the explicit form of $\ket{\varepsilon_+(t)}$ and $\ket{\varepsilon_-(t)}$ can be obtained:
\begin{equation}
\ket{\varepsilon_+(t)} = \ket{\varepsilon_-(-t)} = \prod_k \otimes \left(a_k e^{ig_k t}\ket{0}_k + b_k e^{-ig_k t}\ket{1}_k \right)
\label{eq:spin_entangled_env_states}
\end{equation}

As in Eq.~(\ref{eq:spin_env_reduced_rho_simp}) the reduced density matrix $\rho_{\varepsilon r}$ corresponds to the operator $\sum_i \langle\varepsilon_i | \Psi(t) \rangle \langle\Psi(t)| \varepsilon_i\rangle$, for some environment orthonornal basis set $\{\ket{\varepsilon_i}\}$:
\begin{eqnarray}
\rho_{\varepsilon r} & = & |\alpha|^2 \ket{+} \bra{+}  + |\beta|^2 \ket{-} \bra{-} + \nonumber \\
& & \alpha^* \beta \langle \varepsilon_+(t) | \varepsilon_-(t) \rangle \ket{-} \bra{+}  + \alpha \beta^* \langle \varepsilon_-(t) | \varepsilon_+(t) \rangle \ket{+} \bra{-}  .
    \label{eq:}
\end{eqnarray}
In order to verify whether the above density matrix evolves towards an effectively diagonal one, it is necessary to study the time dependence of the off-diagonal terms.

To check whether the off-diagonal terms decrease in time it is useful to compute the square modulus of $\langle \varepsilon_+(t) | \varepsilon_-(t) \rangle$ ($=_{\mbox{def}}z(t)$):
\begin{equation}
|z(t)|^2 = \parallel \langle \varepsilon_+(t) | \varepsilon_-(t) \rangle \parallel^2 = \prod_{k=1}^N \left\lbrace 1+ \left[\left( |a_k|^2 - |b_k|^2\right)^2 -1 \right]\sin^2{2g_kt} \right\rbrace .
\label{eq:z_t_squared}
\end{equation}
Even without any information on the parameters occurring in the right-hand side of the equation, two facts can be inferred. First, at $t=0$, as expected, $z(t) = 1$. That is, the off-diagonal terms are equal to $\alpha^* \beta$ and $\alpha \beta^*$: information on the spin state phases is present in full.
Second, for $t > 0$ the expression $\sin^2{2g_kt}$ is zero only when $t = n \pi/g_k$.
If the values of the parameters $g_k$ are not all equal, as it should be in a realistic situation, where the interaction strength may depend on distance, for example, then at each instant $t$ there will likely be a large number of factors, in the product for the expression of $|z(t)|^2$, which are smaller than $1$, in contrast to few or no factors equal to $1$. It is straightforward to realize that, for very large $N$, $|z(t)|^2$ quickly becomes very small. More precisely, if $\sin^2{2g_kt}=0$ only for a finite (and realistically small) number of $k$'s, then:
\begin{equation}
\lim_{N\to \infty}|z(t)|^2 = \lim_{N\to \infty} \prod_{k=1}^N \left\lbrace 1+ \left[\left( |a_k|^2 - |b_k|^2\right)^2 -1 \right]\sin^2{2g_kt} \right\rbrace = 0 .
\label{eq:z_t_squared_limit}
\end{equation}

For a thourough understanding of the decoherence phenomenon, it is crucial to notice that, since $z(t)$  is a product of periodic functions, from time to time $\langle \varepsilon_+(t) | \varepsilon_-(t) \rangle$ will become arbitrarily close to $1$. At those times the purity of the reduced density matrix is practically restored.
The mechanism described so far predicts that occasionally the system will recover its quantum superposition state: 
a state of affairs that would appear quite odd and problematic for macroscopic systems.
Such a problem is quickly overcome by observing that the time interval between two instances of a pure superposition state is very long. It can be shown that such time of recurrence, $\tau_{rec}$, is analogous to the Poincar\'e time after which a classical system returns arbitrarily close to the initial state: $\tau_{rec}$ is proportional to $N!$, where $N$ is the number of constituent subsystems \cite{zurek_environment_induced_1982}. It is easy to verify that, for macroscopic environments, the decoherence recurrence time is larger than the lifetime of the universe.

Another crucial and general aspect which can be learned from this bit-by-bit interaction model is the role of the commutation relation between the interaction Hamiltonian and the observable to be measured. 
It is not by chance that $\hat H_{s\varepsilon}$ commutes with $S_z$ ($S_z = 1/2 \left( \ket{+}\bra{+} - \ket{-}\bra{-} \right)$). If $[ \hat H_{s\varepsilon}, S_z ] \neq 0$, then not only the off-diagonal terms of $\rho_{\varepsilon r}(t)$ would be time-dependent, but the diagonal ones as well, in the $S_z$ eigenbasis. The microsystem state would not only lose coherence, but also any relation to the initial coefficients amplitude.

\subsubsection{Spatial decoherence}

Zurek's simple model demonstrates how the application of the four leading ideas, listed at the beginning of the chapter, can account for the decay of the off-diagonal terms in a spin system. The same ideas can be applied to macroscopic systems to try and give an explanation for why naked-eye visible objects do not appear fuzzy.
Such a problem is not only relevant for the emergence of the classical world, but also for the issue of definite outcomes. In fact, most measurement operations ultimately consist in the observation of a non-fuzzy pointer position.

Obviously, it cannot be expected that the decoherence mechanism acting in position basis will provide an explanation for the collapse of the pointer state onto an (approximate) eigenstate of the position operator. What can be expected is an account for the lack of spatial interference effects for the pointer.

For a really macroscopic system, such as a chair, a ball, or the Moon, the condition of isolation is never realized when observed, because, even if the object is in a vacuum, at zero gravity and screened from radiation, it will be hit by a large number of photons when illuminated to allow naked-eye observation.
Consequently one can qualitatively imagine that interference effects, related to the coherent phase of the wave function, are quickly suppressed by the many particles randomly colliding with the object.

Such qualitative prediction has been quantitatively confirmed by the original work of Joos and Zeh \cite{joos_emergence_1985, omnes_interpretation_1994}. As already stressed, a central issue is the commutativity of the interaction Hamiltonian with the observable. In the case of the collision of particles with a macroscopic object it would be overly complicated to write down the explicit interaction Hamiltonian, which would depend on the electromagnetic interaction with the object's constituent particles. It is nevertheless possible to make precise predictions starting from some general considerations.

The first one consists in observing that each single collision of the external environment particle has no directly detectable effect on the center of mass of the macroscopic object. This implies that $\hat X$, the observable associated to the center of mass, is an approximately conserved quantity under such environment interaction. In such an approximation the effect of the collisions will consist in a change of direction of the colliding particles momentum and can thus be described by a scattering operator $\hat S_x$ that only acts on the environment state vectors, but depends parametrically on the center of mass eigenvalues $x$.

The state vector of the whole system can be represented as $\ket{\psi} \otimes \ket{\chi}$, where $\ket{\psi}$ represents the state of the macroscopic object and $\ket{\chi} = \prod_k \otimes \ket{\chi_k}$ the collective state of the environment system.

The collision of an individual particle $j$ will produce the following transition:
\begin{equation}
\ket{\psi} \otimes \ket{\chi} \Longrightarrow \ket{\psi} \otimes \hat S_x(j) \ket{\chi} = \ket{\psi} \otimes \hat S_x(j) \ket{\chi_j} \prod_{k\neq j} \otimes \ket{\chi_k} , 
    \label{eq:scattering_transition}
\end{equation}
with $\hat S_x(j)$ representing the scattering operator acting on the $j$th environment particle.

To study the localization properties of the macroscopic object, it is best to consider the reduced density matrix expressed in the eigenbasis of $\hat X$: $\rho_{r\chi} = \rho_{r\chi}(x^\prime,x^{\prime\prime})$. Since, in the present approximation, the interaction Hamiltonian commutes with $\hat X$, the collision interactions will leave the density matrix diagonal terms unchanged. As explained with the aid of Zurek's model, this is a necessary condition for the decoherence mechanism to work out as expected.

The transition described in Eq.~(\ref{eq:scattering_transition}) produces a corresponding change in the reduced density matrix:
\begin{equation}
\rho_{r\chi,t_0}(x^\prime,x^{\prime\prime}) \Longrightarrow \rho_{r\chi,t_1}(x^\prime,x^{\prime\prime}) = \rho_{r\chi,t_0}(x^\prime,x^{\prime\prime}) g_j(x^\prime,x^{\prime\prime}) .
    \label{eq:rho_scattering_transition}
\end{equation}
The term $g_j(x^\prime,x^{\prime\prime})$ accounts for the change in the scattering particle wave function. It corresponds to the inner product $\langle \chi_{j,x^\prime}|\chi_{j,x^{\prime\prime}}\rangle$, where $\ket{\chi_{j,x}}$ is the $j$th particle's state vector entangled to the state $\ket{x}$ of the macroscopic object in the expansion of $\ket{\psi}$ on the $\hat X$ eigenbasis:
\begin{equation}
\ket{\psi} = \int dx \langle x|\psi \rangle \ket{x} \prod_k \otimes \ket{\chi_{k,x}} .
    \label{eq:psi_x_scattering_expansion}
\end{equation}
As for the case of a spin in a two-level systems environment, for the sake of the argument, one can imagine that initially object and environment are disentangled and the total state vector is the tensor product of the two systems state vectors. In this case the environment particles' states, $\ket{\chi_{k}}$, do not carry the additional object center of mass label, $x$. After the collision with particle $j$ the total state vector becomes an entangled one, as in Eq.~(\ref{eq:psi_x_scattering_expansion}). The $j$th particle's state, relative to the object $\ket{x}$ eigenstate, is identified as $\ket{\chi_{k,x}}$.

Regardless of the detailed form of $\ket{\chi_{k,x}}$, the following relation holds:
\begin{equation}
g_k(x^\prime,x^{\prime\prime}) = \langle \chi_{j,x^\prime}|\chi_{j,x^{\prime\prime}}\rangle \leq 1 .
    \label{eq:scattering_term_decoh}
\end{equation}
Assuming that, in a short time interval $\Delta t$, a relatively large number $n$ of particles, e.g. photons with a random momentum distribution, hit the object, the reduced density matrix $\rho_{r\chi,t_0}(x^\prime,x^{\prime\prime})$ will change into:
\begin{equation}
\rho_{r\chi,t_1}(x^\prime,x^{\prime\prime}) = \rho_{r\chi,t_0}(x^\prime,x^{\prime\prime}) \prod_{k=1}^n g_k(x^\prime,x^{\prime\prime}) .
    \label{eq:rho_many_scattering_transition}
\end{equation}
Given the randomness of the particles momenta it is unlikely that $g_k(x^\prime,x^{\prime\prime}) = 1$ (with $x^\prime \neq x^{\prime\prime}$), if not for a small number of particles. Consequently, after a short interval $\Delta t$, the relation $\prod_{k=1}^n g_k(x^\prime,x^{\prime\prime}) \ll 1$ will likely hold. 
In conclusion, when the reduced density matrix is expressed in the $\hat X$ eigenbasis, the effect of a large number of small random collisions consists in leaving the diagonal terms unchanged and in drastically reducing the magnitude of the off-diagonal ones.

Such an approximate result is particularly enlightening with regard to the interference of large molecules.
Besides the technical difficulties related to the preparation of the molecules initial state and of a suitable diffraction grid, large molecules do not usually produce interference patterns. This is mainly due to the increased likelyhood of interacting with photons or other particles in the environment, because of their higher number of constituent particles and larger size.

One could think that the relatively high number of internal degrees of freedom might themselves constitute the `environment' which decoheres the state of the molecule's center of mass. The interaction between the internal components of an object cannot, however, affect its center of mass motion and thus cannot produce decoherence with respect to the molecule's position.
Such a theoretical prediction is confirmed by the experiments presented in Section~\ref{sec:double_slit_exp}. The authors report that the internal degrees of freedom of the large molecules did not destroy the interference, once the external perturbations were suppressed.
Moreover the interference pattern was consistent with the prediction based on a single particle with the total mass of the molecule.

Once the perturbation by the external environment becomes non-negligible, the classically expected distribution of spots on the screen is found. As already stressed before, while such classical distribution is explained by the scattering mechanism presented above, the reason why an individual molecule leaves a mark on one spot instead of another requires an independent explanation.

The scattering induced spatial decoherence is also useful in explaining the practically irreversible nature of the process. In fact, once an environment particle has collided with the object, it will move away from it, carrying along a piece of information about the object's wave function's phase. At this point it is highly unlikely that the interaction with other particles will bring it back for a second collision which would restore the initial state of the object.

\subsection{Decoherence time}
  \label{decoherence_time}

\subsubsection{Decoherence by scattering}

In their 1985 study, Joos and Zeh \cite{joos_emergence_1985, omnes_interpretation_1994} derived an explicit time dependence for the off-diagonal terms. Assuming that the system is enclosed in a large cubic volume of side $L$, they show that each decoherence factor, $g_k(x^\prime,x^{\prime\prime})$, can be written as:
\begin{equation}
g_k(x^\prime,x^{\prime\prime}) = 1 - \frac{K_k(x^\prime,x^{\prime\prime})}{L^2} - i \frac{\Phi_k(x^\prime,x^{\prime\prime})}{L^2} .
    \label{eq:scattering_decoh_factor}
\end{equation}
The functions $K_k(x^\prime,x^{\prime\prime})$ and $\Phi_k(x^\prime,x^{\prime\prime})$ contain all the complex information about the interaction details. Fortunately it can be shown that, when $|x^\prime - x^{\prime\prime}|$ is large with respect to the de Broglie wavelength, $\lambda_k$, of the environment particles, $K_k(x^\prime,x^{\prime\prime}) \approx \sigma_T$ ($\sigma_T$ being the particle-object total cross section) and $\Phi_k(x^\prime,x^{\prime\prime})$ is negligible.
The condition $|x^\prime - x^{\prime\prime}| \gg \lambda_k$ is easily satisfied under ordinary circumstances. First of all the off-diagonal terms which are assumed to be more strongly suppressed are those farther away from the diagonal, and thus with a larger value of $|x^\prime - x^{\prime\prime}|$. Furthermore, a relatively small value of $\lambda_k$ simply implies a relatively high particle momentum, $p_k$. For example, it is not difficult to verify that the above condition is satisfied for the case of a macroscopic object hit by visible photons.

In this approximation it is possible to derive an explicit time dependence for the decoherence factor $\prod_{k=1}^n g_k(x^\prime,x^{\prime\prime})$ of Eq~\ref{eq:rho_many_scattering_transition}. If $\eta$ is the number of particles per unit volume in the environment, then the number $n$ of particles that hit the object in an interval $\Delta t$ ($=t-t_0$) will be proportional to the number of particles entering the containing volume per unit time:
\begin{equation}
n = C \eta L^2 v \cdot (t-t_0) .
    \label{eq:scattering_particles_number}
\end{equation}
$C$ is a proportionality constant and $v$ the average speed of the particles.
It is now possible to obtain:
\begin{eqnarray}
g(x^\prime, x^{\prime\prime}, t) & = & \prod_{k=1}^n g_k(x^\prime,x^{\prime\prime}) \\ \nonumber
& = & \prod_{k=1}^n \left[ 1 - \frac{K_k(x^\prime,x^{\prime\prime})}{L^2} - i \frac{\Phi_k(x^\prime,x^{\prime\prime})}{L^2} \right] 
\simeq \left[ 1 - \frac{\sigma_T}{L^2} \right]^{C \eta L^2 v (t-t_0)} .
    \label{eq:scattering_decoh_factor_time_dependence}
\end{eqnarray}
Since the side $L$ of the containing volume can be made arbitrarily large, with a good degree of accuracy it is possible to write:
\begin{equation}
\left[ 1 - \frac{\sigma_T}{L^2} \right]^{C \eta L^2 v (t-t_0)} \simeq e^{- C \eta \sigma_T v (t-t_0)}
    \label{eq:scattering_decoh_factor_exponential_decay}
\end{equation}

Though very rough, this simple model predicts an exponential decay of the off-diagonal terms which qualitatively accounts for the difficulty of observing spatial interference of macroscopic objects.
As one would intuitively expect, the  decoherence time, $\tau_d = 1/(C \eta \sigma_T v)$, is inversely proportional to the density of environment particles, to their average speed and to the object's cross section. The latter is not only related to the size of the object, but also to the object-environment interaction strength.

At this stage it is already possible to make a basic comparison between the time it takes for a macroscopic object to dissipate a sizeable amount of its internal energy and the decoherence time. In the above model both energy dissipation and decoherence are produced by the interaction with the external environment. A remarkable difference may arise from the fact that the decoherence time, $\tau_d$, depends on the velocity of the colliding particles and is independent on their masses. One can thus imagine an environment constituted by very light particles with small momenta. While the decoherence time will be the same as for heavier particles, the rate of energy dissipation will be smaller. This elementary consideration may already hint at the higher efficiency of the decoherence process with respect to the energy dissipation.

\subsubsection{Spin interactions}

An explicit formula for the decay of the off-diagonal elements can also be derived for the spin system studied by Zurek. Cucchietti et al. \cite{zurek_gaussian_2003,cucchietti_decoherence_2005} examine the time dependence of the decoherence factor $z(t)$ for various distributions of the coupling constants $g_k$'s.

The simplest case is given by a uniform distribution with $g_k = g$ for all $k$'s. In such a case, with the additional, somewhat unrealistic, further assumption that $\alpha_k = \alpha$ and $\beta_k = \beta$ for all $k$'s, one can write:
\begin{equation}
z(t) = \left( |\alpha|^2 e^{igt} + |\beta|^2 e^{-igt} \right)^n ,
    \label{eq:z_t_uniform_g_k}
\end{equation}
and by using the binomial expansion one obtains:
\begin{equation}
z(t) = \sum_{k=0}^n \binom{n}{k} |\alpha|^{2k} |\beta|^{2(n-k)} e^{-igt(2k-n)} . 
    \label{eq:z_t_uniform_g_k_binomial}
\end{equation}
For a large number of environment two-level systems, $n$, the coefficients of the imaginary exponential terms can be well approximated by a Gaussian as follows:
\begin{equation}
\binom{n}{k} |\alpha|^{2k} |\beta|^{2(n-k)} \simeq \frac{e^{-(k-n|\alpha|^2)^2/(2n|\alpha|^2|\beta|^2)}}{\sqrt{2\pi n|\alpha|^2|\beta|^2}} .
    \label{eq:binomial_gaussian}
\end{equation}
Substituting the above expression in Eq.~(\ref{eq:z_t_uniform_g_k_binomial}), $z(t)$ becomes the Fourier transform of a Gaussian, which is itself a gaussian function of the variable $t$:
\begin{equation}
z(t) = \sum_{k=0}^n \frac{e^{-(k-n|\alpha|^2)^2/(2n|\alpha|^2|\beta|^2)}}{\sqrt{2\pi n|\alpha|^2|\beta|^2}} e^{-igt(2k-n)} . 
    \label{eq:z_t_uniform_g_k_gaussian}
\end{equation}
Some comment on this result is necessary. Not only the decoherence factor is only approximately gaussian, but for large times such an approximation breaks down completely. In fact, the exact $z(t)$, as expressed in Eq.~(\ref{eq:z_t_uniform_g_k}), is a periodic function with period $T = 2\pi/g$. This trivial result is a consequence of the unrealistically simplified assumptions of uniform coupling distribution and identical initial state for all the environment subsystems. Nevertheless, the comparison between the approximate and the exact time dependence allows to highlight a basic aspect of decoherence: the effective, short time, evolution and the unobservable long time behavior. Choosing a very small coupling $g$, it is possible to make the recurrence time $T$ arbitrarily long. Obviously, the decoherence time too will increase indefinitely, but, once $T$ becomes too large with respect to practical observation times, an observer will only notice the first, approximately gaussian, decaying part.

Besides the above trivial case, Cucchietti et al. \cite{zurek_gaussian_2003,cucchietti_decoherence_2005} studied in detail the time dependence of the decoherence factor under the very generic assumption that the couplings $g_k$ are \emph{sufficiently concentrated} around an average value $\bar g$. Again, an approximately gaussian time dependence is found:
\begin{equation}
z(t) \simeq e^{iB_n t} e^{-s_n^2 t^2/2} , 
    \label{eq:z_t_gaussian_gk}
\end{equation}
where $B_n$ and $s_n$ depend on the variance of the $g_k$ distribution and on $n$, the total number of environment subsystems. In this case too the approximations hide the recurrence of coherence predicted by the exact time dependence, but the more realistic assumptions, such as the randomness of the couplings $g_k$ and of the subsystems initial states, allow a much more insightful interpretation of the result.

Since $s_n^2 = \sum_{k=1,n} 4 |\alpha_k|^2 |\beta_k|^2 g_k^2$, it is easy to have a situation where $s_n^2$ is relatively large and provides a very short decoherence time, while the single $g_k$'s are small. In this way, thanks also to the randomness of the couplings, the recurrence time for coherence becomes easily very large.

In support of the explicative power of the decoherence programme, the authors of the work notice that, with different distributions of couplings, different time dependences arise, but the fast suppression of interference terms is achieved for all realistic choices of parameters.

\subsubsection{Harmonic oscillators, dissipation and decoherence times}

A striking demonstration of the effectiveness of the decoherence mechanism comes from the study of a pendulum weakly interacting with an environment constituted by a large number of other harmonic oscillators \cite{k._hepp__1973}. Such an environment can be thought of as modelling the effect of other degrees of freedom of a macroscopic  pendulum, e.g. the molecules of the string or of the spring etc..., rather than the interaction with the air molecules.

The dynamics of the whole system is determined by the total Hamiltonian:
\begin{equation}
\hat H_{T} = \hbar \omega \hat a^\dagger \hat a + \hbar \sum_{k=1}^n \omega_k \hat b_k^\dagger \hat b_k + \hat H_{I},
    \label{eq:pendulum_decoh_Hamiltonian}
\end{equation}
where $\hat a$ is the annihilation operator associated to the pendulum and $\hat b_k$ the one associated to the $k$th environment oscillator.
A reasonable interaction Hamiltonian, $\hat H_{I}$, can be chosen as:
\begin{equation}
\hat H_{I} = \hbar \sum_{k=1}^n \left( \lambda_k \hat a^\dagger \hat b_k + \lambda_k^* \hat b_k^\dagger \hat a \right) .
    \label{eq:pendulum_decoh_int_Hamiltonian}
\end{equation}
Assuming small coupling constants $\lambda_k$, one can expect that, as a good approximation, the pendulum evolution will be dominated by the self-Hamiltonian $\hbar \omega \hat a^\dagger \hat a$ and modified by a slow energy dissipation due to the environment interaction.
As it will become clear later, the crucial effect is the practical irreversibility of the dissipation process. Such irreversibility is taken here as an assumption which will be discussed and motivated later on.

A convenient choice for examining the effect of the environment on a pendulum quantum state is to choose a superposition of two harmonic oscillator coherent states. The occurrence of decoherence should not depend on the details of the initial state, but coherent states offer two advantages: (1) they have convenient mathematical properties, and (2) their evolution shows several classical features, such as the spatial wave function does not spread out and the position expectation value follows the classical equation of motion.

A harmonic oscillator coherent state is defined as an eigenstate of the annihilation operator $\hat a$:
\begin{equation}
\hat a \ket{\alpha} = \alpha \ket{\alpha} .
    \label{eq:coherent_state_1}
\end{equation}
Expanding $\ket{\alpha}$ on the oscillator's energy eigenbasis $\{ \ket{n} \}$ it can easily be found that:
\begin{equation}
\ket{\alpha} = \exp{\left(-\frac{|\alpha|^2}{2}\right)} \sum_{n=0}^\infty \frac{\alpha^n}{(n!)^{1/2}} \ket{n} .
    \label{eq:coherent_state_2}
\end{equation}
For the purpose of decoherence calculations it is useful to evaluate the expression for the scalar product $\langle \alpha | \alpha^\prime \rangle$ between two different coherent states which can be found to be:
\begin{equation}
\langle \alpha | \alpha^\prime \rangle = e^{-\left(|\alpha - \alpha^\prime|^2 +i \phi\right)/2} ,
    \label{eq:coherent_state_scalar_product}
\end{equation}
with $\phi = \mbox{Im}\left(\alpha {\alpha^\prime}^* - \alpha^* \alpha^\prime \right)$.
Finally, expressing the operator $\hat a$ in terms of the position and momentum operators, $\hat a = 1/\sqrt{2m\hbar \omega}\left( m\omega \hat X + i \hat P \right)$, one finds an expression for the eigenvalue $\alpha$ in terms of two parameters $x_0$ and $p_0$ which are the amplitude and initial momentum of the corresponding classical motion:
\begin{equation}
\alpha = \frac{1}{\sqrt{2m\hbar \omega}} \left( m\omega x_0 + i p_0 \right) .
    \label{eq:coherent_state_alpha}
\end{equation}

A remarkable feature of coherent states is that they remain eigenstates of $\hat a$ under the Schr\"odinger evolution, that is they remain coherent, though their eigenvalue changes:
\begin{equation}
i\hbar \frac{d}{dt} \ket{\psi_\alpha} = \hbar \omega \left( \hat a^\dagger \hat a +\frac{1}{2} \right) \ket{\psi_\alpha} ; \hspace{0.5 cm} \ket{\psi_\alpha} = e^{-i\omega t} \ket{\alpha(t)} ; \hspace{0.5 cm} \alpha(t) = \alpha_0 e^{-i\omega t} .
    \label{eq:coherent_state_t}
\end{equation}

If the Hamiltonian is given by Eq.~(\ref{eq:pendulum_decoh_Hamiltonian}), then $\ket{\psi_\alpha}$, with $\alpha(t) = \alpha_0 e^{-i\omega t}$, does not satisfy the Schr\"odinger equation. Fortunately, for $H_I$ given by Eq.~(\ref{eq:pendulum_decoh_int_Hamiltonian}) the solution is well approximated by substituting $\alpha(t)$ with a damped function defined as follows:
\begin{equation}
\alpha_d(t) = \alpha(0) e^{\left(-i(\omega+\Delta \omega)t - \gamma t\right)} + \xi(t) ,
    \label{eq:damped_pendulum_approx_coherent_sol}
\end{equation}
where $\xi(t)$ is a small fluctuating term.
A helpful explanation of the above solution can be found in the book by Omn\`es \cite{omnes_interpretation_1994} and its detailed derivation in the original article by Hepp and Lieb \cite{k._hepp__1973}. For the purpose of the present discussion it is enough to examine it in detail and understand its range of validity.

The parameter $\Delta \omega$ is a small shift in frequency due to the coupling with the environment oscillators and $\gamma$ is a damping coefficient which grows with the couplings $\lambda_k$ and with the density of environment energy eigenstates around the pendulum frequency $\omega$. The use of a continuous  density of states function in the derivation of Eq.~(\ref{eq:damped_pendulum_approx_coherent_sol}), besides the computational convenience, has a crucial physical implication: the infinitesimal gap between energy levels of the environment allows the pendulum to lose energy quanta in a continuous way and to easily disperse such energy in the environment.

For a finite number of environment oscillators the energy spectrum would not be continuous and the damping term, $\exp{\left(-\gamma t \right)}$, would not emerge in the exact solution (assuming one could obtain it) of the Schr\"odinger equation. In this case it can be expected that the pendulum oscillation amplitude, related to $\alpha$ by means of Eq.~(\ref{eq:coherent_state_alpha}), will gradually decrease at a rate which is very close to an exponential decay, but, for very large times, there may be occasional reappearence of large oscillations. Again, the time for such an occurrence is impractically long for realistically large number of environment oscillators, and thus Eq.~(\ref{eq:damped_pendulum_approx_coherent_sol}) is a good approximation for all practical purposes.

With all the above useful background knowledge it is now possible to consider the  time evolution of a pendulum in a superposition of two coherent states. Such a state could be classically described as a pendulum oscillating with two different amplitudes and phases at the same time.

For simplifying the computation it is convenient to assume that the environment oscillators too are initially in coherent states $\{\ket{\beta_k(0)} \}$.
The pendulum-environment state, at time $t=0$, is:
\begin{equation}
\ket{\Psi(0)} = \left( a\ket{\alpha_{1}(0)} + b \ket{\alpha_{2}(0)} \right) \ket{\{ \beta_k(0) \}} ,
    \label{eq:pendulum_env_initial_state}
\end{equation}
where $\ket{\{ \beta_k(0) \}}$ is a shorthand notation for $\prod_{k=1}^n \otimes \ket{\beta_k(0)}$ . 
The state at time $t$ will be:
\begin{equation}
\ket{\Psi(t)} = a \ket{\alpha_{d1}(t)} \ket{\{ \beta_{d1k}(t) \}} + b \ket{\alpha_{d2}(t)} \ket{\{ \beta_{d2k}(t) \}} ,
    \label{eq:pendulum_env_state_t}
\end{equation}
where the label `d' stands for `dissipation' and indicates a function of the type presented in Eq.~(\ref{eq:damped_pendulum_approx_coherent_sol}).

The reduced density operator is given by:
\begin{eqnarray}
\rho_r(t)  & = & |a|^2 \ket{\alpha_{d1}(t)}\bra{\alpha_{d1}(t)} + |b|^2 \ket{\alpha_{d2}(t)}\bra{\alpha_{d2}(t)} \\ \nonumber
& & + a b^* \langle \{ \beta_{d2k}(t) \} | \{ \beta_{d1k}(t) \} \rangle \ket{\alpha_{d1}(t)}\bra{\alpha_{d2}(t)} \\ \nonumber
& & + a^* b \langle \{ \beta_{d1k}(t) \} | \{ \beta_{d2k}(t) \} \ket{\alpha_{d2}(t)}\bra{\alpha_{d1}(t)} .
  \label{eq:pendulum_env_reduced_density_matrix}
\end{eqnarray}
Notice that $\langle \{ \beta_{d2k}(t) \} | \{ \beta_{d1k}(t) \} \rangle = \langle \varepsilon_2 (t) |  \varepsilon_1(t) \rangle$, the scalar product of the environment states entangled to the pendulum state, as in the previous examples.

By making use of Eq.~(\ref{eq:coherent_state_scalar_product}) one finds:
\begin{eqnarray}
\langle \{ \beta_{d2k}(t) \} | \{ \beta_{d1k}(t) \} \rangle & = & \prod_{k=1}^n \langle \beta_{d2k}(t) | \beta_{d1k}(t) \rangle \\ \nonumber
& = & \exp{(- \frac{1}{2} \sum_{k=1}^n |\beta_{d1k}(t) - \beta_{d2k}(t)|^2 + i \Phi)} .
    \label{eq:pendulum_env_scalar_prod_1}
\end{eqnarray}
The function $\Phi$ includes the contribution of all the $\phi_k$ (see $\phi$ in Eq.~(\ref{eq:coherent_state_scalar_product}) and constitutes a fluctuating term.

In order to find the explicit time dependence of the off-diagonal terms it is useful to express the sum over the environment oscillators in terms of the parameter $\alpha_d(t)$ of the pendulum. This can be done with the help of a useful relation found by Weisskopf and Wigner \cite{v._weisskopf__1930}:
\begin{equation}
\alpha_1^* \alpha_2 + \sum_{k=1}^n \beta_{k1}^* \beta_{k2} = C ; \hspace{1 cm} \mbox{$C$ is a constant}.
    \label{Weisskopf_Wigner_relation}
\end{equation}
Such a conservation relation holds for any two exact solutions, $\ket{\alpha_{1}(t)} \ket{\{ \beta_{1k}(t) \}}$ and $\ket{\alpha_{2}(t)} \ket{\{ \beta_{2k}(t) \}}$, of the Schr\"odinger equation with Hamiltonian given by Eq.~(\ref{eq:pendulum_decoh_Hamiltonian}) and Eq.~(\ref{eq:pendulum_decoh_int_Hamiltonian}).
Substituting the sums $\sum_{k=1}^n \beta_{d1k}^*(t)\beta_{d2k}(t)$ and $\sum_{k=1}^n \beta_{d1k}(t)\beta_{d2k}^*(t)$ with the corresponding expressions in terms of $\alpha_{d1}(t)$ and $\alpha_{d2}(t)$, the decoherence factor becomes:
\begin{equation}
\langle \{ \beta_{d2k}(t) \} | \{ \beta_{d1k}(t) \} \rangle =
\exp{\left(- \frac{1}{2} \left[ |\alpha_{d1}(0) - \alpha_{d2}(0)|^2 - |\alpha_{d1}(t) - \alpha_{d2}(t)|^2 \right] + i \Phi \right)} .
    \label{eq:pendulum_env_scalar_prod_2}
\end{equation}
Finally by writing explicitely the time dependence of $\alpha_{d1}(t)$ and $\alpha_{d2}(t)$ the following expression is obtained:
\begin{equation}
\langle \{ \beta_{d2k}(t) \} | \{ \beta_{d1k}(t) \} \rangle =
\exp{\left(- \frac{1}{2} \left[ |\alpha_{d1}(0) - \alpha_{d2}(0)|^2 (1-e^{-2\gamma t}) \right] + i \Phi \right)} .
    \label{eq:pendulum_env_scalar_prod_3}
\end{equation}
With regard to the decay of the off-diagonal terms the relevant expression is the negative exponent. In order to understand its physical significance it is useful to express it in terms of the $x_0$ and $p_0$ parameters of the initial coherent states. Assuming for simplicity $p_{01} = p_{02} = 0$, the exponent becomes:
\begin{equation}
\frac{1}{2} \left[ |\alpha_{d1}(0) - \alpha_{d2}(0)|^2 (1-e^{-2\gamma t}) \right] = \frac{1}{4} \frac{m\omega^2}{\hbar} |x_{01} - x_{02}|^2 (1-e^{-2\gamma t}).
    \label{eq:pendulum_off_diag_exponent}
\end{equation}
It is now instructive to examine the two limits of the above expression, for $t \rightarrow \infty$ and for $t \sim 0$. In the first case one has:
\begin{equation}
\lim_{t\rightarrow \infty} \frac{1}{4} \frac{m\omega^2}{\hbar} |x_{01} - x_{02}|^2 (1-e^{-2\gamma t}) = \frac{1}{4} \frac{m\omega^2}{\hbar} |x_{01} - x_{02}|^2 .
    \label{eq:pendulum_off_diag_exponent_limit_infty}
\end{equation}
This result implies that the decoherence factor does not decay to zero for large times, but it tends to $\exp{(-\frac{1}{4} \frac{m\omega^2}{\hbar} |x_{01} - x_{02}|^2)}$. It is easy to see that for macroscopic systems such a value can be very small. First of all for $m \rightarrow \infty$ the exponential goes to zero. Secondly, a clear macroscopic superposition requires a sizeable value of the difference $|x_{01} - x_{02}|$ between the two coherent states amplitudes. Such a factor, again, produces negligible off-diagonal terms in the reduced density matrix.

A much more interesting information can be found by studying the short time decay rate of the decoherence factor. For $t \sim 0$ the factor $(1-e^{-2\gamma t})$ is well approximated by $2\gamma t$. At short times immediately after the pendulum becomes entangled with the environment, the decoherence factor has the following time dependence:
\begin{equation}
\langle \{ \beta_{d2k}(t) \} | \{ \beta_{d1k}(t) \} \rangle =
g(t) \exp{\left(- 2 K \gamma t\right)} ,
    \label{eq:pendulum_off_diag_limit_0}
\end{equation}
with $K = m\omega^2/(4\hbar) |x_{01} - x_{02}|^2$ and $g(t)$ a fluctuating term. In order to make sense of this result it is helpful to compare it with the time dependence of the pendulum average energy: $\bar E \propto |\alpha(t)|^2 \propto \exp{(-2\gamma t)}$. As for the characteristic energy dissipation time, $\tau = 1/(2\gamma)$, a characteristic decoherence time can be defined as $\tau_d = 1/(2 K \gamma) = \tau/K$. By substituting, in the expression for $K$, values corresponding to a macroscopic system, it becomes clear that $\tau_d$ is easily many orders of magnitude smaller than $\tau$. For example, in the case of a pendulum with $m=1 \mbox{g}$, $T = 1 \mbox{s}$ and $|x_{01} - x_{02}| = 1 \mu\mbox{m}$, one finds the striking result $\tau_d \sim 10^{-20} \tau$ \cite{omnes_interpretation_1994}. This means that, by the time the pendulum has lost a negligible part of its energy, its quantum state has already completely lost its phase coherence.

\subsection{The essence of decoherence}
\label{sec:essence_of_decoherence}
Having acquired some detailed knowledge about the basic implementations of the decoherence programme, it is worthwhile to examine again what are the essential ingredients of the decoherence mechanism and what is its relation with the process of energy dissipation.

At the beginning of this chapter, four key notions have been proposed as an aid to go through the jungle of each physical system's specific details, but now it is possible to identify the two crucial conditions for the occurrence of decoherence: entanglement and an environment with an almost continuous energy spectrum.

It is evident that, without entanglement, the decoherence factor is $\langle\varepsilon_1|\varepsilon_2\rangle = \langle\varepsilon|\varepsilon\rangle = 1$ and the
off-diagonal elements of the reduced density matrix will never disappear.

As soon as system and environment become entangled, the two (or more, depending on the initial superposition) environment relative states, $\ket{\varepsilon_1}$ and $\ket{\varepsilon_2}$, can start to evolve independently, thanks to the linearity of the Schr\"odinger equation. However there are no fundamental laws which force the two states to become orthogonal. In fact, in general they don't. What makes extremely likely that they become orthogonal is the almost continuous energy spectrum of the environment; and what makes it extremely unlikely that their scalar product returns non-negligible is the huge dimension of the environment Hilbert space.

The previous result about the proportionality between the pendulum's decoherence time and the characteristic time of dissipation might induce someone to think that the decoherence process is caused or depends on the process of energy dissipation, but this would not be correct. The two processes are related through some common necessary condition, but they do not share a cause-effect relation.

In order to understand better the environment states orthogonalization process, it is helpful to consider the following simplistic situation: $\ket{\varepsilon} = \prod_{k=1}^n \otimes \ket{\epsilon_k}$, where each particle state $\ket{\epsilon_k}$ is (artificially) assumed to be one of the single particle energy eigenstates $\ket{e_i}$.

The system-environment interaction Hamiltonian, starting from a system superposition state, will in general produce two different environment relative states $\ket{\varepsilon_1}$ and $\ket{\varepsilon_2}$. If the energy gap between $\ket{e_i}$ is extremely small, a tiny transfer of energy from the system to the environment can excite one of the environment particles to a higher energy level. Assuming that in $\ket{\varepsilon_1}$ it is particle $j$ which jumps to a higher energy eigenstate, while in $\ket{\varepsilon_2}$ it is particle $l$, then a negligible energy dissipation from the system to the environment causes the orthogonalization of $\ket{\varepsilon_1}$ and $\ket{\varepsilon_2}$:
\begin{equation}
\langle\varepsilon_1|\varepsilon_2\rangle = \prod_{k=1}^n \langle\epsilon_{1k}|\epsilon_{2k}\rangle = \prod_{k=1}^n \langle e_{1k}| e_{2k}\rangle = \langle e_{1j}| e_{2j}\rangle \langle e_{1l}| e_{2l}\rangle = 0 .
    \label{eq:essence_decoh_orthogonalization}
\end{equation}
Immediate loss of information of the off-diagonal elements of the reduced density matrix follows.

Such an ad-hoc model, though oversimplified to serve any predictive purpose, clarifies the relation between decoherence and energy dissipation. Decoherence occurs at such shorter time scales with respect to energy dissipation because it only requires two quantum states to become orthogonal and such a process, in the case of almost continuous energy spectrum and large number of subsystems, may require very little energy and be very likely to occur.

The huge number of environment subsystems is a necessary condition for the occurrence of both energy dissipation and decoherence. More precisely, it is necessary to make recurrence times impractically long and produce irreversible processes out of a reversible dynamics. 

In this regard, in his 1982 paper \cite{zurek_environment_induced_1982}, Zurek shows that for models of realistic, though finite sized, environments the recurrence times are easily larger than the lifetime of the universe. The fluctuations of the decoherence factor square modulus are of the order of $1/\sqrt{n}$.

The `for all practical purposes' nature of the decoherence irreversibility, should not be considered a weakness of the theory. In fact, if Poincar\`e recurrence times are not considered a problem in classical mechanics, they should not be a problem in quantum mechanics either.

\section{Robust pointer states}
\label{sec:robust_pointer_states}

In the discussion of the spin-1/2 decoherence in Section~\ref{sec:decoherence_mechanism}, it has been noticed that, in order to obtain a diagonal reduced density matrix in the $\hat S_z$ eigenbasis, the system-environment interaction Hamiltonian has to commute with $\hat S_z$. 
Such a requirement carries remarkable consequences with respect to quantum measurement theory, but has not been listed among the essential aspects of decoherence in Section~\ref{sec:essence_of_decoherence}. The reason rests in the effort to disentangle as much as possible the many facets of decoherence theory so that its consequences on the definite outcomes problem can be clearly identified.

In this thesis the term `decoherence' is being used in its narrow meaning of loss of information with regard to the relative phases of a quantum superposition state. In order to understand the different roles that the interaction Hamiltonian commutation properties on one hand, and entanglement with an almost continuous energy spectrum environment on the other hand, play in the decoherence process, it is helpful to consider the usual spin-1/2 system and a generic interaction Hamiltonian $\hat H_I$.

After preparing the system in a $S_z$ superposition state, the interaction with the environment will produce the following transition:
\begin{equation}
\left(\alpha_0 \ket{+} +\beta_0 \ket{-} \right) \ket{\varepsilon} \Longrightarrow \alpha_1(t) \ket{+} \ket{\varepsilon_+(t)} +\beta_1(t) \ket{-} \ket{\varepsilon_-(t)} .
\label{eq:spin_dec_generic_H_I}
\end{equation}
The situation with $\alpha_1 = \alpha_0$ and $\beta_1 = \beta_0$ is realized only if $[\hat S_z , \hat H_I] = 0$. If $\hat H_I$ does not commute with $\hat S_z$, the reduced density matrix will have the general form:
\begin{equation}
\rho_{\varepsilon r} = \left(
    \begin{array}{cc}
|\alpha_1(t)|^2 & \alpha_1(t) \beta_1^*(t) \langle \varepsilon_-(t) | \varepsilon_+(t) \rangle\\
 & \\
 \alpha_1^*(t) \beta_1(t) \langle \varepsilon_+(t) | \varepsilon_-(t) \rangle & |\beta_1(t)|^2
    \end{array}
\right) .
    \label{eq:spin_env_reduced_rho_matrix_generic_H_I}
\end{equation}
As long as the environment system is composed of a large number of subsystems and the gaps in the energy spectrum are small, the scalar product $\langle \varepsilon_-(t) | \varepsilon_+(t) \rangle$ will tend very rapidly to zero, as in all the cases already examined. 
As it is evident, the difference with respect to the $[\hat S_z , \hat H_I] = 0$ case lies in the diagonal terms that do not remain constant. This, obviously, means that the interaction with the environment will modify the spin-1/2 superposition in a way that is not controllable nor recoverable. If a measurement of $S_z$ is performed, it is not clear what is the relation between the measurement outcome and the initial state of the system before the interaction with the environment.

In such a general situation, a deeper understanding of the decoherence mechanism can be obtained by considering the initial state of the spin-1/2 system expressed in the eigenbasis of a hermitian operator that commutes with $\hat H_I$. Calling such an operator $\hat D$ and its eigenvectors $\ket{0}_D$ and $\ket{1}_D$ one has $\left(\alpha_0 \ket{+} +\beta_0 \ket{-} \right) \ket{\varepsilon} = \left(\alpha_{D} \ket{0}_D +\beta_{D} \ket{1}_D \right) \ket{\varepsilon}$.
Since $[\hat D , \hat H_I] = 0$, the interaction with the environment will give:
\begin{equation}
\left(\alpha_{D} \ket{0}_D +\beta_{D} \ket{1}_D \right) \ket{\varepsilon} \Longrightarrow \alpha_{D} \ket{0}_D \ket{\varepsilon_{D0}(t)} +\beta_{D} \ket{1}_D \ket{\varepsilon_{D1}(t)} .
\label{eq:spin_dec_generic_H_I_commuting}
\end{equation}
In the $\hat D$ eigenbasis the reduced density matrix will be:
\begin{equation}
\rho_{\varepsilon r} = \left(
    \begin{array}{cc}
|\alpha_D|^2 & \alpha_D \beta_D^* \langle \varepsilon_{D0}(t) | \varepsilon_{D1}(t) \rangle\\
 & \\
 \alpha_D^* \beta_D \langle \varepsilon_{D1}(t) | \varepsilon_{D0}(t) \rangle & |\beta_D|^2
    \end{array}
\right) .
    \label{eq:spin_env_reduced_rho_matrix_generic_H_I_D_eigen}
\end{equation}
While the off-diagonal terms will decay as usual, the diagonal ones will not be affected by the environment. A set of measurements of the observable $D$ will provide an outcome distribution proportional to $|\alpha_D|^2$ and $|\beta_D|^2$, that is, the correct information about the initial state.

The above discussion shows that the two `crucial conditions' presented in Section~\ref{sec:essence_of_decoherence} are indeed sufficient to produce decoherence, regardless of the commutation properties of the interaction Hamiltonian. The latter, instead, are essential in defining what observables can be reliably measured when the interaction with an environment cannot be neglected.

The issue of reliable measurement records can be well understood with the help of the Von Neumann scheme. An experimentalist who wants to measure the observable $\hat A$ designs a device equipped with a pointer (whatever it may be) that is placed by the system-apparatus interaction Hamiltonian in the eigenstate $p_k$ of the pointer observable $\hat P$ (e.g. an angle, or a position) when the system is in the eigenstate $\ket{a_k}$ of $\hat A$. As in the example of Section~\ref{sec:von_neumann}, this is achieved by designing the apparatus so that the interaction Hamiltonian does commute with $\hat A$ and its pointer state is consistently correlated with the measured system's state.

The above scheme does not include the possible effect of an environment on the measurement process. There are two scenarios: (1) the apparatus interacts with the system before the system-environment interaction becomes relevant, (2) the environment decoheres the system's state before the measurement takes place.

The apparatus, considered as a whole, is always macroscopic. In fact, it actually includes the human observer's eyes (or other sensible organs) and neurons. It is thus impossible to neglect its interaction with an environment, either internal or external.
For this reason, in scenario (1) the (pre-)measurement process, before the eventual state vector reduction, proceeds in two steps:
\begin{eqnarray}
& \left(\alpha_0 \ket{+} +\beta_0 \ket{-} \right) \ket{p_0} \ket{\varepsilon} \Longrightarrow \left(\alpha_0 \ket{+} \ket{p_+} +\beta_0 \ket{-} \ket{p_-} \right) \ket{\varepsilon} \\ \nonumber
& \Longrightarrow \alpha_0 \ket{+} \left( a_1 \ket{p_+} \ket{\varepsilon_{1}} + a_2 \ket{p_-} \ket{\varepsilon_{2}} \right) +\beta_0 \ket{-} \left( a_3 \ket{p_+} \ket{\varepsilon_{3}} + a_4 \ket{p_-} \ket{\varepsilon_{4}} \right) .
\label{eq:spin_measurement_env_no_commutation}
\end{eqnarray}
The state $\ket{p_0}$ is the apparatus `ready' pointer state, while $\ket{p_+}$ and $\ket{p_-}$ are the pointer states that, in the Von Neumann scheme, correspond to the system states $\ket{+}$ and $\ket{-}$ respectively. As it is evident from the second step in Eq.~(4.49),
the environment interaction can in principle destroy the correlation between apparatus pointer eigenstates and system's states. If, for example, the initial state of the spin-1/2 system is $\ket{+}$, the pointer will ultimately end up in an entangled state with the environment: $a_1 \ket{p_+} \ket{\varepsilon_{1}} + a_2 \ket{p_-} \ket{\varepsilon_{2}}$. Even after an eventual state vector reduction, it is not clear what would be the relation of the final pointer state to the spin system state.
As it has been discussed above, this happens if the environment-apparatus interaction Hamiltonian, $\hat H_{\varepsilon A}$, does not commute with the pointer observable $\hat P$.

Eq.~(4.49) shows clearly the need to include the environment interaction commutation properties in the apparatus design. If $[\hat H_{\varepsilon A}, \hat P] = 0$ then the Von Neumann measurement scheme remains valid:
\begin{eqnarray}
& \left(\alpha_0 \ket{+} +\beta_0 \ket{-} \right) \ket{p_0} \ket{\varepsilon} \Longrightarrow \left(\alpha_0 \ket{+} \ket{p_+} +\beta_0 \ket{-} \ket{p_-} \right) \ket{\varepsilon} \\ \nonumber
& \Longrightarrow \alpha_0 \ket{+} \ket{p_+} \ket{\varepsilon_{p_+}} + \beta_0 \ket{-} \ket{p_-} \ket{\varepsilon_{p_-}} .
\label{eq:spin_measurement_env_2steps_robust}
\end{eqnarray}
From the above formulas it is evident that, if the spin-1/2 system is initially in an eigenstate of $\hat S_z$, the pointer will be in the corresponding eigenstate of $\hat P$: the e-e-link is preserved.

An extremely interesting feature of the above measurement process is that the environment itself acts as a reliable measurement apparatus. In fact, given the initial state $\ket{\varepsilon}$, the environment will always become entangled with the same $\ket{\varepsilon_{p_+}}$ and $\ket{\varepsilon_{p_-}}$ relative states. It thus satisfies the e-e-link requirement.
Such a fact, however, should not deceive. While apparatus pointer states are by definition human-readable, different environment states do not present themselves with a clearly identifiable label.

Eq.~(4.50) suggests a general guideline to design a measurement apparatus: once the system-apparatus interaction has been suitably chosen to satisfy the e-e-link, one only needs to make sure that $[\hat H_{\varepsilon A}, \hat P] = 0$. If such commutation relation holds then $\hat H_{\varepsilon A}$ can be expressed as a linear combination of the projectors on the $\hat P$ eigenvectors:
\begin{equation}
\hat H_{\varepsilon A} = \sum_k h_k \ket{p_k}\bra{p_k} ,
    \label{H_int_projectors}
\end{equation}
where the $h_k$'s are real numbers.
The states $\ket{p_k}$ are called \emph{robust} states exactly because they are not affected by the environment interaction.

The two-step (pre-)measurement process sheds also light on the fact that largely different instruments, designed to measure different observables, have the same type of interaction with the environment: this is possible because of the independence of the system-apparatus interaction from the apparatus-environment one.
In all cases where the pointer is an actual indicator of position on a scale, the pointer-environment interaction commutes with pointer position operator, as in the decoherence by scattering example described in Section~\ref{sec:decoherence_mechanism}.

Since for any $\hat H_{\varepsilon A}$ there is an eigenbasis which is dynamically diagonalized by the decoherence mechanism, as shown in Eq.~(\ref{eq:spin_env_reduced_rho_matrix_generic_H_I}), it can be said that it is the environment which selects a preferred basis.

Such environment induced preferred basis is particularly relevant for the understanding of scenario~(2) above: the environment decoheres the system's state before the measurement takes place.
In this case the two-step measurement process can be schematically represented as:
\begin{eqnarray}
& \left(\alpha_0 \ket{+} +\beta_0 \ket{-} \right) \ket{p_0} \ket{\varepsilon} \Longrightarrow \left(\alpha_0 \ket{+} \ket{\varepsilon_+} +\beta_0 \ket{-} \ket{\varepsilon_-} \right) \ket{p_0}  \\ \nonumber
& \Longrightarrow \alpha_0 \ket{+} \ket{p_+} \ket{\varepsilon_{+}} + \beta_0 \ket{-} \ket{p_-} \ket{\varepsilon_{-}} .
\label{eq:spin_env_measurement}
\end{eqnarray}
Here it has been implicitly assumed that the system-environment interaction, $\hat H_I$, commutes with $\hat S_z$. If that were not the case, the spin-1/2 superposition coefficients would have been modified before the measurement had taken place.
This is actually what happens in many naturally occurring physical systems.

For example, when studying atoms in a gas, regardless of which observable is measured, the outcomes statistical distribution usually agrees with what one would obtain if the atoms states were a classical mixture of various energy eigenstates. This can be understood by observing that often the atom-environment interaction commutes approximately with the atom self-Hamiltonian. Regardless of the `initial' state of an atom, such an interaction rapidly diagonalizes the reduced density matrix in the energy eigenbasis thus transforming it in a mixture.
When a set of measurements is performed on a sample of such atoms, the corresponding statistics is found.

On the contrary, for macroscopic objects, or even for dust particles, the environment interaction Hamiltonian usually (approximately) commutes with the center of mass position operator. It is thus not surprising that such systems are usually found in position eigenstates.

\section{Implications for different interpretations}
\label{sec:implications_interpretations}

In Chapter~\ref{ch:interpretations} some connections between the decoherence mechanism and each particular interpretation have been already addressed.
The purpose of the present section is to present the mutual implications in a more comparative manner.

Decoherence theory attempts to account for the emergence of classical physics from the quantum mechanical formalism, without relying on some particular interpretation's specific postulates.
As already explained in Section~\ref{subsec:general_decoherence_approach} the central mathematical operation in the decoherence formalism is the trace over the unobserved degrees of freedom. As long as the chosen interpretation provides a meaning for such operation, that is, it connects the trace operation to the statistical average over the measurement outcomes, the decoherence programme can be carried on. Such a condition consists in having the Born rule as part of the theory, either as a postulate or as a theorem.

Collapse theories, such as Copenhagen and GRW, and hidden variables, such as de Broglie-Bohm, provide the necessary framework for the decoherence theory. In hidden variables and Copenhagen type theories, Born rule is postulated, while in the GRW theory, it is derived.

As discussed in Section~\ref{sec:relative_state_interpretation}, relative state interpretations do not offer a straightforward way to derive the Born rule and thus in such theories the applicability of the decoherence results is less clear.
In this regard a promising proposal by Zurek \cite{zurek_probabilities_2005} will be presented in the following section.

The basic result of decoherence theory consists in the prediction of a classical statistical distribution of outcomes for quantum systems which, isolated, are supposed to produce a non-classical statistics. 
From this point of view, every quantum mechanics interpretation, that aims to satisfactorily account for the empirical observations, should provide the means for the implementation of the decoherence programme.

Within the framework of relative states theories, the loss of relative phase coherence is a particularly important phenomenon because it suggests a reason for the apparent lack of interfence between different worlds or minds.

A central aspect of quantum mechanics interpretations is the role of the observer. In this regard the decoherence programme is perfectly applicable to all of them. In fact, the process which leads from a pure to a mixed quantum state does not depend on the presence of an observer. Viceversa, the presence of a human observer immediately introduces a not-completely-controlled macroscopic system in the measurement process, thus providing the sufficient condition, a large interacting environment, for the emergence of classical properties which often appear in spite of the quantum mechanical nature of the systems.

In Section~\ref{decoherence_time} the time dependence of the off-diagonal elements of the reduced density matrix has been investigated. The factors affecting the decoherence time are the strength of the system-environment interaction and the size of the environment Hilbert space. Though extremely fast, the loss of coherence occurs in a finite time which, putting special care on the experimental setup, can be extended so as to allow the study of the pure quantum state. The experiments of Section~\ref{sec:macroscopic_superpositions} are examples of such control of the environment perturbation.

The ability of controlling decoherence opens up the possibility of putting to the test some of the interpretations' assumptions. It has already been shown in Sections~\ref{sec:macroscopic_superpositions}~and~\ref{Copenhagen_interpretation} how Bohr's view of two autonomous physical domains, a quantum and a classical one, does not stand the test of experiments on macroscopic superpositions.
Objective collapse theories, as discussed in Section~\ref{objective_collapse_theories}, allow, in principle, the design of experiments to test the wave function collapse by studying situations where the expected decoherence time is longer than the theory's localization time. 

Decoherence theory apparently poses no problem to the de Broglie-Bohm theory, which, on the contrary, provides a perfectly self-contained framework for the implementation of the decoherence programme. Such a situation is due to the realist assumption regarding particles positions. From such an assumption immediately follows the meaning of the trace operation, $\mbox{Tr}(\hat A \rho)$, as statistical average over an ensemble of particles. This, as a consequence, gives a clear and straightforward meaning to the reduced density matrix.

With regard to relative state theories it is not easy to identify experiments aimed at falsification, both because of the probably non-falsifiable nature of the many-worlds (minds) postulate, and because of the still unclear status of probabilities within the theory. The latter makes it difficult to derive predictions based purely on a relative state theory, without adding probabilities from the outside.
Assuming that probabilistic outcomes can be explained within the framework of such theories, a particularly relevant result of decoherence is the emergence of environment selected robust pointer states.

As explained in Section~\ref{sec:robust_pointer_states}, the commutation properties of the apparatus-environment interaction Hamiltonian dynamically selects a basis whose vectors remain unaffected by the environment interaction. Expressed in such a basis, the state of the universe acquires the factorized form:
\begin{equation}
\ket{\Psi} = \sum_k \ket{\chi_S(k)}\ket{\chi_O(k)}\ket{\chi_\epsilon(k)} , 
\end{equation}
where the terms of the sum have the following crucial properties: (1) they represent a definite state for the system and for the observer, (2) are approximately orthogonal and thus (3) do not interfere with each other. This latter feature may be chosen as a criterion to solve the pointer basis ambiguity, typical of relative state theories as explained in Section~\ref{sec:relative_state_interpretation}.

Contrary to some claims \cite{schlosshauer_decoherence_2005,zurek_decoherence_2003}, the pointer basis ambiguity has no consequence on the measurement process in other interpretations. In fact, such a problem is certainly not present in Bohmian mechanics and in stochastic collapse theories, thanks to real localization and suitable apparatus design.

If, for example, the initial state of system-observer-environment is a completely generic superposition such as:
\begin{equation}
\ket{\Psi} = \sum_{ijk} \ket{\chi_S(i)}\ket{\chi_O(j)}\ket{\chi_\epsilon(k)} , 
\end{equation}
the measurement removes any ambiguity, either because there was already no underlying ambiguity, i.e. in Bohmian mechanics, or because the macroscopic apparatus-observer wave function very rapidly localizes around a definite position.

In Copenhagen style interpretations too there is no pointer basis ambiguity, for two reasons:
\begin{enumerate}
\item either the apparatus can only record the value of one observable by design: e.g. the fluorescent screen in a $z$-direction Stern-Gerlach setup. In this case the spots on the screen will always be along the z-axis due to the spatial evolution of the wave function,
\item or, if the apparatus is such that it can record the value of more than one observable (like the two levels atom of the 1982 paper by Zurek \cite{zurek_environment_induced_1982}), then, according to the collapse postulate and the e-e-link intrinsic in the apparatus design, the pointer will indicate $S_x =+1$ or $S_z=+1$ depending on whether the spin is in an eigenstate of $\hat S_x$ or $\hat S_z$ (or $\hat S_y$...).

Such an apparatus would be analogous to a dial with a pointer with four positions (or more): two vertical ones and two horizontal ones. When the pointer is ``up'', not only we do know that the state of the spin is $\ket{S_z=+1}$ , but we also know that the spin is in a well defined superposition of $\ket{S_x=+1}$  and $\ket{S_x=-1}$. There is nothing wrong or strange with such a state of affairs in the context of the Copenhagen interpretation.
\end{enumerate}

The fundamental importance of a preferred pointer basis, selected by the interaction with the environment, lies in the stability of its basis vector. Knowledge of the selection mechanism provides essential information on how to design an apparatus so that a stable measurement record can be obtained. Alternatively, it explains why reliable measurement devices do exist at all, despite the continuous random perturbation by the environment.

\subsection{Environment assisted invariance and the Born rule}

In order to be applicable to relative state theories, the decoherence formalism needs an independent derivation of the Born rule. In Section~\ref{sec:relative_state_interpretation} one recent attempt by 
Saunders~\cite{saunders_derivation_2004} and Wallace \cite{wallace_everettian_2003} has been introduced.
A more recent proposal is the one by Zurek \cite{zurek_decoherence_2003, zurek_probabilities_2005}, which has been well outlined by Schlosshauer in his 2005 review article \cite{schlosshauer_decoherence_2005}.

Zurek's proposal, being based purely on properties of entangled states, integrates nicely within the decoherence programme.
The derivation of the Born rule is based on a particular symmetry of some entangled states under a class of transformations and on four almost self-evident assumptions.
Being the Born rule strictly related to measurement, and measurement to macroscopic devices, in such context it makes sense to take into account quantum systems entanglement with the ubiquitous environment.
From such observation follows the term `\emph{environment assisted invariance}' for the type of symmetry which is going to be presented.

If $\ket{\Psi_{S\varepsilon}}$ is the collective state of a system $S$ and an environment $\varepsilon$, then $\ket{\Psi_{S\varepsilon}}$ is said to be `\emph{envariant}' under the unitary operator $\hat U_S$, which acts only on the state of $S$, if and only if there exists a unitary operator $\hat U_{\varepsilon}$, which acts only on the state of the environment $\varepsilon$, such that:
\begin{equation}
\hat U_{\varepsilon}\hat U_{S}\ket{\Psi_{S\varepsilon}} = \ket{\Psi_{S\varepsilon}} .
    \label{eq:envariance}
\end{equation}
More precisely, $\hat U_{S} = \hat u_S \otimes \hat I_\epsilon $ and $\hat U_{S} = \hat I_s \otimes \hat u_\epsilon $, where $\hat I_\epsilon$ and $\hat I_S$ are the identity operators on the Hilbert spaces $\cal{H}_S$ and $\cal{H}_\varepsilon$ respectively.

Clearly envariance can be present only in entangled states and thus it is a uniquely quantum property. In fact, the change produced on a quantum state by an envariant transformation which acts on a subsystem can be undone by a corresponding transformation which acts on the complementary subsystem, without the need for the two subsystems to interact.

Considering the familiar spin-1/2 system with an environment, it is possible to examine some important example of envariance. A typical entangled state is given by:
\begin{equation}
\ket{\Psi_{S\varepsilon}} = \frac{1}{\sqrt{2}}\left( e^{i\phi_1}\ket{+}\ket{\epsilon_+} + e^{i\phi_2}\ket{-}\ket{\epsilon_-} \right) ,
    \label{eq:envariance_entangled_state}
\end{equation}
where the states $\ket{\epsilon_+}$ and $\ket{\epsilon_-}$ are exactly orthogonal (i.e. they do not necessarily correspond to the environment relative states of the decoherence examples).

Two simple envariant transformations for the state in Eq.~(\ref{eq:envariance_entangled_state}) are the \emph{phase transformation} and the \emph{swap transformation}.
The phase transformation is given by an operator $\hat u_S$ defined as follows:
\begin{equation}
\hat \Phi_{S}(\theta_1,\theta_2) = e^{i\theta_1}\ket{+}\bra{+} + e^{i\theta_2}\ket{-}\bra{-} .
    \label{eq:envariance_phase_transformation}
\end{equation}
Such a transformation is ``undone'' by a $\hat u_\varepsilon$:
\begin{equation}
\hat \Phi_\varepsilon(\theta_1,\theta_2) = e^{-i\theta_1}\ket{\epsilon_+}\bra{\epsilon_+} + e^{-i\theta_2}\ket{\epsilon_-}\bra{\epsilon_-} .
    \label{eq:envariance_phase_transformation_undone}
\end{equation}
The swap transformation is defined as:
\begin{equation}
\hat X_{S}(\theta_1,\theta_2) = e^{i\theta_1}\ket{+}\bra{-} + e^{i\theta_2}\ket{-}\bra{+} ,
    \label{eq:envariance_swap_transformation}
\end{equation}
and its change on $\ket{\Psi_{S\varepsilon}}$ is reversed by:
\begin{equation}
\hat X_{\varepsilon}(\theta_1,\theta_2) = e^{-i\theta_1}\ket{\epsilon_+}\bra{\epsilon_-} + e^{-i\theta_2}\ket{\epsilon_-}\bra{\epsilon_+} .
    \label{eq:envariance_swap_transformation_undone}
\end{equation}
It is necessary to stress that a unitary operator $\hat U_S$ may be an envariant transformation on a particular state $\ket{\Psi_{S\varepsilon}}$, but not on another state $\ket{\Phi_{S\varepsilon}}$.

The Born rule can be deduced from the following four basic assumptions \cite{zurek_probabilities_2005,schlosshauer_decoherence_2005, schlosshauer_zurek_2005}, through the application of the phase and swap transformations to the entangled state of Eq.~(\ref{eq:envariance_entangled_state}):
\begin{enumerate}[(i)]
    \item The state of system $S$ is not changed by a unitary transformation of the form $\hat I_S \otimes \hat u_\varepsilon$.
    \item Information on all measurable properties of $S$, including outcome probabilities, is completely described by the state of $S$.
    \item The state of $S$ is completely specified by $\ket{\Psi_{S\varepsilon}}$.
    \item The product states $\ket{+}\ket{\epsilon_+}$ and $\ket{-}\ket{\epsilon_-}$, in the state vector expansion, imply a perfect correlation between the outcomes of measurements performed on $S$ and on $\varepsilon$.
\end{enumerate}
For simplicity, the whole argument, including the above assumptions, is being presented for the bidimensional Hilbert space of a spin-1/2 system, but everything can be extended to arbitrary quantum systems.

Some consideration about the assumptions is necessary. First, probabilities of measurement outcomes are considered a property of the system's state. This is a highly non-trivial, if reasonable assumption.
Second, Assumption (iv) implies that if a measurement of the observable associated to $\hat O_S = (\lambda_1 \ket{+}\bra{+} + \lambda_2 \ket{-}\bra{-})$ finds system $S$ in state $\ket{+}$, then a measurement of $\hat O_\varepsilon = (\gamma_1 \ket{\epsilon_+}\bra{\epsilon_+} + \gamma_2 \ket{\epsilon_-}\bra{\epsilon_-})$ immediately afterwards will find system $\varepsilon$ in state $\ket{\epsilon_+}$ with certainty.

Zurek's strategy is to first prove that, when measuring $S_z$ on the state $\ket{\Psi_{S\varepsilon}}$ of Eq.~(\ref{eq:envariance_entangled_state}), the probability, $p(\ket{+},\ket{\Psi_{S\varepsilon}})$, of having outcome $+1$ ($S$ in $\ket{+}$) is equal to the probability of having outcome $-1$ ($S$ in $\ket{-}$).
To this end, thanks to the envariance of $\ket{\Psi_{S\varepsilon}}$ under swap transformations $\hat X_{S}(\theta_1,\theta_2)$, it holds that:
\begin{equation}
p(\ket{+},\ket{\Psi_{S\varepsilon}}) =  p(\ket{+},\hat X_{\varepsilon}\hat X_{S}(\theta_1,\theta_2)\ket{\Psi_{S\varepsilon}}) .
    \label{eq:envariance_Born_rule_1}
\end{equation}
The operator $\hat X_{\varepsilon}(\theta_1,\theta_2)$ does not affect the properties of system $S$ (Assumption (i)) and thus the following equality holds:
\begin{equation}
p(\ket{+},\hat X_{\varepsilon}\hat X_{S}(\theta_1,\theta_2)\ket{\Psi_{S\varepsilon}}) =  p(\ket{+},\hat X_{S}(\theta_1,\theta_2)\ket{\Psi_{S\varepsilon}}).
    \label{eq:envariance_Born_rule_2}
\end{equation}
From Assumption (iv), taking into account the effect of the swap transformation, it follows that:
\begin{equation}
p(\ket{+},\hat X_{S}(\theta_1,\theta_2)\ket{\Psi_{S\varepsilon}}) = p(\ket{\epsilon_-},\hat X_{S}(\theta_1,\theta_2)\ket{\Psi_{S\varepsilon}}).
    \label{eq:envariance_Born_rule_3}
\end{equation}
Again, Assumption (i) implies that a swap transformation on $S$ does not affect the probabilities regarding system $\varepsilon$:
\begin{equation}
p(\ket{\epsilon_-},\hat X_{S}(\theta_1,\theta_2)\ket{\Psi_{S\varepsilon}}) = p(\ket{\epsilon_-},\ket{\Psi_{S\varepsilon}}).
    \label{eq:envariance_Born_rule_4}
\end{equation}
Finally, from Assumption (iv) and for transitivity of the identity relations up to Eq.~(\ref{eq:envariance_Born_rule_1}), one obtains:
\begin{equation}
p(\ket{\epsilon_-},\ket{\Psi_{S\varepsilon}}) = p(\ket{-},\ket{\Psi_{S\varepsilon}}) = p(\ket{+},\ket{\Psi_{S\varepsilon}}).
    \label{eq:envariance_Born_rule_5}
\end{equation}
For superposition states with coefficients of equal magnitude, the probabilities of the different outcomes are equal.

Following Zurek's derivation, the case of superpositions with unequal coefficients, of rational square modulus, can be addressed as above by means of a counting method. The general case with non-rational relative states weights is derived by using a continuity argument \cite{zurek_probabilities_2005}.

With respect to other approaches to the derivation of the Born rule, Zurek's proposal may be considered more appealing for two reasons: it is based on simmetry considerations and environment entanglement.
This makes it more independent from observer's subjectivity, contrary to the case of Saunders~\cite{saunders_derivation_2004} and Wallace~\cite{wallace_everettian_2003} proposals, and almost completely contained within the quantum mechanical basic framework.

Clearly the envariance based derivation of the Born rule is particularly relevant for relative state theories. By providing a way to introduce probabilities in an overall deterministic framework, it allows to implement the whole decoherence programme within Everett-type theories.
Zurek also shows that it is possible to demonstrate the emergence of robust pointer states from envariance independently from the decoherence approach \cite{zurek_decoherence_2003, zurek_relative_2007}.

Nevertheless, the status of probabilities within a complete and purely deterministic description of physical phenomena, such as that implied by a relative state theory, remains somewhat puzzling.

\section{Limitations of the decoherence programme}

By now it should be clear that, if the term `decoherence programme' is meant to indicate the (generally successful) attempt to account for the emergence of classical physics from the formalism of quantum mechanics, then such a programme does not appear to have intrinsic limitations.

It is however necessary to bear in mind that, currently, all implementations of the decoherence programme rely heavily on the application of the Born rule. The latter one cannot be derived from decoherence arguments, at least to avoid circularity.
The Born rule is crucial in order to relate the reduced density matrix formalism to the statistics of experimental outcomes. 

As it emerges from the analysis in Sections~\ref{sec:decoherence_mechanism}~and~\ref{sec:essence_of_decoherence} however, from a technical standpoint, what produces the loss of coherence in the subsystem state is the orthogonalization of the environment relative states. Moreover, the physical process which brings the entangled environment states to rapidly become almost orthogonal does not depend on the reduction postulate, nor on the Born rule. This fact suggests that the Born rule is perhaps not fundamental for the decoherence process, but it is clearly necessary to account for the probabilistic nature of measurement outcomes.

\subsection{Common criticism}

The most common criticism to the decoherence approach consists in its ``for all practical purposes'', non-fundamental, nature.

The argument goes as follows: the decoherence mechanism explains the quick disappearance of observable interference in macroscopic objects, but (1) the overall composite system state remains a pure superposition, and (2) the theory predicts the occasional reappearance of observable interference.
These two issues can actually be shown to be false problems.

The first one is sometimes wrongly attributed to J.S.~Bell \cite{omnes_interpretation_1994}, who, instead, acknowledged the relevance of the decoherence mechanism for deriving a classical picture of the macroscopic world \cite{bell_1975}, while correctly pointing out that the problem of the state vector reduction remains.

The fact that the state of the composite system remains a superposition seems particularly unconvenient in the context of relative state theories. In this case decoherence does not prevent two different worlds to interfere with each other: an event which may appear too loaded of ontological implications to be comfortable with.
However, as long as one maintains that aim of scientific investigation is to provide a consistent, and possibly predictive, description of what is empirically observed, then arguments which appeal to unobservable ``strange'' phenomena cannot be upheld as crucial for correctness.

With regard to the possibility of observing the effects of the latent superposition, still present in the state vector of the composite system after decoherence of the subsystem to be measured, J.S.~Bell showed that in principle it is always possible to find a suitable observable, actually an hermitean operator, for which the effects of coherence are still noticeable \cite{bell_1975}.
Typically, an observer measures some property of the specific system she is interested in, neglecting to observe the other interacting subsystems, the so called environment. The operator associated to the observable of interest does not act on the Hilbert subspace of the other degrees of freedom and the reduced density matrix formalism can be applied to obtain the familiar fast suppression of the off-diagonal terms. If, however, one chooses to measure a property of a larger subsystem, say system of interest, apparatus pointer, and part of the environment, then the unobserved degrees of freedom are fewer and the suppression rate of the off-diagonal terms is lower.

Bell shows that, at fixed time intervals, it is possible, in principle, to choose larger and larger observables, i.e. hermitean operators which depend on more and more degrees of freedom, so that at each time the effects of the initial superposition are still noticeable. In Bell's words: ``\emph{While for any given observable one can find a time for which the unwanted interference is as small as you like, for any given time you can find an observable for which it is as big as you do {\bf not} like.}''

With regard to Bell's observations two remarks are in order. First of all, the fact that quantum theory predicts the possibility of observing macroscopic superpositions, even after long times, should not be considered a shortcoming of the theory. At worst it represents one more opportunity to experimentally test the predictions of quantum mechanics. If macroscopic systems superpositions were to be found to persist for long times, there would be no argument about bizarre predictions that could hold against empirical observation.

The second remark is due to R.~Omn\`es \cite{omnes_interpretation_1994} who shows that it is actually not possible to measure every theoretically definable observable. The first reason is that an actual measurement requires a suitable interaction between system and apparatus, but not all imaginable interactions can be implemented in a measurement instrument. The second and main reason relies in Omn\`es demonstration that the size of the apparatus, necessary to measure properties of a larger and larger system as in Bell's argument, grows exponentially with the number of degrees of freedom of the system to be observed. If the environment, in which global phase coherence is still preserved, is constituted by the macroscopic apparatus components, the number of its degrees of freedom is of the order of $10^{27}$. In this case, according to Omn\`es estimation, the ``super-apparatus'' necessary to measure the apparatus should be constituted by a number of particles of the order of $10$ to the power $10^{18}$. As the author remarks, such number is larger than the number of atoms within the visible horizon and, in any case, it is unclear how an instrument this large could work given the limitations posed by the theory of relativity.

Finally, the second common criticism to the decoherence programme lies in the presence of a coherence recurrence time. It has already been shown, in Section~\ref{sec:essence_of_decoherence}, that such an issue is completely analogous to the Poincar\`e recurrence times in classical mechanics. As such it is an issue only for those who consider problematic the reversibility of classical mechanics in spite of the empirical validity of the second law of thermodynamics.

\subsection{The problem of the state vector reduction remains}
\label{sec:problem_remains}

It has been already stressed that the appearance of definite outcomes in single measurements does not follow from the decoherence mechanism, but, in light of the many contradictory or misleading statements present in the relevant literature, the author feels that such a point is never highlighted enough.
In fact, up until the first half of the 2000's, it was not uncommon to find such statements as: ``\emph{...In particular `reduction of the wave packet', postulated by Von Neumann to explain definiteness of an outcome of an individual observation, can be explained when a realistic model of an apparatus is adopted}'' \cite{zurek_environment_induced_1982}; or ``\emph{...the word `decoherence' which describes the process that used to be called `collapse of the wave function'...}'' \cite{p.w._anderson_science:_2001}.

Some of the authors, expressing the view that the decoherence programme has somehow solved the definite outcomes problem, actually base their position on sophisticated assumptions which, unfortunately, they often fail to expound. Some others, such as for example W.~Zurek, later on acknowledged the incorrectness of their statement and clarified which particular interpretation they were adopting, even proposing new solutions \cite{zurek_decoherence_2003}.

Given the apparent confusion about the issue on the one hand, and the competence of the authors on the other hand, the following words of J.S.~Bell \cite{bell_1975} are more appropriate then ever to clarify the crux of the problem: ``\emph{The continuing dispute about quantum measurement theory is not between people who disagree on the results of simple mathematical manipulations. Nor is it between people with different ideas about the actual practicality of measuring arbitrarily complicated observables. It is between people who view with different degrees of concern or complacency the following fact: so long as the wave packet reduction is an essential component, and so long as we do not know exactly when and how it takes over from the Schr\"odinger equation, we do not have an exact and unambiguous formulation of our most fundamental physical theory.}'' 

Since the author of this thesis belongs to the group of people who view with some concern, and without any complacency, the above fact, some more clarifications will be presented here.

Following S.~Adler's discussion of the issue \cite{adler_why_2003}, it can be said that, starting from a state of the form $\alpha \ket{0}_S \ket{0}_A \ket{a}_\epsilon + \beta \ket{1}_S \ket{1}_A \ket{a}_\epsilon$, the decoherence mechanism explains the transition to the state:
\begin{equation}
\alpha \ket{0}_S \ket{0}_A \ket{a^{\prime}}_\epsilon + \beta \ket{1}_S \ket{1}_A \ket{a^{\prime\prime}}_\epsilon , \hspace{0.5 cm} \mbox{with} \hspace{0.5 cm} \langle_\epsilon a^{\prime} \mid a^{\prime\prime} \rangle_\epsilon \simeq 0 ,
    \label{eq:decoherence_collapse_problem_remains}
\end{equation}
while the definite outcomes problem requires explaining the transition to either the state
\begin{equation}
\ket{0}_S \ket{0}_A \ket{a^{\prime}}_\epsilon ,
    \label{eq:collapse_problem_1}
\end{equation}
or to the state
\begin{equation}
\ket{1}_S \ket{1}_A \ket{a^{\prime\prime}}_\epsilon .
    \label{eq:collapse_problem_2}
\end{equation}
Even more generally, for the above two-state example, the difference between the apparatus state after a measurement, as predicted by the decoherence theory, and as it actually occurs, consists in the difference between the following reduced density matrices: (1) fully decohered ($t \rightarrow \infty$) apparatus pointer state,
\begin{equation}
\rho_{dec} = \left(
\begin{array}{cc}
    |\alpha|^2 & 0 \\
    0 & |\beta|^2
\end{array}
\right) ,
    \label{eq:decohered_density_matrix}
\end{equation}
and (2) apparatus state after an actual measurement, 
\begin{equation}
\rho_{m} = \left(
\begin{array}{cc}
    1 & 0 \\
    0 & 0
\end{array}
\right) , \hspace{0.5 cm} \mbox{or} \hspace{0.5 cm} \rho_{m} = \left(
\begin{array}{cc}
    0 & 0 \\
    0 & 1
\end{array}
\right) .
    \label{eq:density_matrix_definite_outcome}
\end{equation}
It is perfectly clear that the decoherence programme, by itself, does not provide any mechanism to explain the transition to the states of Eq.~(\ref{eq:density_matrix_definite_outcome}).

In concluding this discussion it seems fitting to mention a work by Jona-Lasinio and Claverie \cite{giovanni_jona-lasinio_symmetry_1986}, which is sometimes incorrectly presented as an example of how decoherence produces localization starting from a superposition of two localized states \cite{omnes_interpretation_1994, joos_decoherence_2003}.

The two authors concentrate their attention on systems that present a double-well symmetric potential, such as, for example, the case of the nitrogen atom in $\mbox{NH}_3$, and address the question of why this type of molecules are found in a chiral configuration, even though the two lowest energy eigenstates are symmetric and anti-symmetric with respect to the double-well potential.

The issue is dealt with by observing that a very small perturbation of the symmetric potential, near the central maximum, is enough to modify the two lowest energy eigenstates, making them well localized inside one of the wells \cite{jona-lasinio_new_1981, jona-lasinio_semiclassical_1981}.
Specifically, the ratio of the wave function values at the two minima is estimated as:
\begin{equation}
\frac{\psi_0(m_l)}{\psi_0(m_r)} \approx \frac{\psi_1(m_l)}{\psi_1(m_r)} \approx \exp{\left\{-\frac{1}{\hbar}\int_{a_1}^{a_2} \sqrt{2mV_0(x)} dx \right\}} ,
    \label{eq:jona-lasinio_ratio}
\end{equation}
with $V_0(x)$ the double-well potential, $m_l$ and $m_r$ the left and right minima, $m$ the mass of the atom, and $a_1$ and $a_2$ the two points within which the small perturbing potential is different from zero.
The energy gap between the first excited state and the ground state is also found to increase exponentially with respect to the unperturbed situation.

A remarkable aspect of the above findings is the robustness of such a localization: it is almost independent from the strength of the perturbation and is quite stable due to the large energy gap with the first excited state.

Assuming that the conditions, required by the above mechanism to work, are actually realized in nature, then 
Jona-Lasinio and Claverie proposal accounts, for the special systems for which it is applicable, exactly for the transition from a coherent superposition of two orthogonal states to only one of the two, as in Eq.~(\ref{eq:density_matrix_definite_outcome}).

Now, such a proposal has very little to do with the decoherence mechanism, if not for the fact that both depend on the presence of a disordered environment.
The question of why certain molecules are normally found in chiral states, instead of the energy lowest eigenstates, can certainly be addressed by means of the decoherence theory, through the environment selected basis states, but such an approach, if successful, can only lead to a situation of the type described by Eq.~(\ref{eq:decohered_density_matrix}).

In this regard, it is pertinent to notice that the authors of the above mentioned work consider the example of the chiral molecules a particular case of a more general phenomenon of symmetry breaking. In the decoherence theory no symmetry breaking is ever involved.

\chapter{Alternative approaches}
\label{ch:future}

In Chapter~\ref{ch:interpretations} the most commonly adopted interpretations have been presented. All of them assume that, in order to account for the experience of definite outcomes, it is necessary to add some postulate to the `minimal', unitarily evolving, quantum mechanics. 
Even relative state theories, which actually maintain the unitary evolution at all times, require the addition of a heavy ontological framework.

Obviously, there is nothing wrong in adding postulates, as long as the theory does not conflict with experimental findings. From this point of view all consistent interpretations are equally legitimate. 
However, a theory, which could satisfactorily account for the definite outcomes problem, perhaps in a `For All Practical Purposes' manner, without the need of ad-hoc ingredients, would have the advantage of greater simplicity and of appearing more natural.

In the present chapter possible approaches to a `real', FAPP, collapse of the wave function will be introduced and their limitations discussed.

\section{Environment induced collapse}

The proof that, starting from a state as $\alpha \ket{0}_S \ket{r}_A \ket{a}_\epsilon + \beta \ket{1}_S \ket{r}_A \ket{a}_\epsilon$ (where $\ket{r}_A$ stands for the apparatus `ready' state), and within the unitary stage of the von Neumann measurement scheme, it is not possible to obtain one of the states in Eqs.~(\ref{eq:collapse_problem_1})~and~(\ref{eq:collapse_problem_2}), is straigthforward.
In fact, if $t_0$ and $t_1$ are the starting and ending measurement times respectively, and $\hat U(t_0, t_1)$ the time evolution operator during the measurement, the standard theoretical scheme requires:
\begin{eqnarray}
\hat U(t_0, t_1) \ket{0}_S \ket{r}_A \ket{a}_\epsilon & = & \ket{0}_S \ket{0}_A \ket{a^\prime}_\epsilon \hspace{0.5 cm} \mbox{and} \nonumber \\
\hat U(t_0, t_1) \ket{1}_S \ket{r}_A \ket{a}_\epsilon & = & \ket{1}_S \ket{1}_A \ket{a^{\prime\prime}}_\epsilon .
    \label{vN_measurement_scheme}
\end{eqnarray}
Due to the linearity of $\hat U(t_0, t_1)$, one obtains again the von Neumann measurement problem, regardless of the presence of the environment:
\begin{equation}
\hat U(t_0, t_1) \left( \alpha \ket{0}_S \ket{r}_A \ket{a}_\epsilon + \beta \ket{1}_S \ket{r}_A \ket{a}_\epsilon \right) = \alpha \ket{0}_S \ket{0}_A \ket{a^\prime}_\epsilon + \beta \ket{1}_S \ket{1}_A \ket{a^{\prime\prime}}_\epsilon .
    \label{vN_measurement_problem}
\end{equation}

Notwithstanding the correctness of the result, the relevance of the von Neumann scheme for actual measurement methods is quite questionable. In fact, most measurement settings present the following features:
\begin{itemize}
    \item the initial quantum state of the apparatus is not known exactly,
    \item there is a large number of different apparatus pointer states which are macroscopically indistinguishable by the observer,
    \item macroscopically different pointer states are not necessarily exactly orthogonal,
    \item due to the environment interactions, the perfect correlation between system and apparatus states may be unattainable.
\end{itemize}
One may argue that, once the above points are taken into account, the simple impossibility proof above cannot be obtained and a FAPP explanation for the occurrence of definite outcomes cannot be ruled out.

More general proofs have been given by removing some of the von Neumann requirements \cite{wigner_problem_1963, despagnat_conceptual_1971, fine_insolubility_1970}, such as, for example, that the system or the apparatus are in a pure quantum state, but the requirement that the apparatus states corresponding to different measured values be orthogonal was upheld. Bassi and Ghirardi \cite{bassi_general_2000}, and Gr\"ubl \cite{grubl_quantum_2003} have given much more general proofs that will be presented in next section.

Based on the above four points, a very general measurement scheme may be formulated.
The first step is to consider the basic essential requirements for a measurement apparatus to perform its function:
\begin{itemize}
    \item Every time the system is in the same particular eigenstate $\ket{o_k}$ of the observable to be measured, the apparatus, upon measurement, will present the same macroscopic pointer position.
    \item At different eigenstates $\ket{o_i}$ correspond different, macroscopically distinguishable, pointer positions.
\end{itemize}
There is a whole set of apparatus states that correspond to a macroscopic pointer position associated to a certain eigenvalue $o_k$. Even more generally, there is a set of states of the system-apparatus-environment composite system which correspond to macroscopic apparatus configurations which are indistinguishable for the human observer. Such set of composite system states is associated to the same outcome: the eigenvalue $o_k$.

The second requirement, obvious if the apparatus has to be useful at all, implies that the quantum states of system-apparatus-environment, corresponding to different measurement outcomes, have to be nearly orthogonal. In fact, the wave functions of two naked-eye-distinguishable pointer positions are necessarily peaked around the two positions respectively and are very small elsewhere, thus overlapping very little. Viceversa, if the two wave functions were to overlap significantly, it would be impossible to unambiguously distinguish the two pointer positions.

It is important to notice that it is not required that the measurement process leaves the observed system in an eigenstate of the measured observable. In fact, in many practical cases this does not happen and it is debatable whether, in real measurement processes, it is rigorously possible at all.

In order to more precisely formulate such a tolerant measurement scheme, it is convenient to consider the familiar spin-1/2 system.

The $S_z = +1$ and $S_z = -1$ eigenstates are indicated by $\ket{\uparrow}$ and $\ket{\downarrow}$ respectively. A state of the apparatus-environment composite system, corresponding to the macroscopic ready position for the pointer, is represented by a vector $\ket{r_i, \vec{\alpha}_i}$, where $r_i$ is a collective variable associated to the pointer position and $\vec{\alpha}_i$ are associated to the environment (all else) degrees of freedom. The index $i$ labels the different microscopic pure states corresponding to the same macroscopic `ready' apparatus configuration.

The spin-apparatus-environment states corresponding to the apparatus reading `spin up' are represented by vectors $\ket{s_{Ui}, U_i, \vec{\alpha}_{Ui}}$, while for the `spin down' reading the state vectors are $\ket{s_{Di}, D_i, \vec{\alpha}_{Di}}$. The variables $s_{Ui}$ and $s_{Di}$ represent the spin degree of freedom.
As before, there are more than one microscopic pure states associated to the same measurement outcome

With the above definitions the first essential requirement for a general measurement scheme corresponds to the following transitions:
\begin{eqnarray}
\ket{\uparrow} \otimes \ket{r_i, \vec{\alpha}_i} & \Longrightarrow & \ket{s_{Ui}, U_i, \vec{\alpha}_{Ui}} \nonumber \\
\ket{\downarrow} \otimes \ket{r_i, \vec{\alpha}_i} & \Longrightarrow & \ket{s_{Di}, D_i, \vec{\alpha}_{Di}} .
    \label{eq:generalized_measurement_scheme}
\end{eqnarray}
It is important to notice that, in principle, the final states don't need to be factorized. After the measurement, not only the spin can end up in a superposition state, but the whole system can be in a completely entangled state: as long as, starting from the spin up state $\ket{\uparrow}$ (spin down state $\ket{\downarrow}$), such a state belongs to the set $V_U$ ($V_D$) of states which provide the macroscopic `pointer up' reading (`pointer down'), the apparatus works as it should.

Within such a general measurement scheme, it is not obvious that, starting from a spin superposition state such as $\alpha \ket{\uparrow} + \beta \ket{\downarrow}$, it is impossible that the composite system evolves unitarily towards a state belonging either to the set $V_U$ or to $V_D$.

The above considerations provide support to some authors suggestion \cite{jr._can_1997, joos_decoherence_2003} that it might be possible, or at least worth investigating, an environment induced, effective, wave function collapse. In particular, as suggested by O.~Pessoa \cite{jr._can_1997}, the effective (FAPP) collapse could be associated to localization due to the random interactions with the environment particles. 
Such an approach shares two aspects with the GRW theory: the random collapse and the interpretation of the wave function as a real entity. The difference here lies in the non-spontaneous localization of the wave function, which is instead determined by interaction with the environment. The probabilistic aspect of the localization is related to our ignorance of the detailed state of the environment particles. The overall evolution remains fundamentally deterministic.

With regard to the definiteness of the measurement outcomes, two already introduced experimental observations support the hypothesis that they are strictly related to the localizing effect of many small random interactions, typical of an environment microscopic components.
The mentioned experiments are the one by Tanaka et al. \cite{tanaka_dc-squid_2002, takayanagi_readout_2002} on single shot measurement of a macroscopic current superposition (Section~\ref{sec:macroscopic_superposition_realization}), and the experiments on chiral molecules polarization properties examined by Jona-Lasinio and Claverie \cite{giovanni_jona-lasinio_symmetry_1986} (Section~\ref{sec:problem_remains}).
The first one demonstrates that a macroscopic system \emph{can} be found in a non classical state with a single measurement, without causing its collapse. The trick is that what the human observer actually sees is the well localized pointer (whatever it is) of the measurement apparatus that, under those specific experimental conditions, are theoretically known to correspond to the current superposition state. 

In Tanaka et al. experiment the system-apparatus interaction can be tuned such that, starting from a current superposition state, if the coupling is small, the von Neumann measurement scheme predicts a superposition pointer state, while for strong coupling it predicts a definite pointer state.
As a matter of fact, in the weak coupling regime the pointer is found to be well localized (as usual) and consequently the system macroscopic current acquires a definite value, even though it was initially in a superposition state.
On the contrary, in the strong coupling regime the pointer localization does not destroy the current superposition. All this suggests that the localization of macroscopic objects might be at the origin of the emergence of definite outcomes.
In this regard it is important to notice that, even though an apparatus pointer is not always constituted by a mechanical arm, the measurement reading by a human is always mediated by some macroscopic object.

Jona-Lasinio and Claverie explanation for atom localization in certain symmetric molecules may hint at a general mechanism which might be applicable to pointer localization as well. In fact, the nitrogen atom in the $\mbox{NH}_3$ molecule can be thought of as the pointer of a measurement apparatus. An isolated molecule can present the pointer in a stable up-down superposition. In this case it is actually the most stable state. It is the small random perturbations by the environment that make the pointer-atom localized in an up or down position, thus breaking the underlying symmetry of the apparatus.

All the above considerations, by themselves, are clearly only interesting ideas which may guide towards the search for an alternative explanation of the definite outcomes. A precise mechanism for the apparent collapse of the wave function should be found and rigorously described.

Unfortunately, as it will be proved in the next section, the interaction with an environment alone cannot explain the occurrence of definite measurement outcomes, not even in a FAPP manner and by means of a tolerant measurement scheme as described above.

Nevertheless the ideas presented above remain still relevant for any theory which involves an effective wave function collapse, be it a stochastic collapse as in the GRW style theories or gravity induced collapse as in proposals that will be presented in Section~\ref{sec:gravity_induced_collapse}.

\section{Impossibility theorems}

In a little known 2000 paper \cite{bassi_general_2000} Bassi and Ghirardi proved that, as long as the state vector evolution is linear, it is not possible to account for the experience of definite outcomes, even within a very general measurement scheme such as the one described in the previous section.

The crucial condition for the proof lies in the second of the two basic requirements for a meaningful measurement scheme: to different eigenstates $\ket{o_i}$ of the system to be measured correspond different, macroscopically distinguishable, pointer positions. This implies the near orthogonality of the system-apparatus-environment states which produce different pointer positions. Such condition, again in the case of a spin-1/2 system, can be conveniently expressed with the following inequality:
\begin{equation}
\inf \parallel \ket{s_U,U,\alpha_U} - \ket{s_D,D,\alpha_D} \parallel \ge \sqrt{2}-\epsilon \hspace{1 cm} \epsilon \ll 1 ,
\label{eq:non_ambiguity_condition}
\end{equation}
for any $\ket{s_U,U,\alpha_U}$ belonging to the set $V_U$ of states associated to the `up' macroscopic pointer position and any $\ket{s_D,D,\alpha_D}$ belonging to the set $V_D$ of states associated to the `down' position.

For $\epsilon = 0$ the case of strictly orthogonal states is obtained. Such a strict requirement is unnecessary, since two perceptively distinct macroscopic configurations may correspond to wave functions with a small but non-zero overlap. On the other hand it is necessary that $\epsilon \ll 1$, both for avoiding measurement reading ambiguity and for empirical consistency, since in actual experiments the pointer does not appear fuzzy.

If one prepares the spin system in the state $1/\sqrt{2} (\ket{\uparrow}+\ket{\downarrow})$, the generalized scheme of Eq.~(\ref{eq:generalized_measurement_scheme}) and the linearity of the time evolution operator $\hat U(t_0,t_1)$ will give:
\begin{eqnarray}
& & \ket{\Psi_{bf}} = 1/\sqrt{2} (\ket{\uparrow}+\ket{\downarrow}) \otimes \ket{r_k, \vec{\alpha}_k} \Longrightarrow \nonumber \\
& & \Longrightarrow \ket{\Psi_{aft}} = \hat U(t_0,t_1) \ket{\Psi_{bf}}  \nonumber \\
& & = \frac{1}{\sqrt{2}}\left( \ket{s_{Uk}, U_k,\alpha_{U_k}} + \ket{s_{Dk}, D_k,\alpha_{Dk}} \right) .
\label{eq:generalized_superposition_measurement}
\end{eqnarray}

If, due to the random environment interactions, the state $\ket{\Psi_{aft}}$ of Eq.~(\ref{eq:generalized_superposition_measurement}) happens to produce a perceptively definite pointer position, say `up', then such a state belongs to the set $V_U$. As such it has to satisfy the non-ambiguity condition of Eq.~(\ref{eq:non_ambiguity_condition}):
\begin{equation}
\inf \parallel \ket{\Psi_{aft}} - \ket{s,D,\alpha} \parallel \ge \sqrt{2}-\epsilon \hspace{1 cm} \epsilon \ll 1 .
\label{eq:non_ambiguity_condition_2}
\end{equation}
The above inequality has to hold for any state $\ket{s,D,\alpha} \in V_D$, thus also for $\ket{s_{Dk}, D_k,\alpha_{Dk}}$. Computing the distance between the two states one obtains:
\begin{eqnarray}
& & \parallel \ket{\Psi_{aft}} - \ket{s_{Dk}, D_k,\alpha_{Dk}} \parallel \nonumber \\
& & = \parallel \frac{1}{\sqrt{2}}\left( \ket{s_{Uk}, U_k,\alpha_{U_k}} + \ket{s_{Dk}, D_k,\alpha_{Dk}} \right) - \ket{s_{Dk}, D_k,\alpha_{Dk}} \parallel \nonumber \\
& & = \parallel \frac{1}{\sqrt{2}} \ket{s_{Uk}, U_k,\alpha_{U_k}} - (1-\frac{1}{\sqrt{2}}) \ket{s_{Dk}, D_k,\alpha_{Dk}} \parallel \le \frac{1}{\sqrt{2}} + 1 - \frac{1}{\sqrt{2}} = 1 \nonumber \\
& & 
\label{eq:inequalities}
\end{eqnarray}
Putting together the results of Eqs.~(\ref{eq:non_ambiguity_condition_2})~and~(\ref{eq:inequalities}) one obtains the following contradiction:
\begin{equation}
1 \ge \parallel \ket{\Psi_{aft}} - \ket{s_{Dk}, D_k,\alpha_{Dk}} \parallel \ge \sqrt{2}-\epsilon .
\label{eq:impossibility_contradiction}
\end{equation}

It is clear that the crucial ingredient of the proof is the linearity of $\hat U(t_0,t_1)$.
Remarkably, the result not only implies $\ket{\Psi_{aft}} \notin V_U$, but also that the final state does not correspond to any perceptively definite pointer position. In fact, if $\ket{\Psi_{aft}} \in V_X$, with $X$ some definite pointer position different from `up' and `down', the non-ambiguity condition with $\ket{s_{Dk}, D_k,\alpha_{Dk}}$ should still be required and the contradiction of Eq.~(\ref{eq:impossibility_contradiction}) would result anyway.

The authors also stress that the above proof is not affected by the additional hypothesis of occasionally faulty apparatus behaviour, since, for an acceptable measurement instrument, it is necessary that the erroneous measurement events be a very small percentage of the total. Attributing the apparent state vector reduction to the random misbehaviour of the apparatus would require a totally unreliable measurement instrument!

To make the impossibility proof even more general, Gr\"ubl extended it \cite{grubl_quantum_2003} to the case of impure initial states both for the microsystem to be measured and for the apparatus-environment system. This addresses possible objections related to the fact that it may not be possible, or at least it may be very difficult, to prepare the microsystem in such a way that the initial state of the overall composite system is strictly factorized, i.e. completely not entangled. Consequently, in most cases the initial microsystem on the one hand, and the apparatus-environment on the other, are not in a pure state and are more accurately described by impure reduced density matrices.

The above ab-absurdo proofs show in the most general way that, no matter how complex and sophisticated, it is not possible to construct a consistent measurement scheme that accounts for the experienced definite outcomes without introducing some non-linear term in the time evolution operator.
In light of this result, even an open-system approach as suggested by O.~Pessoa \cite{jr._can_1997},
as long as it presents a linear, though not unitary, evolution, would not work.

Finally, it is important to be aware that the above impossibility proofs, do not apply to the theories presented in Chapter~\ref{ch:interpretations}. In fact, the standard interpretation and the objective collapse theories explicitely break unitarity. On the contrary, in hidden variable theories unitary evolution is preserved, but it is not necessary to associate nearly orthogonal states to different pointer positions. The second hypothesis is thus not satisfied and the theorem does not follow.
Relative state theories do not require state vector reduction, either strict or effective, and consequently are not affected by the impossibility theorems.

\section{Gravity induced collapse}
\label{sec:gravity_induced_collapse}

In recent years an alternative approach to localization and definite outcomes has started to be considered more seriously: a gravity induced collapse. Such an approach clearly suffers from the fact that a satisfactory theory that unifies general relativity and quantum mechanics is still lacking. Consequently, any proposal about the effects of gravity on the quantum evolution can only be based on educated hypothesis and reasonable approximations.

Nevertheless, exploring the possibility that the localization properties of macroscopic systems are due to the interaction with the gravitational field or, equivalently, to symmetry properties of space-time, is worthwhile for at least two reasons: it would still constitute a solution within the known laws of physics, and it would be in principle experimentally testable.

Before introducing the main proposals, it is essential to notice that the previous section's result provides a clear-cut test for the plausibility of a physical collapse theory: the presence of some non-linear modification to the unitary Schr\"odinger dynamics. A gravity effect which does not introduce a non-linearity falls into the range of validity of Bassi and Ghirardi negative proof.

One of the first works on a gravity induced collapse is due to Di\`osi \cite{diosi_models_1989} who, inspired by an earlier work by K\'arolyh\'azy on the possible role of gravity, developed a theory analogous to the GRW objective collapse, by replacing the \emph{universal position localization} term with a \emph{universal density localization} term. Whereas a GRW-style theory predicts random transitions according to the following rule:
\begin{equation}
\Psi \Longrightarrow e^{-\frac{1}{2} \alpha \sum_n (\hat q_n - \overline q_n)^2 } \Psi ,
    \label{eq:grw_style_upl}
\end{equation}
where $\overline q_n$ are the random localization positions, Di\`osi proposes that the wave function undergoes a localization as follows: 
\begin{equation}
\Psi \Longrightarrow e^{-\frac{1}{2} \alpha \parallel \hat f - \overline f \parallel^2_G } \Psi ,
    \label{eq:diosi_udl}
\end{equation}
where $\hat f(\vec{r})$ is the mass density operator, $\overline f(\vec{r})$ is a random density function, and 
\begin{equation}
\parallel f \parallel^2_G = G \int \int \frac{f(\vec{r_1})f(\vec{r_2})}{\vec{r_{12}}} d\vec{r_1}d\vec{r_2} .
\label{eq:diosi_udl_norm}
\end{equation}
The effect of gravity enters the theory through the gravitational potential energy dependence given by Eq.~(\ref{eq:diosi_udl_norm}). The euristic idea behind such a model is that, if gravity induces wave function collapse, it is the mass density the quantity that undergoes localization, not directly the particle's position. The latter becomes localized as a consequence.

The author shows \cite{diosi_models_1989} that the above collapse model produces very rapid localization of spatial superpositions and classical trajectories for massive objects.

Di\`osi model explicitely breaks the wave function unitary evolution and is not affected by Bassi and Ghirardi's impossibility result.

More recently Penrose \cite{penrose_gravitys_1996} developed arguments in support of gravity based localization theories. In the authors' own words his considerations do not \emph{give any clear indications of the mathematical nature of the theory that would be required to incorporate a plausible gravitationally induced spontaneous state-vector reduction}. Nevertheless such considerations can provide the basis for constructing a phenomenological theory, such as the one by Di\`osi. 

The main concepts in Penrose examination of gravity's role in quantum state reduction are the instability of superposition of spatially separated states, and the dependence of such instability on the mass density.

The analysis starts from considering that a superposition of two localized states of a particle implies also a superposition of the gravitational field state, assuming a quantum description of gravity is possible. Thus the state vector of the particle-field system would be:
\begin{equation}
\ket{\Psi} = \lambda \ket{\psi} \ket{G_{\psi}} + \mu \ket{\chi} \ket{G_{\chi}} ,
    \label{eq:penrose_superposition}
\end{equation}
where $\ket{\psi}$ and $\ket{\chi}$ represent the localized states of the particle, while $\ket{G_{\psi}}$ and $\ket{G_{\chi}}$ represent the corresponding states of the gravitational field (still assuming a quantum theory of gravity).
For a state to be stationary it has to be an eigenstate of the time-translation operator $\hat T$, and thus also an eigenstate of the total energy operator. In the case of non-entangled states as $\ket{\psi} \ket{G_{\psi}}$ and $\ket{\chi} \ket{G_{\chi}}$, the energy of the particle-field system is well defined. In the case of an entangled state, such as $\ket{\Psi}$ in Eq.~(\ref{eq:penrose_superposition}), the ``\emph{superposition of two different space-times}'' makes the notion of a time-translation operator ill-defined. From this Penrose argues that the entangled state is unstable and should have a lifetime inversely proportional to the energy uncertainty of the state:
\begin{equation}
\tau \simeq \frac{\hbar}{\Delta E_{\Psi}} .
    \label{eq:penrose_halftime}
\end{equation}
The uncertainty on the energy, $\Delta E_{\Psi}$, is estimated as:
\begin{equation}
-4 \pi G \int \int \frac{(\rho_1(\vec{r}) - \rho_2(\vec{r})) (\rho_1(\vec{r^\prime}) - \rho_2(\vec{r^\prime}))}{|\vec{r}-\vec{r^\prime}|} d\vec{r} d\vec{r^\prime} ,
    \label{eq:penrose_uncertainty}
\end{equation}
with $\rho_1(\vec{r})$ and $\rho_2(\vec{r})$ the mass densities of the two localized states in the superposition.
Except for the $- 4 \pi$ factor, the expression in Eq.~(\ref{eq:penrose_uncertainty}) is formally identical to Eq.~(\ref{eq:diosi_udl_norm}).

The general idea behind the above considerations is that, starting from an unstable superposition of two or more localized states, the state vector will evolve towards a stable well localized state with a well defined time-translation operator and such a transition will occur on average in a time $\tau$ given by Eq.~(\ref{eq:penrose_halftime}). No precise theory of how such a process can occur is given and the above arguments present several weaknesses \cite{gao_diosi-penrose_2010}. Penrose is however able to provide some estimations for the lifetime of superposition of separated spatial locations, depending on the object's mass: few million years for a proton, the order of an hour for a water nano-droplet, about a millionth of a second for a $10\mu\mbox{m}$ water droplet. Such estimations indicate at which mass and time scales deviations from the linear Schr\"odinger dynamics could be observed and can serve as the basis for wave function collapse experiment design.

The proposed experiments by Marshall et al. \cite{marshall_towards_2003}, and by van Wezel and Oosterkamp \cite{van_wezel_nanoscale_2012} follow exactly from the above considerations. Both experimental proposals aim at testing possible deviations from the unitary evolution by controlling, and thus filtering out, the decoherence process in mesoscopic systems.

The proposal by Marshall et al. \cite{marshall_towards_2003} consists in entangling a micromechanical oscillator to a photon by means of a tiny mirror attached to one end of the oscillator. This is achieved by means of a Michelson interferometer with high finesse cavity in each arm and the mirror-oscillator at the end of one of the two arms. Such a set-up was used by O'Connell et al. \cite{oconnell_quantum_2010} for their experiment presented in Section~\ref{quantum_state_control}.

The central idea of the experiment is a negative test on the occurrence of a gravity induced collapse. The mirror and the photon (actually the quantized electromagnetic field) are initially in a disentangled state, with the photon in the superposition state $\ket{0}_A\ket{1}_B+\ket{1}_A\ket{0}_B$ ($A$ and $B$ being the two arms of the interferometer). The system set-up is such that the mirror-photon state becomes entangled, but, if decoherence is kept out, the evolution is periodic and the state periodically returns to its initial factorized form.
The return to the disentangled state can be verified by observing the reappearance of photon interference.
If such an interference revival is observed it means not only that decoherence has been kept under control, but also that in the meantime no collapse has occurred. In this sense the experiment constitutes a negative test.

Bassi et al. \cite{bassi_towards_2005} have analyzed and discussed the above proposal and found that, while it can provide a useful test for the Penrose estimations, it is not suitable for testing objective collapse models such as for example GRW theory, at least within the experimental parameters values currently achievable.

Van Wezel and Oosterkamp \cite{van_wezel_nanoscale_2012} propose a different experimental set-up that aims at discriminating between different collapse models. The experimental configuration consists of a tiny pendulum, constituted by a thin gold plate attached to a nanowire, interacting with a single electron spin through the coupling with small spherical magnet. The authors show how such a configuration can be realized by means of recent experimental achievements. Regardless of the apparatus details, the main idea is to propose an \emph{experimental protocol}, composed of two stages, to distinguish the contribution to the decay of interference terms due to decoherence from the one due to some genuine collapse mechanism.

An interesting part of the work is the simulation of the time evolution of the system in the hypothesis of a gravity induced collapse. Following Di\`osi \cite{diosi_models_1989} and Penrose \cite{penrose_gravitys_1996}, it is suggested that the effect of gravity on quantum dynamics is to introduce a very small non-linearity which in turn, in the thermodynamic limit, can produce a spontaneous breaking of time-translation symmetry \cite{van_wezel_broken_2010}.
Instead of modelling gravity influence with some non-linear term, as in Eq.~(\ref{eq:diosi_udl}), the authors propose a modification of the Schr\"odinger equation with a randomly fluctuating term that should account for the ill-definedness of \emph{any quantity that can be used as a measure of distance between locations in different components of a space-time superposition}, according to general relativity. The modified dynamics of the quantum system is given by:
\begin{equation}
\frac{\partial}{\partial t} \psi = -\frac{i}{\hbar} \left[ \hat H -iG \frac{m^2}{2L^3} (\hat x - \xi)^2 \right] \psi ,
    \label{eq:van_wezel_schrodinger}
\end{equation}
where $m$ is the mass of the oscillating plate and $L$ its width.
An interesting aspect of the above equation is that, while it produces an non-unitary evolution due to the anti-hermitean additional term, it is still linear. It would appear that the modified quantum dynamics of Eq.~(\ref{eq:van_wezel_schrodinger}) could still fall in the domain of applicability of the Bassi and Ghirardi's impossibility argument. However this is not the case due to the random time dependence of the fluctuating variable $\xi$ which, starting from the same initial state, and during a time interval $\Delta t$, produces different final states depending on the initial time $t_0$. This prevents the derivation of the inequalities in Eq.~(\ref{eq:inequalities}) and the final contradiction of Eq.~(\ref{eq:impossibility_contradiction}).

In concluding this short presentation of gravity induced collapse proposals it is worthwhile to stress a central aspect shared with objective collapse theories and even with hidden variable theories: the localization approach to the measurement problem. The idea that the definite outcomes problem ultimately reduces to the localization properties of macroscopic systems constitutes perhaps the main difference with respect to other approaches such as different versions of the standard interpretation and the various relative state theories.

\chapter{Conclusion}

The aims of this thesis have been stated in the Introduction as: (1) to clarify the relation between decoherence and definite outcomes, (2) to dispel common misconceptions on the measurement problem, and (3) to present recent alternative approaches to the issue of state vector reduction.

In regard to the first aim, formalism, mechanism, and time dependence of decoherence have been examined separately (Chapter~\ref{ch:decoherence}). It has been shown unequivocally that the current formalism, based on reduced density matrices, requires an independent account of the Born rule. It follows that state vector reduction is a pre-condition for the interpretation of results based on such a formalism.
On the other hand, orthogonalization of environment states, which is the process that produces decoherence, does not depend on the Born rule: decoherence can occur without state vector reduction, but its effects are observed through the statistics of the reduced states of the measured system.

The distinct nature of decoherence and state vector reduction has been further stressed by showing that wave function collapse can occur before decoherence takes place (Section~\ref{sec:facts_and_misconceptions}).

Particular attention has been devoted to the analysis of recent experiments aimed at controlling decoherence and at realizing superposition states in nearly macroscopic systems (Section~\ref{sec:macroscopic_superpositions}). The results show no evidence for a possible break up of the laws of quantum physics and for the existence of a distinct classical domain as in Bohr's view.

The experiments analysis served also to point out some misconceptions as in the second aim mentioned above. One of these consists in confusing loss of interference due to `which-way' detection by some component of the apparatus-environment system and genuine decoherence where localization does not necessarily occur before detection.
It has been noticed that such confusion occasionaly shows up even in excellent articles as for example the $\mbox{C}_{60}$ interference paper by Arndt et al. \cite{arndt_wave-particle_1999}.

A remarkable and little known experiment with SQUID devices \cite{tanaka_dc-squid_2002, takayanagi_readout_2002}, which constitutes \emph{the first direct observation of a macroscopic quantum superposition}, has been examined. The difference between direct, single measurement, observation and `indirect observation', reconstructed from an average value over a large number of measurements, has been stressed. The experiment by Tanaka et al. demonstrates that, under specific conditions, single measurements can be meaningful also for quantum systems and offers new possibilities for practical quantum computation and information.

In the context of a possible environment-induced effective wave function collapse, an incorrect interpretation of Jona-Lasinio and Claverie work \cite{giovanni_jona-lasinio_symmetry_1986} on localization in chiral molecules has been pointed out (Section~\ref{sec:problem_remains}). While that article is sometimes presented as an example of localization produced by decoherence, it is shown that the arguments and mechanism presented by the two authors are unrelated to the decoherence mechanism, except for the appeal to the random interaction with environment particles.

Another clarification concerns the relation between preferred basis and measurement problem (Section~\ref{sec:implications_interpretations}). Contrary to several statements in the scientific literature, it is shown that the freedom in the basis of the state vector expansion produces an ambiguity, in the measured observable, only in the framework of a relative state interpretation.
Given this fact, the specific relevance of an environment-selected preferred pointer basis is examined for each quantum mechanics interpretation.

Finally, in regard to the third aim, it has been discussed how a scheme could be formulated to try to account for the appearance of definite outcomes, by appealing only to the interaction with the many degrees of freedom of the environment and the unitary Schr\"odinger evolution (Chapter~\ref{ch:future}). Following the argument by Bassi and Ghirardi \cite{bassi_general_2000} it is proved that such a scheme cannot work and that some non-linear modification of the quantum dynamics is necessary for a physical collapse theory.

In light of this result, recent gravity collapse proposals have been presented and the common aspects highlighted. In spite of their still very approximate nature, such gravity collapse theories predict deviations from the unitary evolution that are in principle within reach of most recent experimental techniques.

In concluding, this thesis has tried to offer clear analytical arguments in support of a programme for a physics based approach to the definite outcomes problem. It is not claimed that such a programme is particularly promising by itself, but that it is worthwhile pursuing in contrast to the two common and opposite positions: that decoherence has solved the measurement problem, and that there is no solution within the already known laws of physics.


\bibliographystyle{unsrt}
\bibliography{decoherence_and_def_out}{}

\end{document}